%% file: hc_letter.tex
\documentclass[preprint,12pt]{emulateapj}

\shorttitle{Time-Series Spectroscopy of HD 98922 and V1295 Aql}
\shortauthors{Aarnio et al.}

\usepackage{longtable}
\usepackage{graphicx}
\usepackage{aas_macros}

\slugcomment{}
\accepted{20 August 2017}
\journalinfo{The Astrophysical Journal}

\begin{document}

\title{High Cadence, High Resolution Spectroscopic Observations of Herbig Stars HD 98922 and V1295 Aquila}

\author{Alicia N.\ Aarnio\altaffilmark{1}\altaffilmark{*}, John D.\ Monnier\altaffilmark{1}, Tim J. Harries\altaffilmark{2}, Stefan Kraus\altaffilmark{2}, Nuria Calvet\altaffilmark{1}, David Acreman\altaffilmark{2} \& Xiao Che\altaffilmark{1}}

\altaffiltext{1}{Department of Astronomy, University of Michigan, 311 West Hall, 1085 S. University Ave., Ann Arbor, MI 48109}
\altaffiltext{2}{Department of Physics and Astronomy, University of Exeter, Stocker Road, Exeter EX4 4QL, UK}
\altaffiltext{*}{Now a Hale Fellow at the University of Colorado Boulder, 8665 Discovery Drive, Boulder, CO 80303}

\begin{abstract}
Recent observational work has indicated mechanisms for accretion and outflow in Herbig Ae/Be 
star-disk systems may differ from magnetospheric accretion as it is thought to occur in 
T Tauri star-disk systems.
In this work, we assess the temporal evolution of spectral lines probing accretion and mass 
loss in Herbig Ae/Be systems and test for consistency with the magnetospheric accretion 
paradigm. For two Herbig Ae/Be stars, HD 98922 (B9e) and V1295 Aql (A2e), we 
have gathered multi-epoch ($\sim$years) and high cadence ($\sim$minutes) high resolution optical 
spectra to probe a wide range of kinematic processes. 
Employing a line equivalent width evolution correlation 
metric introduced here, we identify species co-evolving (indicative of common line origin) via novel visualization. 
We interferometrically constrain often problematically degenerate parameters, inclination and inner disk 
radius, allowing us to focus on the structure of the 
wind, magnetosphere, and inner gaseous disk in radiative transfer models. 
Over all timescales sampled, the strongest variability 
occurs within the blueshifted absorption components of the Balmer series lines; 
the strength of variability increases with the cadence of the observations. 
Finally, high 
resolution spectra allow us to probe substructure within the Balmer series' blueshifted absorption 
components: we observe static, low-velocity features and time-evolving features at higher velocities. 
Overall we find the 
observed line morphologies and variability are inconsistent with a scaled-up T Tauri magnetospheric accretion scenario. 
We suggest that as magnetic field structure and strength change dramatically with increasing stellar mass from
T Tauri to Herbig Ae/Be stars, so too may accretion and outflow processes.
\end{abstract}

\section{Introduction}\label{s_intro}

Herbig Ae/Be (HAeBe) stars are the high mass ($\sim$2-10 M$_{\odot}$) pre-main sequence 
counterparts of T Tauri stars (TTS). Classically defined, they are spectral type A or earlier, 
show emission lines in their spectra, lie in obscured regions, and illuminate surrounding nebulosity 
\citep{Herbig:1960}. Being similar in evolutionary state and observed phenomena, these systems have often
been treated as scaled-up TTS, magnetospherically accreting via strong, 
extended dipolar magnetic field components. Indeed, in some cases, magnetospheric accretion (MA) has 
been able to successfully reproduce observed spectral line profiles \citep{Muzerolle:2004}.
Recent spectropolarimetric surveys, however, have found a very low ($\sim$10\%) detection rate 
for magnetic fields on Herbig Ae/Be stars
\citep{Wade:2007,Alecian:2013a}, and those detected have weaker dipole components than observed 
in TTS ($\sim$100 G in HAeBe, as opposed to $\sim$1 kG for TTS). Indeed,
moving to higher mass in the Hertzsprung-Russell diagram, the dipole moment is seen 
to weaken, and higher order moments dominate the overall field structure \citep{Gregory:2012};
this effectively truncates the ``reach'' of the stellar magnetosphere in star-disk interaction 
and distributes the accretion footpoints about the stellar surface in a less-ordered manner. 

At the same time, it has been shown that the few HAeBe stars with magnetic field detections are 
slower rotators than their non-magnetic counterparts, indicating effective braking has taken place 
somehow \citep{Alecian:2013b}. In higher mass HAeBe systems ($>$5 M$_{\odot}$), \citet{Alecian:2013b} 
showed substantial angular momentum loss occurs as they contract to the main sequence; the authors 
suggest increased wind mass loss as the driver of this. 
\citet{Mendigutia:2011Nov} showed a clear break in the dependence of accretion indicators (Balmer 
excess, H$\alpha$ line width) as a function of stellar mass, with Ae stars resembling TTS and Be 
stars not. 
Spectrointerferometric results have indicated that for late-type HAeBe systems, Br$\gamma$ 
line emission is originating from within a few stellar radii of the star, consistent with a 
magnetospheric accretion paradigm \citep{Kraus:2008a,Kraus:2008b,Ellerbroek:2015}.
For earlier-type 
Herbigs, however, the origin of line emission is $\sim$1-6 au, consistent with a disk or X-wind \citep[e.g.,][]{Weigelt:2011,Kurosawa:2016}. 
Unfortunately, only a few systems have been observed via this technique.

While this is suggestive that later-type Herbigs accrete magnetospherically while earlier types 
accrete via different mechanisms \citep[supported by spectropolarimetric observations of][]{Vink:2002}, ambiguity remains: \citet{Kraus:2008b} observe Br$\gamma$ emission 
from within the dust sublimation radius in all of their targets regardless of spectral type, 
and these radii could be consistent with magnetospheric accretion, a stellar wind, or an X-wind. 
In only one case do \citet{Kraus:2008b} suggest the compact Br$\gamma$ emission line region, in 
concert with accretion indicators, to be likely indicating magnetospheric accretion: HD 98922.
The modeling work of \citet{Caratti:2015} is able to successfully reproduce the Br$\gamma$ emission as originating from a 
disk wind launched over $\sim$0.1-1 au and a mass loss rate of $2$x$10^{-7}$M$_{\odot}$ yr$^{-1}$.

These observations counter the scaled up T Tauri magnetospheric accretion paradigm of a strong, 
extended dipole interacting with circumstellar material, and in concert with a lack of magnetic 
field detections in these objects, indicate a need to expand upon our picture of star-disk 
interaction in HAeBe systems.

Observations done in the time domain, sampling a range of dynamical timescales across the system, 
are critical for disentangling episodic, periodic (linked to rotation), and 
large-scale (disk-driven) processes' effect 
on observables, especially given the degeneracy in line formation mechanisms.
Using high resolution, high signal-to-noise optical spectra, 
we compare the mass flux within two star-disk systems on both short and long timescales 
with the goal of determining whether relationships exist between accretion and outflow processes, 
on what timescales these happen, and whether the observations are consistent with the 
magnetospheric accretion paradigm as envisaged for lower mass pre-main sequence stars.

In Section \ref{s_obs} we describe our target selection and observations, and in Section \ref{s_analysis} we 
describe our analysis of the data. Sections \ref{s_res} and \ref{s_discn} share results 
and conclusions, and in Appendices \ref{s_appendixU}, \ref{s_appendixT}, and \ref{s_appendixS},
we present an atlas of our spectra, 
radiative transfer models, and further details of our correlation analysis methods.

\section{Target Selection and Observations}\label{s_obs}

From 2006 to 2013, we conducted an optical spectroscopic observational campaign of a sample of 
$\sim$60 HAeBe stars. Our target 
sample consists of stars for which we either already have interferometric data or targets that 
would be suitable for interferometric observation (i.e., bright; $\sim$$V<14$, $H<8$). These 
targets are ideal for modeling as the interferometric data break degeneracies 
that create ambiguity in interpretation
(e.g., inclination and inner disk radius). Over the course of gathering data, we observed 13 
objects at multiple epochs; two of these, V1295 Aql and HD 98922, were followed up on in 
2013 with high cadence time series observations. 
V1295 Aql and HD 98922 were chosen for follow up observation because they
occupy and extend into the range of stellar masses (2.9 and 5.0M$_{\odot}$, respectively)
  in which trends in mass accretion rate and line luminosity break from those observed in TTS \citep{Mendigutia:2011Nov}.
  Also of particular interest, these stars 
represent magnetic V1295 Aql and non-magnetic HD 98922 examples to compare and contrast.

Our optical spectra were taken with MIKE \citep{Bernstein:2003}, the \facility{Magellan} Inamori Kyocera Echelle, a cross-dispersed 
double \'{e}chelle spectrograph. The blue side of MIKE covers a $\sim$3200$-$5000\AA\ range, 
and the red chip spans $\sim$4900$-$10000\AA; using the narrowest slit, 0.''35, the resolving 
powers of the blue and red sides are R$\sim$80,000 and R$\sim$65,000, respectively. Within this 
wavelength range, we simultaneously capture many well-studied MA and outflow tracing lines, including 
the Balmer discontinuity, Hydrogen Balmer series lines, Na D doublet, He \textsc{I} 5876\AA\ line, Ca \textsc{II} infrared 
triplet, and [O \textsc{I}] 6300\AA\ line. 

\begin{figure}
\centering
\includegraphics[width=0.5\textwidth,clip,trim=2cm 0 1.5cm 0]{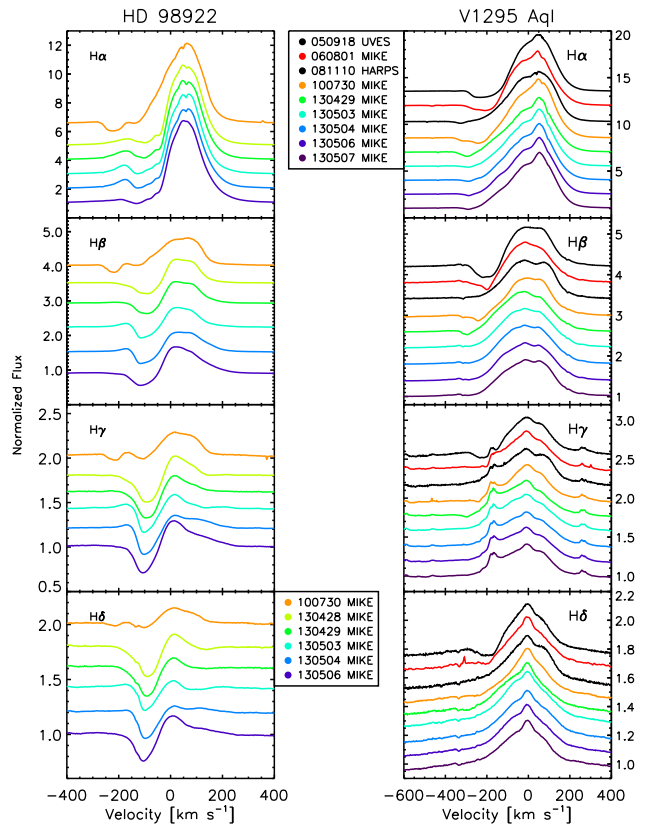}
\caption{Continuum-normalized spectra for HD 98922 and V1295 Aql: evolution of the Balmer series lines. 
Stellar photospheric absorption has been removed from the spectra. Mean spectra are shown in progressing date 
order from top to bottom; all available epochs from both our MIKE data (color-coded by date) and the literature
(black) plotted here with arbitrary offsets for display purposes. UT dates in the legends are in YYMMDD format.
V1295 Aquila's H$\gamma$ line shows contamination from four strong Fe I lines at $-$157, 1.2, 90, and 277 km s$^{-1}$.}
\label{f_balmer}
\end{figure}

For six nights in a ten-day period, we observed the targets in $\sim$1 
hour long visits, taking spectra at as high temporal resolution as possible while obtaining 
adequate signal. Our final data set is comprised of 111 blue and 132 red spectra for HD 98922, and 
75 blue and 95 red spectra for V1295 Aql. The cadence of these observations is $\sim$3 minutes for 
HD 98922 (read-out limited), and $\sim$4.5 minutes for V1295 Aql. The log of our observations is 
reported in Table \ref{t_timeseries}.
We show the mean profiles for each night, plus earlier epochs of observation with MIKE, in Figure 
\ref{f_balmer}.

\input{table_obslog_cropped.txt}

\section{Analysis}\label{s_analysis}

For the basic spectral reduction process, flat fielding, trimming and bias subtraction, 
background removal, and boxcar extraction, we employed the MIKE IDL pipeline 
\citep{Bernstein:2015}.
Post-pipeline tools written by us further removed the blaze and continuum, and merged the 
\'{e}chelle orders. Our blaze/continuum removal procedure fit a weighted surface to the
extracted \'{e}chelle orders to preserve broad line wings. 

To obtain radial velocities, we cross-correlated mean spectra from each night with synthetic 
templates from \citet{Munari:2005}. We used Fe \textsc{II}, Ti \textsc{II}, Cr \textsc{II},
and Mg \textsc{II} lines as described in \citet{Mora:2001}
to determine the $v$sin$i$ for each star. Finally, we removed stellar photospheric absorption 
by subtracting a continuum-normalized, high-resolution synthetic spectrum \citep{Munari:2005}.

\begin{figure} 
\centering
\includegraphics[width=0.236\textwidth,clip,trim=1.5cm 3cm 1.5cm 3cm]{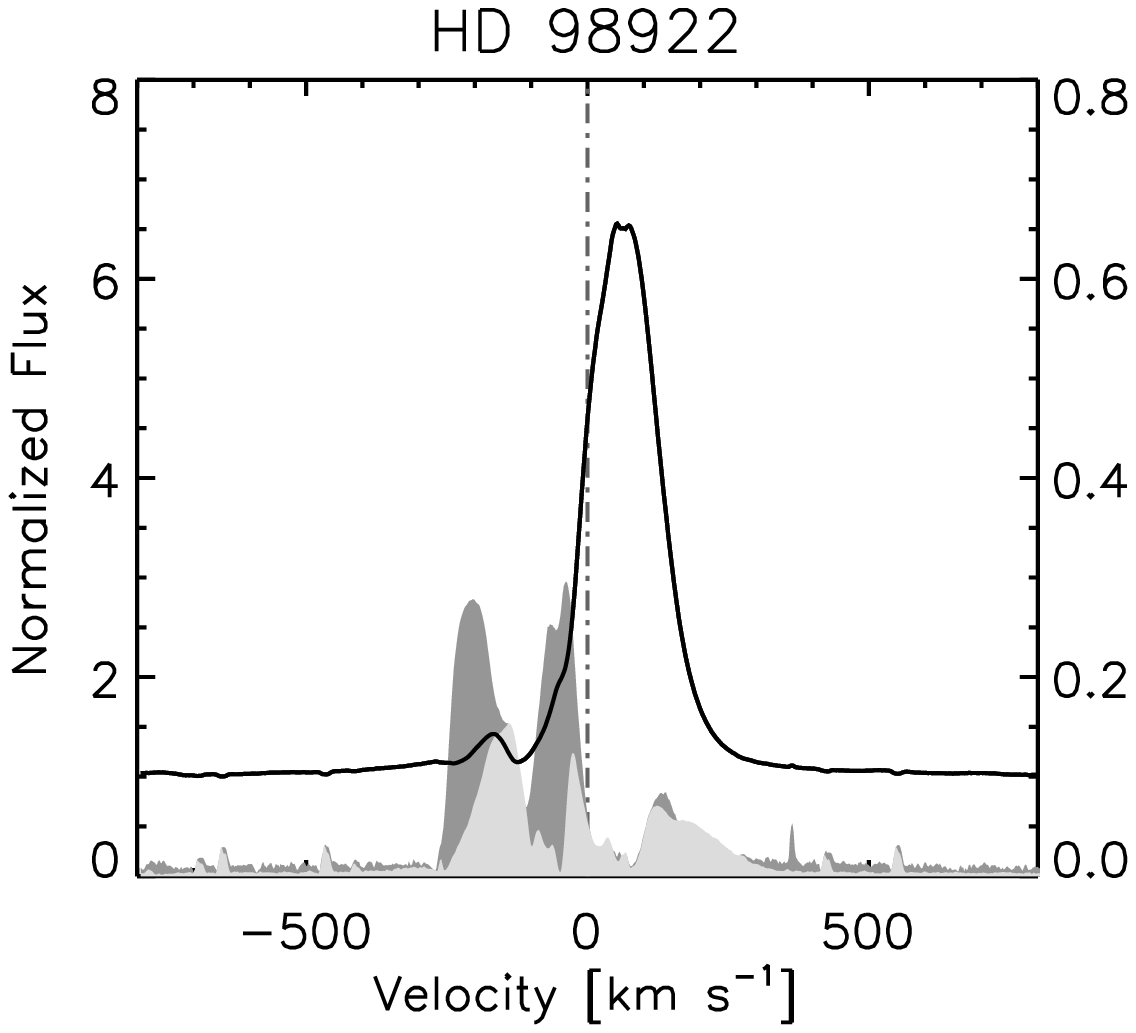}
\includegraphics[width=0.236\textwidth,clip,trim=2cm 3cm 1cm 3cm]{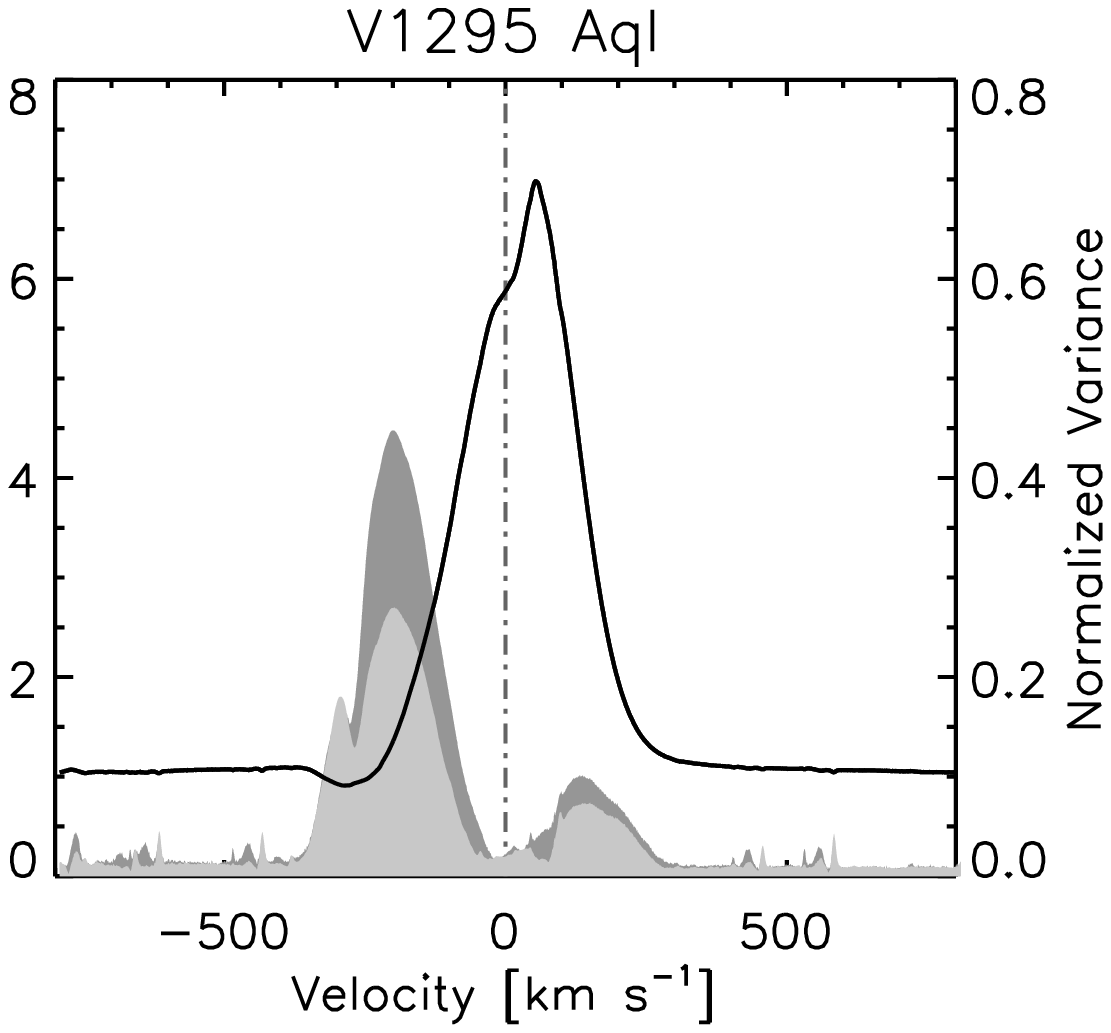}
\caption{H$\alpha$ variability: mean profiles (solid lines) and normalized variance profiles (filled) for 
HD 98922 (left) and V1295 Aql (right). Dark shaded regions are the variance profiles over all epochs of observation, 
while the lighter filled areas represent the variance profiles over our five nights of observation in 2013. 
The left axis of ordinates denotes the normalized flux of the mean H$\alpha$ profiles (solid, black lines), 
while the right axis of ordinates marks the variance profiles' (filled, solid) value.}
\label{f_allepochnvps}
\end{figure}

To quantitatively assess line variability, we calculated normalized variance profiles \citep{Johns:1995}, 
which measure the variance in a given line's velocity bins across all spectra observed. In 
Figure \ref{f_allepochnvps} we show the variance of HD 98922 and V1295 Aql's H$\alpha$ lines 
with respect to the mean line profile over all epochs of available data (spanning years) as well 
as with respect to the mean line profile across the entire week of observation in 2013.
In both cases, the variance is greater over the longer time baselines sampled. V1295 Aql
  is more strongly variable than HD 98922 over both short and long-term observing windows;
we note in Section \ref{s_res} physical 
changes observed in V1295 Aql over this timescale that could be responsible.

To assess contributions of distinct components, we employed PAN, Peak ANalysis\footnote{PAN was written by 
Rob Dimeo at the NIST Center for Neutron research and based upon activities supported by the National Science 
Foundation under Agreement No. DMR-0086210.}, to fit our line profiles with a series of Gaussian functions. 
In principle, a model with arbitrarily many components could fit the data well; to guard against overfitting
the line profiles, we calculated the F-statistic of each fit and did not add additional 
components if they did not statistically significantly improve the fit. 
In the case of H$\alpha$, the line complexity and strength made fits with relatively few components 
indistinguishable from one another; in these cases, we proceeded with as few components as possible to obtain 
a qualitatively good fit. As an arbitrarily large number of Gaussian components could
  certainly reduce the $\chi^{2}$ value of the fit, we rely on the F-statistic to ensure we are not overfitting 
and only refer to the emission and absorption components in general terms.
Within the PAN gui, in order to avoid unrealistic and spuriously large (though statistically equivalent)
  component fits, we set limits on the Gaussian components' widths, areas, and central wavelengths; in
  cases where components weakened to non-detection, these components' areas are allowed to go to 0. If
  any parameter reached the imposed limit, we re-expanded the allowed range and re-fit.
We present examples of fits to the H$\alpha$ profiles of HD 98922 and V1295 Aql in Appendix \ref{s_appendixS}.

Having fit our spectral lines, we then measured the equivalent widths (EWs) of the emission and wind absorption 
components of 26 species and calculate the correlation of their variability relative to one another:  
should we see strongly correlated or anti-correlated behaviors, it could be indicative of a common timescale for 
evolution of the line feature and thus potentially a common origin. 
The correlation coefficient we employ here is Spearman's $\rho$ \citep{Press:1992}; we note that this statistic 
gauges linear relationships between monotonically increasing quantities; this is not always the case with our 
data. When the data are significantly non-monotonic, however, the coefficient approaches zero and the two-sided 
significance is large (here, we use 10$^{-3}$ as a cutoff), meaning it is probable that unrelated variables could 
produce this level of correlation and thus it is not statistically significant. These results are shown in Figures 
\ref{f_v1295corrmap} and \ref{f_hd98922corrmap}.

Finally, we have run preliminary fiducial radiative transfer models using the TORUS code 
\citep[Transport Of Radiation Under Sobolev/Stokes; ][]{Harries:2000}. We employ a simple dipolar 
field geometry, a split-monopole style disk wind \citep[][]{Knigge:1995,Long:2002}, and calculate 
line transport in the Balmer series to compare to observations. We describe details of the parameter choices and model physics in Appendix \ref{s_appendixT} and discuss general findings in the following sections.

\section{Results}\label{s_res}

Over the broad wavelength range observed with MIKE, we have detected and measured the permitted transitions 
of H$\alpha$,
H$\beta$,
H$\gamma$,
H$\delta$,
He \textsc{I} 5875\AA,
Na \textsc{I} 5889\AA, 
Na \textsc{I} 5895\AA,
He \textsc{I} 6678\AA,
O \textsc{I} 8446\AA,
Ca \textsc{II} IR triplet (8498\AA, 8542\AA, 8662\AA),
O \textsc{I} 7772\AA,
Fe \textsc{II} (42) multiplet (4923\AA, 5018\AA, 5169\AA), and 
Ca \textsc{II} H and K (3933\AA, 3968\AA).
We have also detected and measured the forbidden transitions of 
[O \textsc{I}] 5577\AA,
[O \textsc{I}] 6300\AA,
[O \textsc{I}] 6363\AA,
[Ca \textsc{II}] 7323\AA,
[Fe \textsc{II}] 7155\AA. 
At MIKE's high resolution, we note a wealth of substructure in the blueshifted line absorption wings observed,
including three distinct velocity minima within the P Cygni profile in H$\alpha$ for HD 98922 (we explore this
substructure further in Section \ref{s_hd98922}).
Plots of these transitions can be found in Appendix \ref{s_appendixU}.

Assuming the observed Balmer jump excess and Br$\gamma$ emission are produced via MA, accretion rates have 
been estimated for both V1295 Aql \citep[$\sim$10$^{-5}$ M$_\odot$ yr$^{-1}$, ][]{Mendigutia:2011Nov} 
and HD 98922 \citep[$\sim$10$^{-6}$M$_\odot$ yr$^{-1}$, ][]{GL:2006}.
In contrast to accreting TTS, we do not see any redshifted absorption; \citet{Hartmann:1994} note that at 
the high velocities found in accretion flows H$\alpha$ can thermalize, thereby filling in the telltale inverse 
P Cygni absorption. If present, however, we should observe redshifted absorption in the higher order Balmer 
lines as they do not thermalize; for neither object do we see redshifted absorption indicative of
magnetospheric accretion-related infall in 
any transition we observed. \citet{Cauley:2014} observed 
the He 10830\AA\ transition in V1295 Aql and also noted only blueshifted absorption. In our 
fiducial TORUS models for both stars, we are unable to produce any line profile invoking MA that does not 
have an inverse P Cygni profile in all of the Balmer lines (see Appendix \ref{s_appendixT} for further detail). 

We do not observe a high velocity component of the [O \textsc{I}] 6300\AA\ line frequently seen in TTS systems 
\citep[e.g.,][]{HEG:1995}, and thus have not estimated wind mass loss rates. In HAeBe systems, this feature 
is rare: \citet{Acke:2005} pointed out that in their sample of 49 HAeBe stars, only 3 showed any high 
velocity [O \textsc{I}] 6300\AA\ feature. 

Variability in spectral lines is related to the emitting or absorbing volume being probed by the line and its 
dynamical properties.
At the shortest timescales, of order minutes to hours, we are probing regions
close to the stellar surface: accretion flows, the launching region of a stellar wind, episodic mass loss 
and the rotational modulation of any of these events. 
Longer timescales, of order days to weeks, probe regions between the star and disk, from corotation to fractions 
of an au. In these regions, we may observe variability related to extended magnetospheric structure, hot gaseous 
inner disk components, and magnetospheric ejections \citep[cf.][]{Ferreira:2006,Ferreira:2013}. The longest 
timescales, of order months to years, probe inner disk Keplerian orbital times.

While we do note low, $\sim$5\% level, variability over the course of our hour-long observations, the 
strongest variability we observe occurs on timescales of days or longer; years for V1295 Aql. 
In both objects, the most variability on any timescale occurs in the blueshifted absorption components. 
In contrast to TTS variability observations, we see little change 
in the overall emission level or the red wings of our lines; what variability we observe in the red line 
wings appears to be due to modest overall change in the width of the line \citep[consistent with the
 results of][]{Mendigutia:2011May}. The most dramatic 
spectrum-to-spectrum changes occur on the longest time baselines available to us, over years, and are 
dramatic shifts in the blueshifted absorption component velocity.
Studies of line profile variability in TTS and IMTTS \citep[e.g.,][]{Johns:1995,Costigan:2014,Sousa:2016}
have found substantial amounts of variability across the line profile, including at redshifted 
velocities. \citet{Nguyen:2009ApJ} and \citet{Costigan:2014} find variability in TTS and Herbig Ae stars saturates at the 
rotation period.

\begin{figure}
\centering
\includegraphics[width=0.48\textwidth,clip,trim=0.4cm 0 2.6cm 0.2cm]{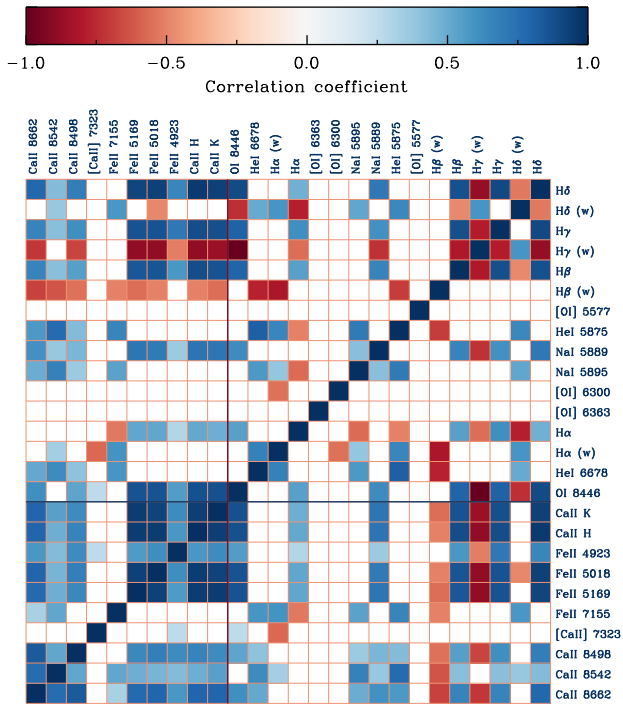}
\caption{Correlations of time-series line EWs for V1295 Aql, 2013 high 
cadence spectra only were used here. Lines are grouped by ionization state 
(singly ionized the left and below dividing lines, neutral to the right of 
and above) and within those groups, sorted by increasing excitation energy 
to the upper right corner.}
\label{f_v1295corrmap}
\end{figure}

Quantitatively assessing variability over the full set of observations, Figures \ref{f_v1295corrmap} and 
\ref{f_hd98922corrmap} readily allow for comparison between V1295 Aql and HD 98922. Each square in the diagram
represents how any given two line 
components evolve in time with respect to each other; for reference, we show specific examples in Appendix \ref{s_appendixS}. 
In both stars, singly ionized species show more variability, both anti- and correlated with other singly ionized species, than
the neutrals during the $\sim$10 day long period sampled. This implies 
the singly ionized species originate in a compact region evolving on short timescales. 
We also see that in both stars species in the same multiplet, e.g., Fe \textsc{II} (42), have positively correlated
time evolving behavior, as expected under an assumption of a common origin of these lines.
Similarly, lines of the same multiplet share common time-series evolution when compared with other lines. 
  Interestingly, in V1295 Aql, the emission and absorption components of lines with P-Cygni profiles
  show anticorrelated time evolution, whereas the components evolve together for HD 98922. Unlike what is seen
  in HD 98922, for V1295 Aql we note little correlation between forbidden line evolution and the neutral and
singly-ionized species selected; we note that our method is insensitive to potential time delays, this method is biased 
  toward short timescale co-evolution. In forthcoming work, we will use radiative transfer modeling to explore
  further the conditions that would cause anticorrelated versus correlated evolution of components in the same
line.

\begin{deluxetable*}{l|cc|cc}
\tabletypesize{\scriptsize}
\tablecaption{Stellar parameters\label{t_params}}
\tablewidth{0pt}
\startdata
 & \multicolumn{2}{c}{V1295 Aql} & \multicolumn{2}{c}{HD 98922} \\ \hline
Spectral type               & A2e             &                & B9e              &       \\
T$_{eff}$ [K]                & 9250$\pm$250    & 2              & 10600            & 3     \\ 
Mass [M$_{\odot}$]           & 2.9$\pm$0.5     & 4              & 5               & 5     \\
Radius [R$_{\odot}$]         & 3.6$\pm$0.5      & 4              & 9.1             & 3      \\ 
Inclination [$^{\circ}$]     & 34$\pm$2         & 6              & 20-45           & 7, 8   \\ 
$v_{r}$ [km s$^{-1}$]        & -1.2$\pm$1.3     & 1              & -4.9$\pm$4.4  & 1      \\
$v$sin$i$ [km s$^{-1}$]      & 3.2$\pm$2.5   & 1               & 39$\pm$5.3   & 1      \\
P$_{rot}$ [d]                & 32$\pm$25    & 1              & 4-8             & 1      \\ 
Magnetic field strength [G] & $\sim$100        & 9, 4, 10, 11   & Non detection   & 11     \\
\enddata
\tablecomments{References: 
1. Derived in this work with MIKE spectra, rotation periods calculated as P= 2$\pi$R$_{\rm star}$sin$i$ / $v$sin$i$.
    Errors cited incorporate the error on radius, $v$sin$i$, and $i$; V1295 Aql's uncertainty is dominated by uncertainty on
    $v$sin$i$; a range is given for HD 98922, as a large range of inclination values are reported in the literature.
2. \citet{Acke:2004}, using spectral type and best visually-matched synthetic spectrum; 
3. \citet{Kraus:2008b}, R calculated using T$_{\rm eff}$ and L; 
4. \citet{Catala:2007}, $<$B$>$ from spectropolarimetry and mass, radius, from PMS track placement; 
5. \citet{Wade:2007}, masses, radii derived via PMS track placement;
6. \citet{Lazareff:2017}, H-band interferometry assuming shared star and disk inclination;
7. \citet{Hales:2014}, inclination from combined $^{12}$CO and SED modeling; 
8. \citet{Blondel:2006}, inclination from fitting shape of Lyman-$\alpha$ profile; $<$B$>$ from
  spectropolarimetry in
9. \citet{Hubrig:2006},
10. \citet{Hubrig:2009}, and
11. \citet{Alecian:2013a}.}
\end{deluxetable*}

\subsection{V1295 Aql}\label{s_v1295}

In Table \ref{t_params}, we summarize stellar parameters for V1295 Aql. Its magnetic field, 
$\sim$100G, has been detected and confirmed by multiple authors over several years 
\citep{Alecian:2013a,Hubrig:2006,Hubrig:2009,Catala:2007} and one non-detection was reported by \citet{Wade:2007}.
As described in Section \ref{s_analysis}, from our MIKE spectra we measure $v_{r}=$ -1.2$\pm$1.3 km s$^{-1}$ 
and $v$sin$i=$3.19$\pm$2.45 km s$^{-1}$, consistent with \citet{Alecian:2013a}.
With these parameters and assuming star and disk share an inclination of 34$\pm$2$^{\circ}$
\citep{Lazareff:2017}, we derive a rotation period of $\sim$32d.
As \citet{Alecian:2013b} note, the handful of magnetic HAeBe stars observed are also slow rotators, 
consistent with V1295 Aql's apparently very slow rotation rate.

For the observations spanning multiple years, we see 
the P Cygni absorption is consistently growing weaker. In our first nights of data, in 2005 and 2006, blueshifted absorption 
is detectable in H$\alpha$, H$\beta$, H$\gamma$ and H$\delta$, but can only be seen in H$\alpha$ 
and H$\beta$ thereafter (Fig. \ref{f_balmer}); simultaneously, the emission component's EW is also growing weaker. 
The Ca \textsc{II} infrared triplet is also dramatically weakening, as are the Na I D lines. 
The Ca \textsc{II} H and K lines, notable for their complex structure \citep[cf.][and references therein]{Pogodin:2005}, 
are not seen to vary much on night-to-night timescales, or within any given night (this is consistent with previous 
observations; in In Appendix \ref{s_appendixU} we show plots of all of these lines).
\citet{Alecian:2013_v1295} note spectropolarimetric changes indicating V1295 Aql had a stable, 
aligned dipole over observations from 2004-2009, but in 2011-2012, variability was observed, indicating a change in 
the star's magnetic field structure. In 2005 and 2006 data, we see broad P-Cygni absorption, and in 2008 data,
  we only note a weak, higher velocity absorption feature. In 2010, a broad, multi-trough component appears, only to
  have disappeared again by our 2013 data taken for this work. The correlation between magnetic field structure and the
  H$\alpha$ profiles is unclear, but we note in the epochs with weak absorption, the emission component's width has
  broadened, possibly filling in underlying absorption.

[O \textsc{I}] 6300\AA\ in particular is of interest, as its driving mechanism remains unclear. 
Using multi-epoch data, 
\citet{Corcoran:1998} noted a relationship between the [O \textsc{I}] emission and presence of near-IR excess; 
they suggested the IR excess was indicative of accretion. 
Line component analysis of our time-series data finds the H$\alpha$ wind 
absorption component is weakly anticorrelated with the [O \textsc{I}] emission: over the period of observation, the 
[O \textsc{I}] 6300\AA\ emission line 
grew stronger as the wind absorption in H$\alpha$ grew weaker.
To quantitatively assess the line components' evolution, we calculated Spearman's correlation coefficient for
time-series equivalent widths: comparing [O \textsc{I}] emission and blueshifted absorption in H$\alpha$, the
test statistic $\rho$ is -0.39 with a significance of 8x10$^{-5}$; this is to say the likelihood of a spurious
correlation is very small, and this is a statistically significant result. In contrast, $\rho$ for the emission
component of H$\alpha$ and the [O \textsc{I}] EW is -0.09 with a significance of 0.37, which does not 
fall below the adopted 10$^{-3}$ threshold for significance.

The timescales on which the blueshifted absorption in H$\alpha$ and H$\beta$ varies, $\sim$1 day, are
more consistent with regions close to the star; the velocity center of the absorption is consistent with 
the (de-projected) stellar escape velocity, $\sim$300 km s$^{-1}$.
Indeed, earlier authors have suggested 
the narrow line widths and line symmetry suggest heating at the base of a 
stellar wind launching region as the origin of metallic line emission 
\citep{Catala:2007}. Asymmetries in these metallic lines are attributed to 
obscuration of the emitting 
region by the wind \citep{Pogodin:2005}. At the relatively low system 
inclination, a high degree of collimation/small opening angle would be 
necessary in order for material to be present along the line of sight; this 
is easier to accomplish with a stellar wind than a disk wind. Indeed, our initial 
radiative transfer models, which include only a disk wind, fail to reproduce 
blueshifted absorption for the interferometric system inclination (models detailed in Appendix \ref{s_appendixT}).
Interestingly, we observe that as the wind absorption weakens, the [O \textsc{I}] emission strengthens. 
We suggest one possible physical interpretation for this anticorrelated EW evolution 
invoking star-disk 
interaction. Assuming the absorption is done by a compact wind originating near the star 
and that the observed [O \textsc{I}] emission originates in the disk surface layer, as the wind 
becomes less loaded with mass (weakening absorption), the emitting 
layer of the disk can remain low enough density that [O \textsc{I}] is able to spontaneously de-excite 
rather than being continually collisionally excited (thus strengthening emission). 
At velocities of a few hundred km s$^{-1}$, wind 
material would take $\sim$half a day to travel $\sim$0.1 au, making this scenario
consistent with our observations.

While we do see significant variability within a single night of our observations of V1295 Aql, 
unfortunately, our data do not cover a sufficiently long time baseline to establish whether there is 
any periodicity to the variability that could be attributed to rotational modulation 
\citep[periodicity has been seen in CTTS, cf.][and references therein]{Johns:1995}. We observe persistent, weak absorption and emission components 
within the blueshifted absorption in H$\alpha$, not attributable to any telluric features, and have used a local minimum and 
maximum finding algorithm to quantify their velocities and evolution during the 
observation window (see Appendix \ref{s_appendixS}). We find that the lower velocity structures (between $\sim$-190 
to -234 km s$^{-1}$) remain stationary while the higher velocity structures 
($\le$-265 km s$^{-1}$) accelerate. The highest velocity transient absorption 
feature both accelerates and decelerates during our observation window; potentially, given the short
  timescale of significant velocity evolution, we are observing structures entrained and corotating in a stellar wind.

\subsection{HD 98922}\label{s_hd98922}

HD 98922 is a B9e star between 20-45$^{\circ}$ inclination \citep{Blondel:2006,Hales:2014}; 
it has no detected magnetic field \citep{Alecian:2013a}. From our MIKE spectra, we measure $v$sin$i$ to be 38.9$\pm$5.31 km s$^{-1}$, and the V$_r$ is $-$4.88$\pm$4.43 km s$^{-1}$.
Varying the inclination within the reported range of values from the literature, we obtain rotation periods between 4-8 days; thus our observations likely span a full rotation period. 
We see the greatest variability on the order of day to year timescales, however, substantial variability 
is observed over the course of our observations in 2013 (Fig. \ref{f_allepochnvps}, left panel). 
The shorter timescales are most consistent with rotational modulation, but we do not detect periodicity on $\sim$day timescales in either object.

\begin{figure}
\centering
\includegraphics[width=0.48\textwidth,clip,trim=0.4cm 0 2.6cm 0.2cm]{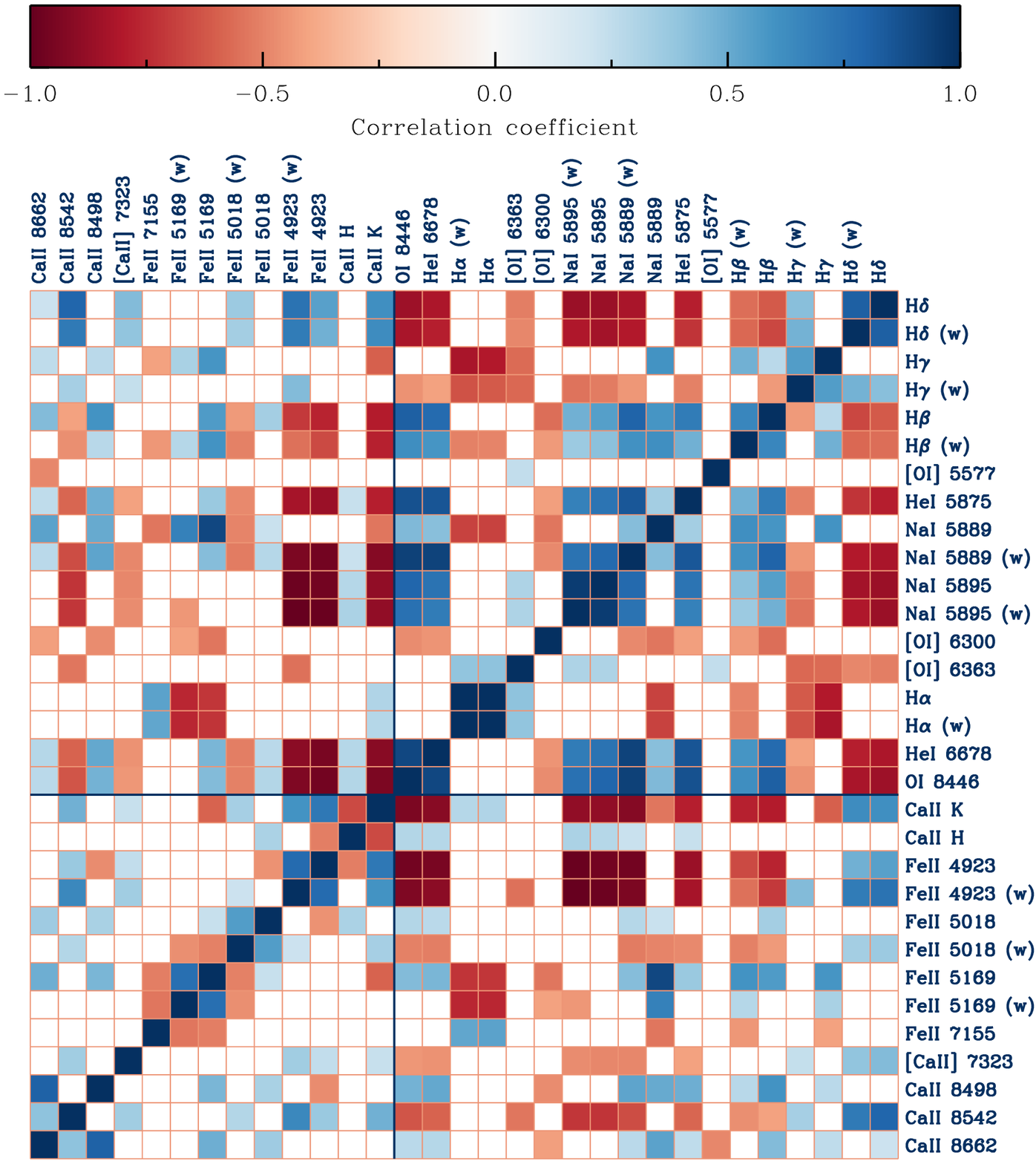}
\caption{Correlations of time-series line EWs for HD 98922.}
\label{f_hd98922corrmap}
\end{figure}

In Figure \ref{f_hd98922corrmap} we show the EW(t) correlation matrix as described in Section \ref{s_v1295}.
We see strong anticorrelation of the evolution of Na \textsc{I} D line widths and the higher order
Balmer series lines, including both emission and absorption components; in contrast, the Na \textsc{I}
  D line width evolution is weakly correlated with H$\beta$ and any correlation or anticorrelation with
  H$\alpha$ is not statistically significant. Anticorrelation would not be expected in the TTS case, 
where H$\alpha$ and Na \textsc{I} D emission have been linked to an accretion flow 
\citep[e.g., the models of ][]{Muzerolle:2004}, but it is unclear what differing levels of correlation
with different Balmer series lines could be indicating. In contrast to 
the same figure for V1295 Aql, we see more strongly related line profile 
evolution in HD 98922 generally, possibly simply due to our observations sampling a full rotation period.
  Despite the coverage, we searched for periodicity to line profile evolution and did not find any (we show our
equivalent width time series measurements in Appendix \ref{s_appendixU}).

The wind, as probed through the line minimum in the H$\alpha$ P Cygni component appears to be 
getting faster over the course of observation. We note, however, the difficulty in modeling the wind as 
our high-resolution data allow us to see at least three distinct velocity minima in the P Cygni profile. Similar to the distinct weak 
absorption features in V1295 Aql, the lower velocity components 
($\sim$-50 to -100 km s$^{-1}$) 
in HD 98922 are static during our observations while the higher velocity components
($\sim$-125 to -175 km s$^{-1}$) accelerate.
When the velocity of the component doing the strongest absorbing is slow, 
during our 2013 MIKE observations, we see P Cygni absorption in the Na D 
lines; in the earlier epoch of MIKE data from 2010, when the wind is a factor of 2 faster
(Fig. \ref{f_balmer}, upper left panel), there is no blueshifted absorption in the Na D lines. This is
potentially probing mass loading in a wind and corresponding changes in the density 
gradient in the acceleration region: when the wind is slower, the density gradient is steeper 
toward the base of the launch region, causing the observed absorption in Na \textsc{I}. \citet{Powell:2012} noted this in FU Ori, observing a 1.8d lag in variability between 
Na \textsc{I} and the blueshifted absorption in H$\alpha$; unfortunately, we lack temporal coverage in 
the earlier epoch of V1295 Aql data to search for a similar effect.

\section{Discussion}\label{s_discn}

We have conducted a high cadence, time-series survey of two Herbig Ae/Be stars. Combining these new 
data with earlier observations by our team and archival spectra, we find strong variability in 
spectral line 
profiles on the longer timescales sampled: years, months, and days. We see very weak modulation on 
the shortest timescales sampled, $\sim$minutes. In all timescales, the strongest variability occurs in 
the blueshifted absorption components of the spectral lines. The lack of strong variability in lines' emission 
components as well as an absence of inverse P Cygni profiles is in distinct discord with our modeling 
efforts based on a dipolar magnetospheric accretion geometry. 
Given the lack of magnetic field detections in Herbig Ae/Be stars as well as 
the evidence for higher-order moments dominating global magnetic field structure as stellar mass 
increases \citep{Gregory:2012}, the observed accretion in 
both systems may be taking place via higher-order magnetic field structure, a boundary layer, or
at a novel magnetospheric inclination with respect to the rotation axis (i.e., misaligned).

In the case of V1295 Aql, the system inclination and variability on shorter than inner disk Keplerian timescales 
point to a stellar or X-wind originating close to the star, not a farther-out disk wind, as playing a role 
in generating the observed spectral lines. \citet{Catala:2007} suggest the acceleration region at the base of 
a stellar wind as the origin of narrow metallic emission lines. Our magnetospheric accretion+disk wind
model and the stellar wind model of \citet{Catala:2007} demonstrate that a compact stellar wind
acceleration region and a magnetosphere can 
produce the same emission profile; difficulty arises in how the blueshifted absorption is produced. 
Interferometric evidence for low system inclination makes it highly unusual that we observe as 
much variability in the blueward absorption components; indeed, our disk wind model cannot reproduce 
the observed absorption at such low system inclination. Observations by \citet{Alecian:2013_v1295} during 
overlapping epochs indicate the structure of the stellar magnetic field itself has changed, lending 
further support to material closer to the star being responsible for variations in line morphology. 
Potentially further evidencing a stellar wind, we
observe anticorrelated evolution of the blueshifted absorption in 
H$\alpha$ and [O \textsc{I}] 6300\AA\ emission (i.e., as the H$\alpha$ blueshifted absorption weakens,
  the [O \textsc{I}] emission strengthens)
that may indicate a decrease in outflow sufficiently lowers the density in upper disk layers to
  enhance forbidden line emission; 
correlation is not seen when comparing the emission component of H$\alpha$ to [O \textsc{I}].
Further, the Fe II (42) emission line, suggested by \citet{Catala:2007} to be forming in a
  stellar wind, has positively correlated evolution with the emission components of all Balmer series lines.
Lastly, the velocity structure of the blueshifted absorption in H$\alpha$ is consistent with 
obscuring material moving at or faster than stellar escape velocity. 

Our observations span the full rotation period of HD 98922; over this timescale, one may observe 
accretion flows moving in and out of the line of sight, modulating observed flux in spectral lines. We 
do observe strong variability in line profiles over this time, but do not find periodicity in 
the evolution of line equivalent widths. As in the case of V1295 Aql, the lack of any inverse P Cygni 
profile or variability on that side of the line, despite our full sampling of the stellar rotation 
period, is problematic for a canonical magnetospheric accretion paradigm. Unlike V1295 Aql, HD 98922 
is at an intermediate inclination and the blueshifted absorption can be readily reproduced via disk 
wind. We see anticorrelation of the Na \textsc{I} D lines' evolution and H$\delta$ and H$\gamma$,
  but weak correlation with H$\beta$ and no significant correlation with H$\alpha$; the significance of
  this is unclear, but positive correlation would be expected were all the lines to originate in an accretion
flow.

Models of disk winds have recently begun to take into account the disk magnetic field. Of particular import to 
the HAeBe case, the models of \citet{Bai:2013} show that in the presence of a strong FUV field, the wind mass loss rate 
is boosted; an FUV flux increase of an order of magnitude increases the mass loss rate by a factor of 3. In developing 
a picture of accretion and outflow for intermediate to higher mass protostars, the winds--both stellar and disk in origin--will likely be 
critical to understand, playing a dominant role in the system mass flux.

For the two systems studied here, it is another interesting point of contrast that in one case,
  the time evolution of emission and absorption components in the same line profile are apparently unrelated
(V1295 Aql) while in the other, they are strongly correlated (HD 98922).
Broadly, the differing evolution of lines in the same multiplet or series that would be expected
to co-evolve underscore the need for rigorous definitions of line components due to the potential for  
overlapping temporal and spatial properties.
The ambiguity of line profiles, e.g., the degeneracy 
in emission line production by an extended magnetosphere versus a pure P-Cygni spherical wind (i.e., 
our modeling of the H$\alpha$ line of V1295 Aql) calls for more rigorous, multi-component models 
that include asymmetry. 
High resolution, high cadence spectroscopy reveals spectral line structure 
on fine velocity/spatial and temporal scales
while $\sim$nightly visits and multi-epoch data demonstrate rotational and longer-term system
evolution, both of which are necessary  to inform models of evolving modes 
of accretion and outflow with increasing stellar mass.

\acknowledgments
A.~N.~A. would like to acknowledge the support of NASA Origins of Solar Systems grant 
NNX09AB87G to J.~D.~M. and NSF AAG 1311698 to A.~N.~A. and J.~D.~M. The authors thank B.~P.
Brown for computational contributions; 
models were calculated with support of the NASA High End Computing (HEC) Program through the NASA 
Advanced Supercomputing (NAS) Division at Ames Research Center on Pleiades with allocation GID s1647.
S.~K. acknowledges support
from a STFC Rutherford fellowship (ST/J004030/1) and ERC Starting Grant (Grant Agreement No. 639889).
We also express appreciation for observations taken for this program by 
Ajay Tannirkulam, Xiao Che, Stefan Kraus, Mario Mateo, Keren Sharon and Traci Johnson.
Finally, the authors thank an anonymous referee for thorough comment and suggestions that
improved the manuscript.

\clearpage

\appendix

\section{A. Spectral atlas}\label{s_appendixU}

\begin{figure}[ht]
  \includegraphics[angle=90,width=0.9\textwidth]{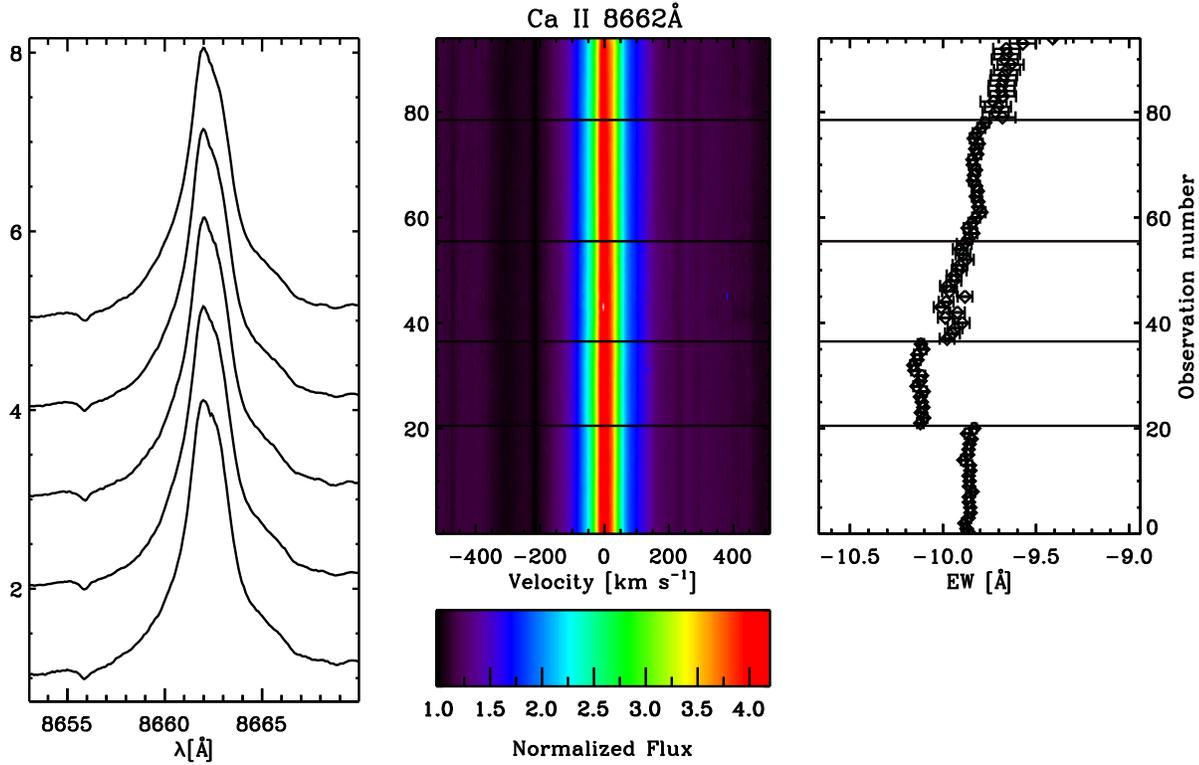}
  \caption{V1295 Aql Ca II 8662\AA.
    Left panel: Average line profiles for each of the five nights
    of high cadence observation, increasing in time from bottom to top:
    UT130429, UT130503, UT130504, UT130506, UT130507. Spectra are continuum
    normalized, their photospheric components removed, and arbitrary
    y-axis offsets are applied for display purposes. 
    Middle panel: surface plot showing every spectrum observed, ascending
    time order from bottom to top. Spectra are continuum normalized, their
      photospheric components removed, and a color
    scale is shown in bar below figure. Each night is separated by a horizontal
    line. Right panel: measured equivalent widths as a function of time.
    Unless otherwise indicated, total line equivalent widths are shown (including
    both emission and absorption components). Abscissa reversed to indicate decreasing
    emission line strength to the right. Error bars indicate RMS of measurements
    within a single night.}\label{f_v1295atlas1}
\end{figure}

\begin{figure}
  \includegraphics[angle=90,width=0.9\textwidth]{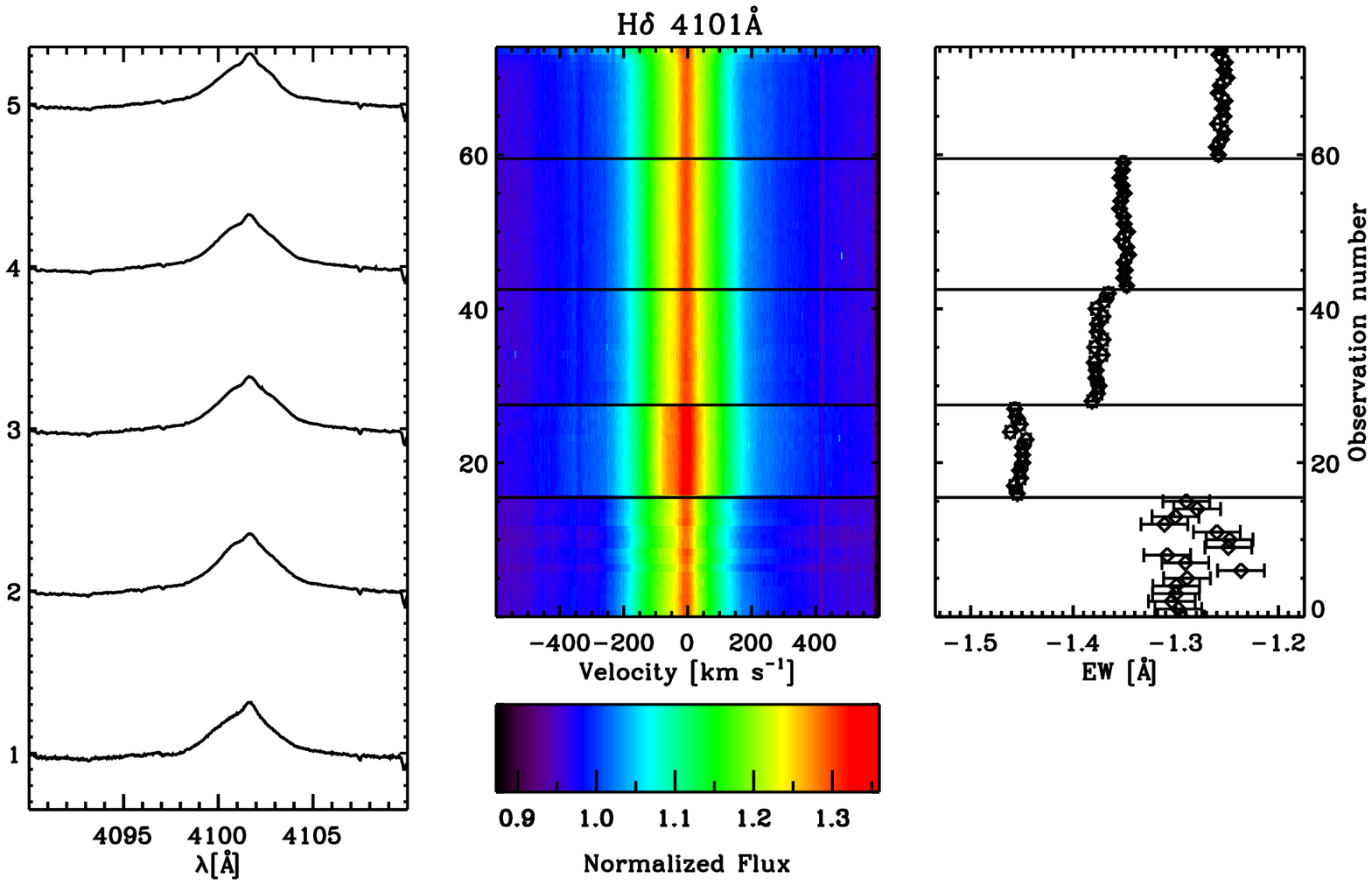}
  \caption{V1295 Aql H$\delta$. As in Fig. \ref{f_v1295atlas1}. Here, we
  show just the fit emission components' EW for plot scaling purposes.}\label{f_v1295atlas_last}
\end{figure}

\begin{figure}[ht]
  \includegraphics[angle=90,width=0.9\textwidth]{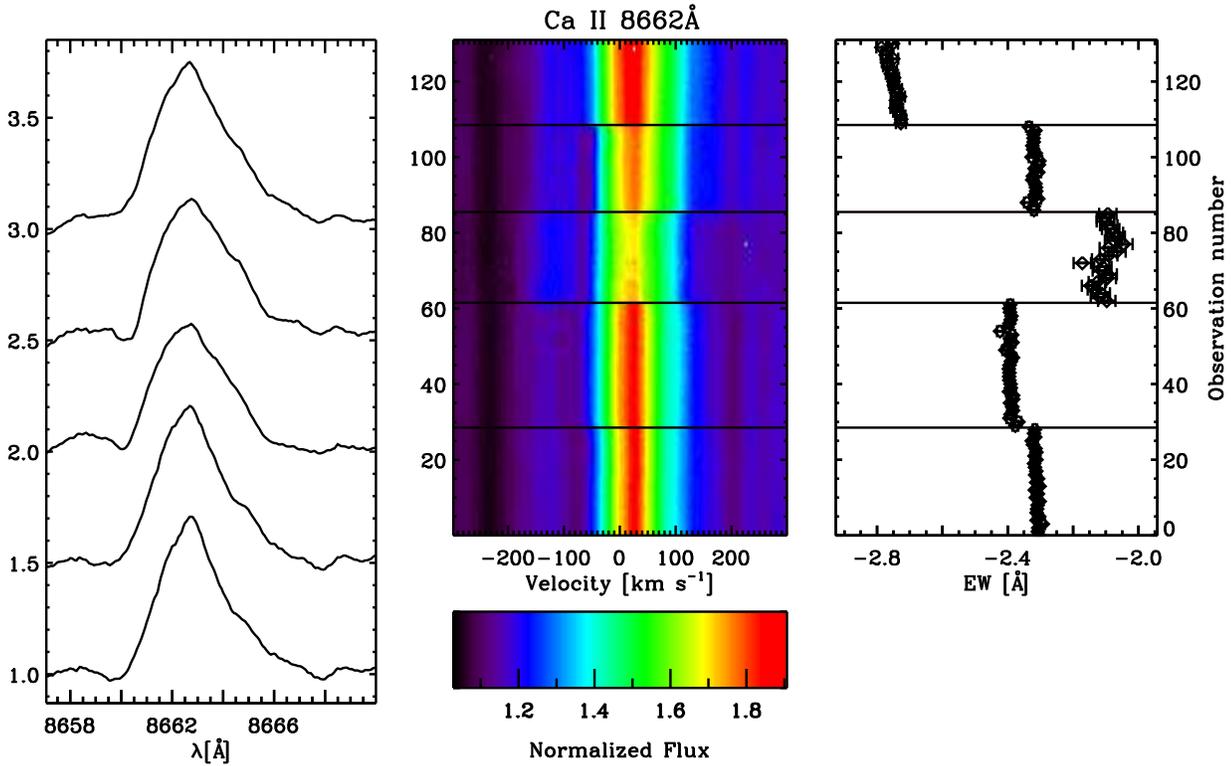}
  \caption{Left panel: Average line profiles for each of the five nights
    of high cadence observation, increasing in time from bottom to top:
    UT130428, UT130429, UT130503, UT130504, UT130506. Spectra are continuum
    normalized, their photospheric components removed, and
    arbitrary y-axis offsets are applied for display purposes. 
    Middle panel: surface plot showing every spectrum observed, ascending
    time order from bottom to top. Spectra are continuum normalized, their
      photospheric components removed, and a color
    scale is shown in bar below figure. Each night is separated by a horizontal
    line. Right panel: measured equivalent widths as a function of time.
    Unless otherwise indicated, total line equivalent widths are shown (including
    both emission and absorption components). Abscissa reversed to indicate decreasing
    emission line strength to the right. Error bars indicate RMS of measurements
    within a single night.}\label{f_hd98922atlas1}
\end{figure}

\begin{figure}
  \includegraphics[angle=90,width=0.9\textwidth]{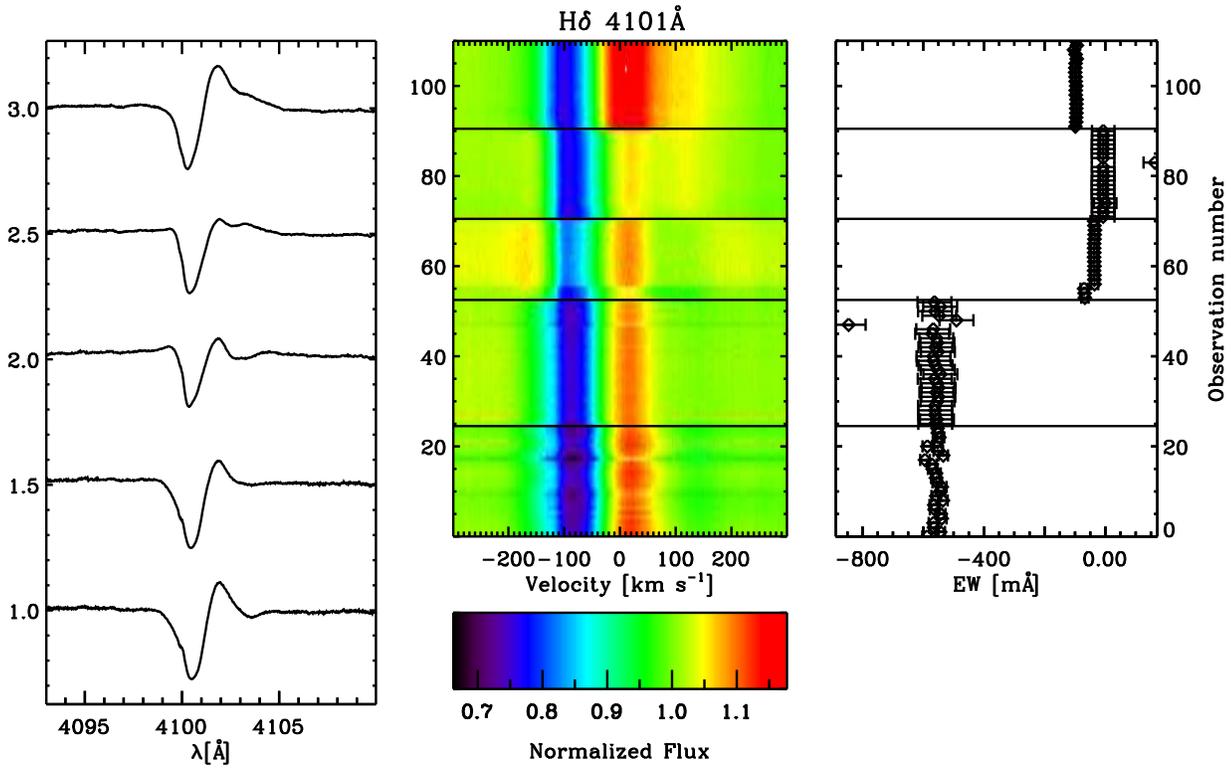}
  \caption{HD 98922 H$\delta$. As in Fig. \ref{f_hd98922atlas1}. Here, we
  show just the fit emission components' EW for plot scaling purposes.}\label{f_hd98922atlas_last}
\end{figure}

Here we show spectra of all species discussed in the text for V1295 Aql (Figs. \ref{f_v1295atlas1}-
  \ref{f_v1295atlas_last} and HD 98922 (Figs. \ref{f_hd98922atlas1}-\ref{f_hd98922atlas_last}). The figures
  are in order of appearance in the correlation matrices (Figs. \ref{f_v1295corrmap} and \ref{f_hd98922corrmap}):
  singly ionized species first and neutrals second, sorted in order of increasing excitation potential within
  those groups.

\subsection{V1295 Aquila}

\begin{figure}
  \includegraphics[angle=90,width=0.9\textwidth]{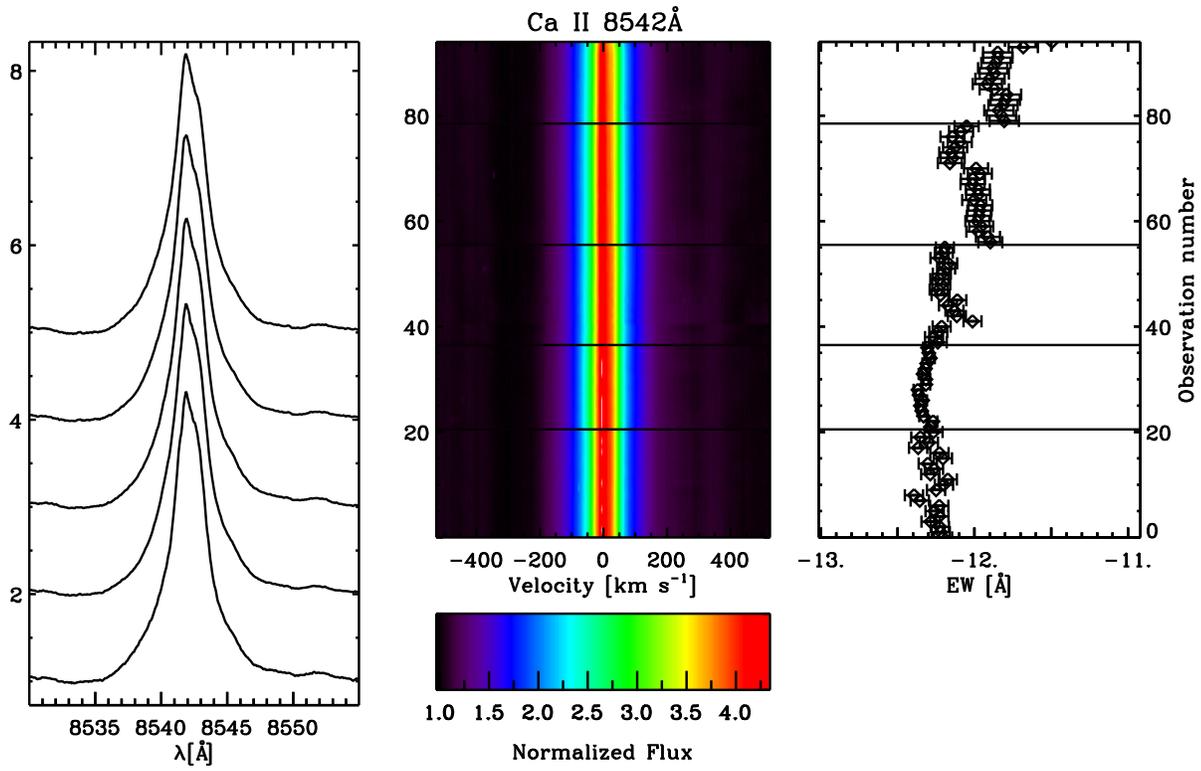}
  \caption{V1295 Aql Ca II 8542\AA. As in Fig. \ref{f_v1295atlas1}.}
\end{figure}
\clearpage

\begin{figure}
  \includegraphics[angle=90,width=0.9\textwidth]{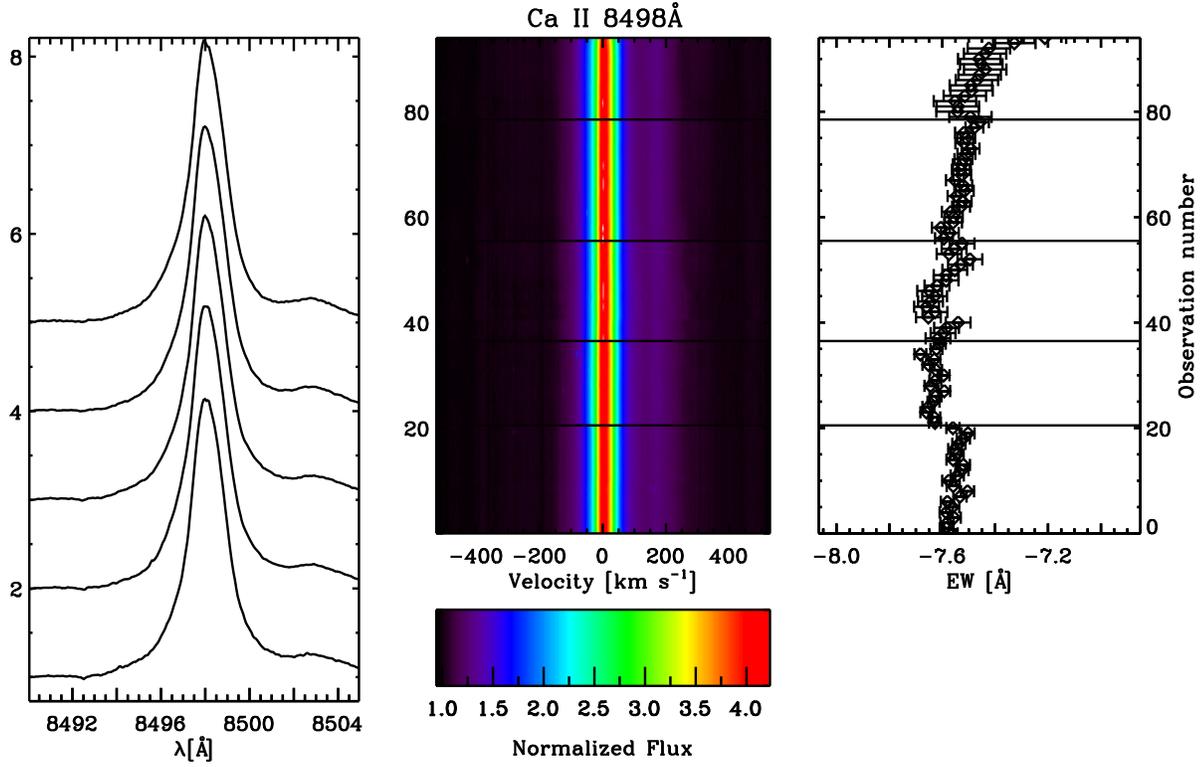}
  \caption{V1295 Aql Ca II 8498\AA. As in Fig. \ref{f_v1295atlas1}.}
\end{figure}

\begin{figure}
  \includegraphics[angle=90,width=0.9\textwidth]{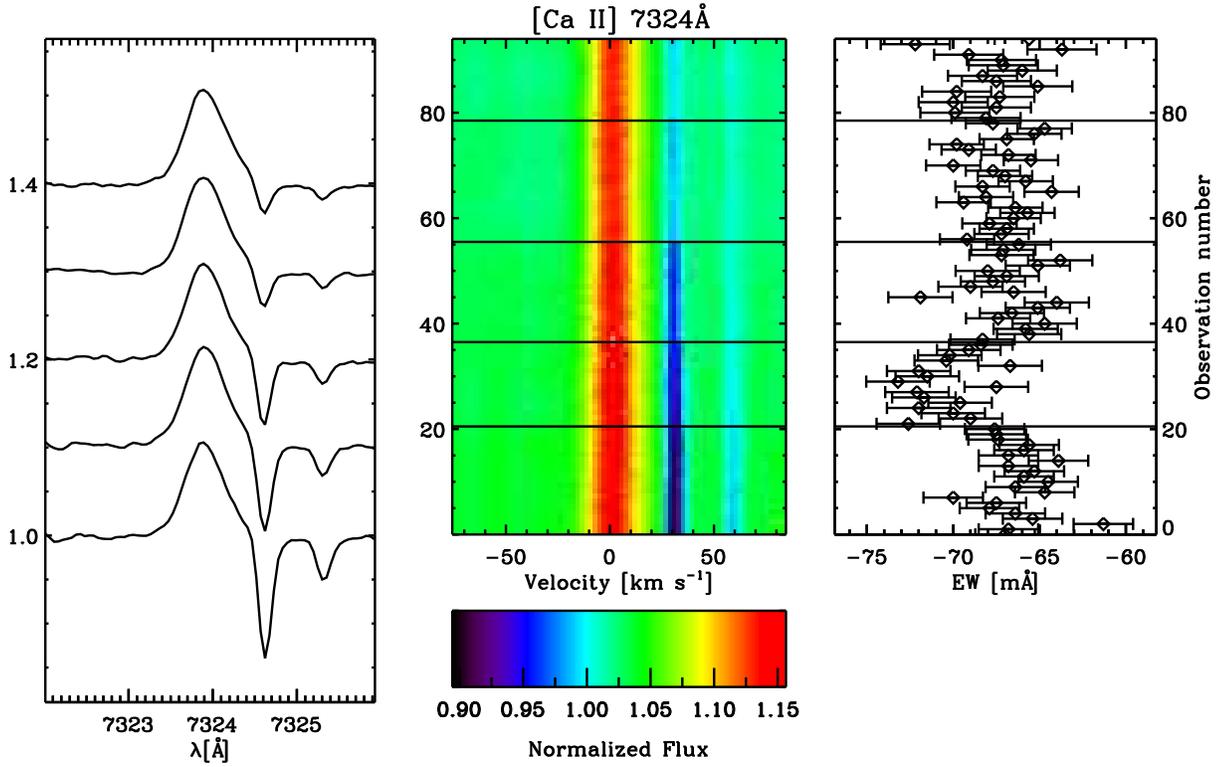}
  \caption{V1295 Aql [Ca II] 7324\AA. As in Fig. \ref{f_v1295atlas1}. A telluric
    absorption line at 7324.14\AA\ is overpowered here by the emission, but two
    water vapor lines are seen redward of the emission line \citep{Kurucz:2005}.}
\end{figure}
\clearpage

\begin{figure}
  \includegraphics[angle=90,width=0.9\textwidth]{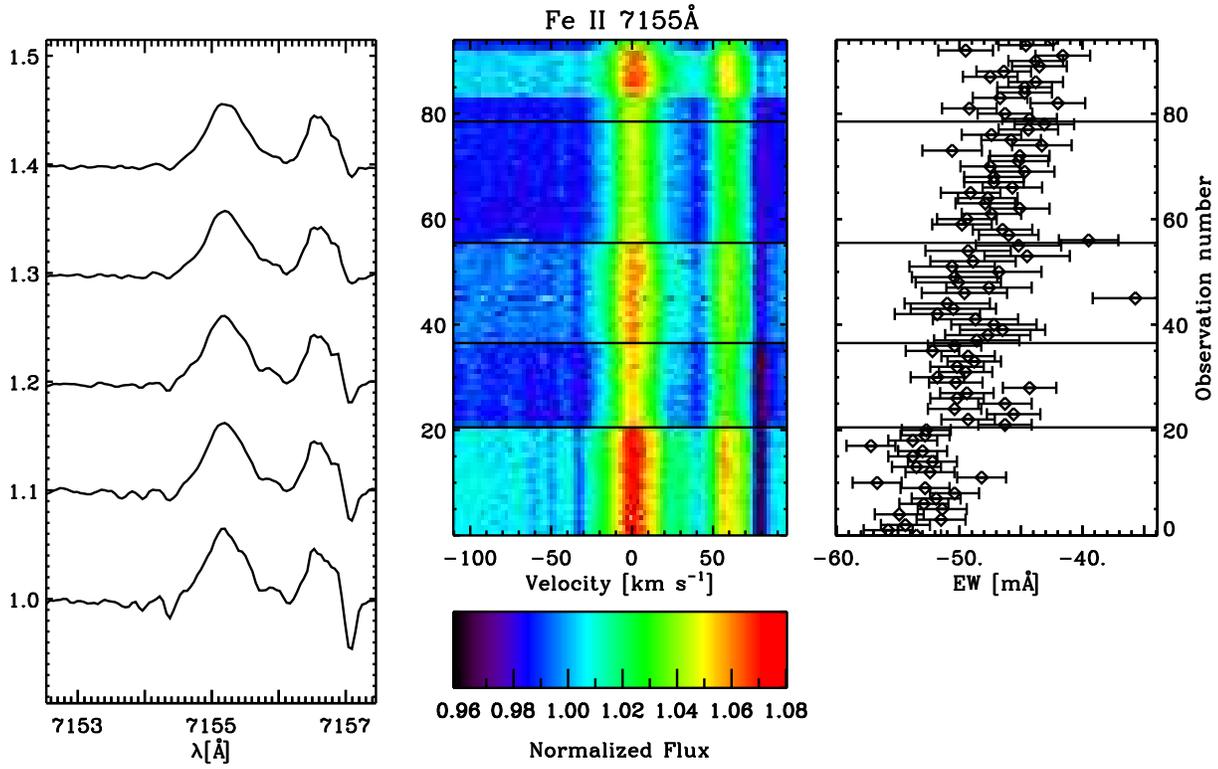}
  \caption{V1295 Aql Fe II 7155\AA. As in Fig. \ref{f_v1295atlas1}. A telluric
    line at 7156.3\AA\ is clearly seen to vary from night to night
    \citep{Catanzaro:1997}.}
  
\end{figure}

\begin{figure}
  \includegraphics[angle=90,width=0.9\textwidth]{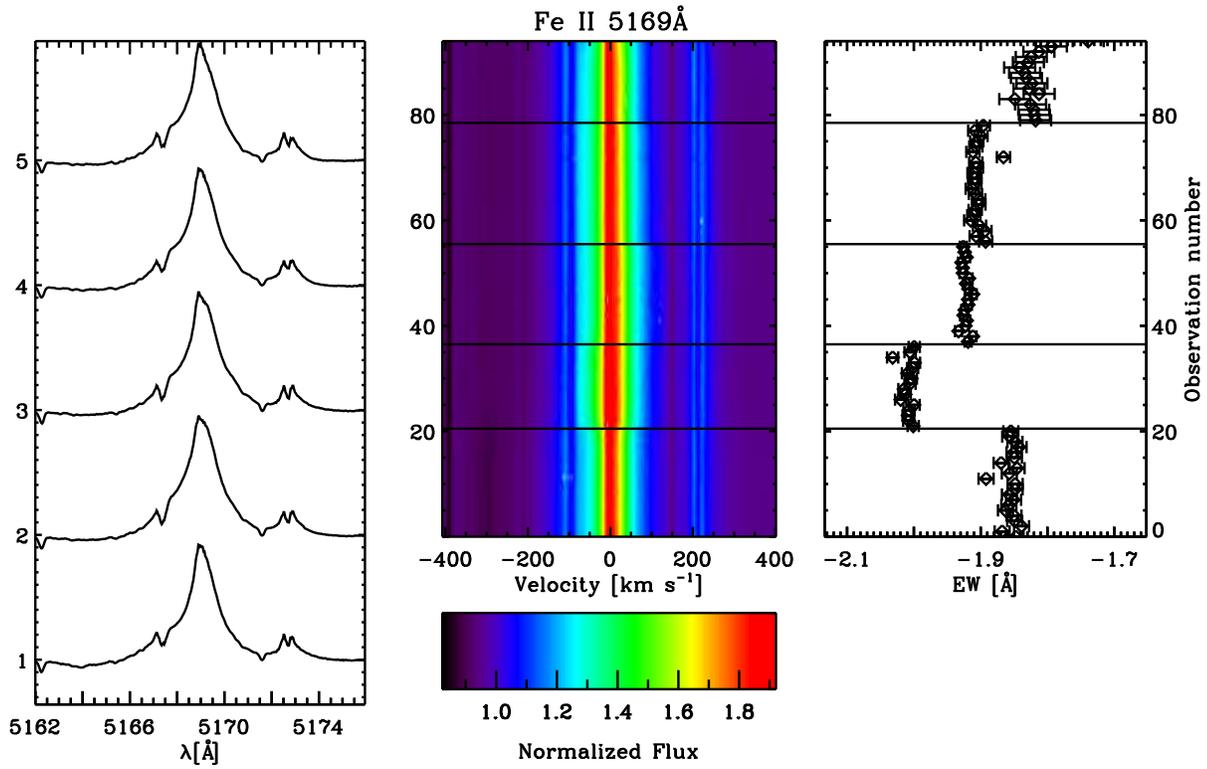}
  \caption{V1295 Aql Fe II 5169\AA. As in Fig. \ref{f_v1295atlas1}.}
\end{figure}
\clearpage

\begin{figure}
  \includegraphics[angle=90,width=0.9\textwidth]{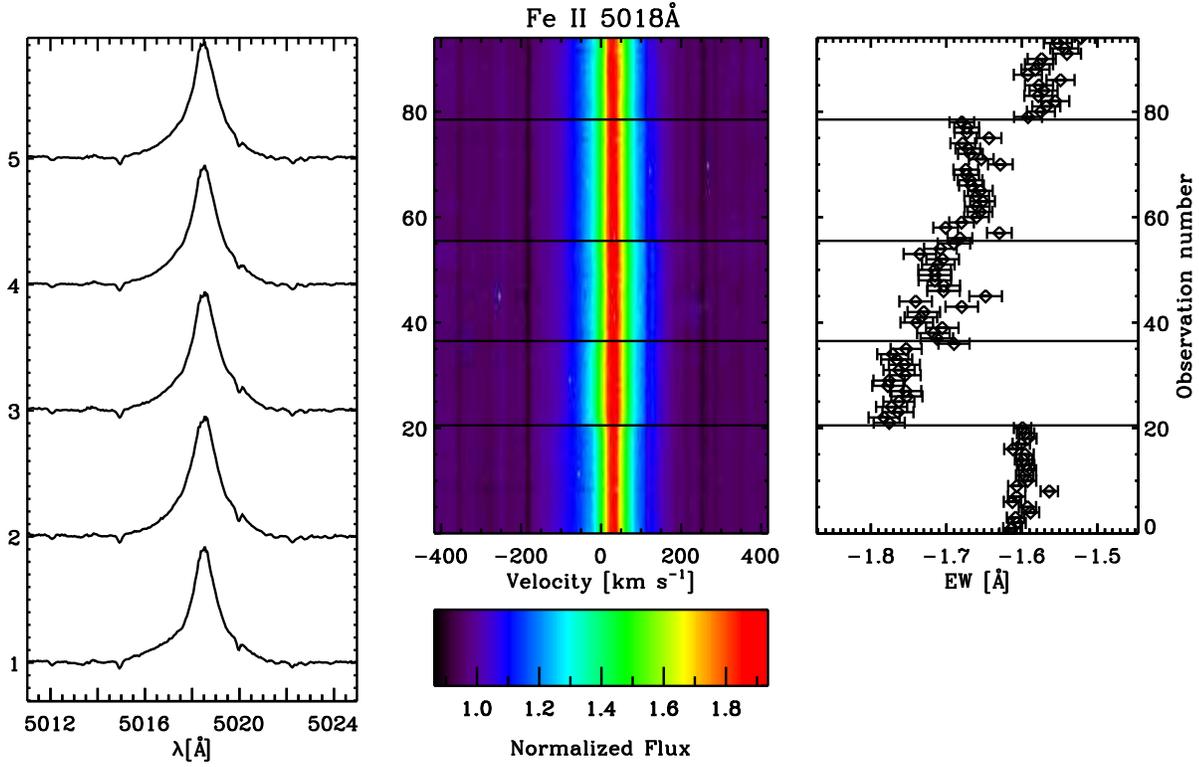}
  \caption{V1295 Aql Fe II 5018\AA. As in Fig. \ref{f_v1295atlas1}.}
\end{figure}

\begin{figure}
  \includegraphics[angle=90,width=0.9\textwidth]{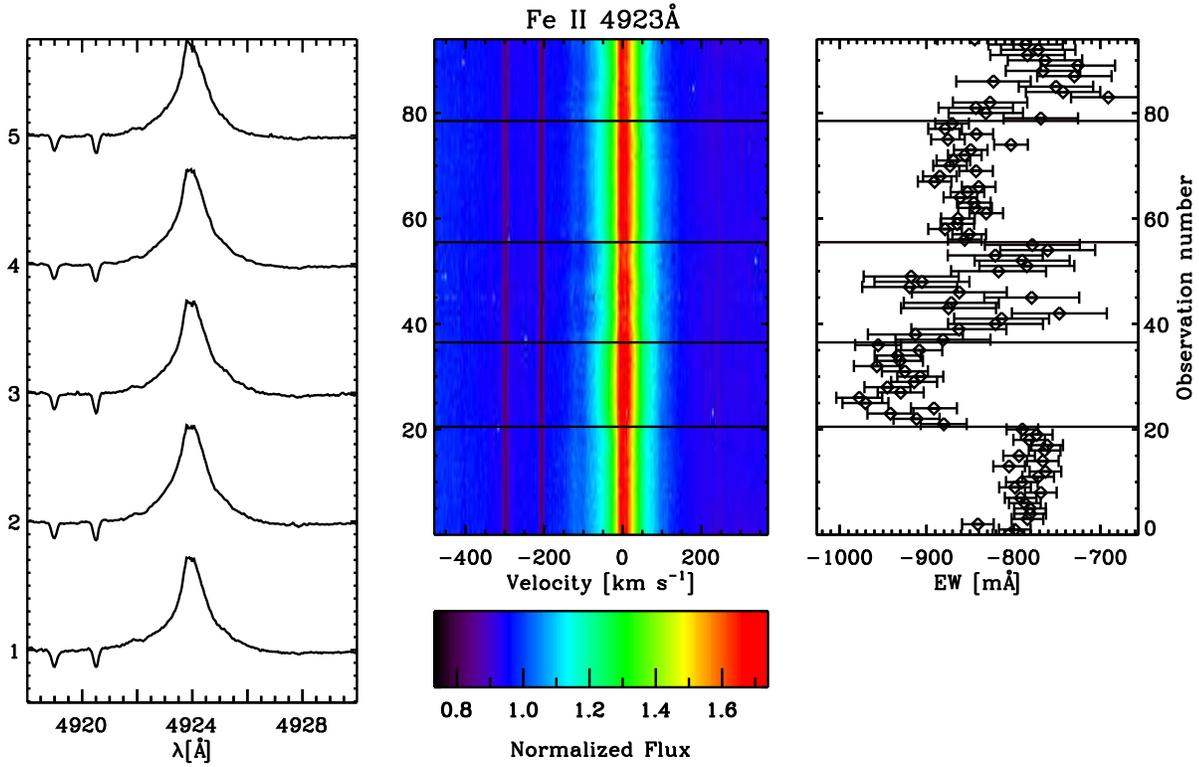}
  \caption{V1295 Aql Fe II 4923\AA. As in Fig. \ref{f_v1295atlas1}.
    Poor seeing on the third night, as well as the line being on the edge
    of the MIKE red chip, contribute to the scatter in EW.}
\end{figure}
\clearpage

\begin{figure}
  \includegraphics[angle=90,width=0.9\textwidth]{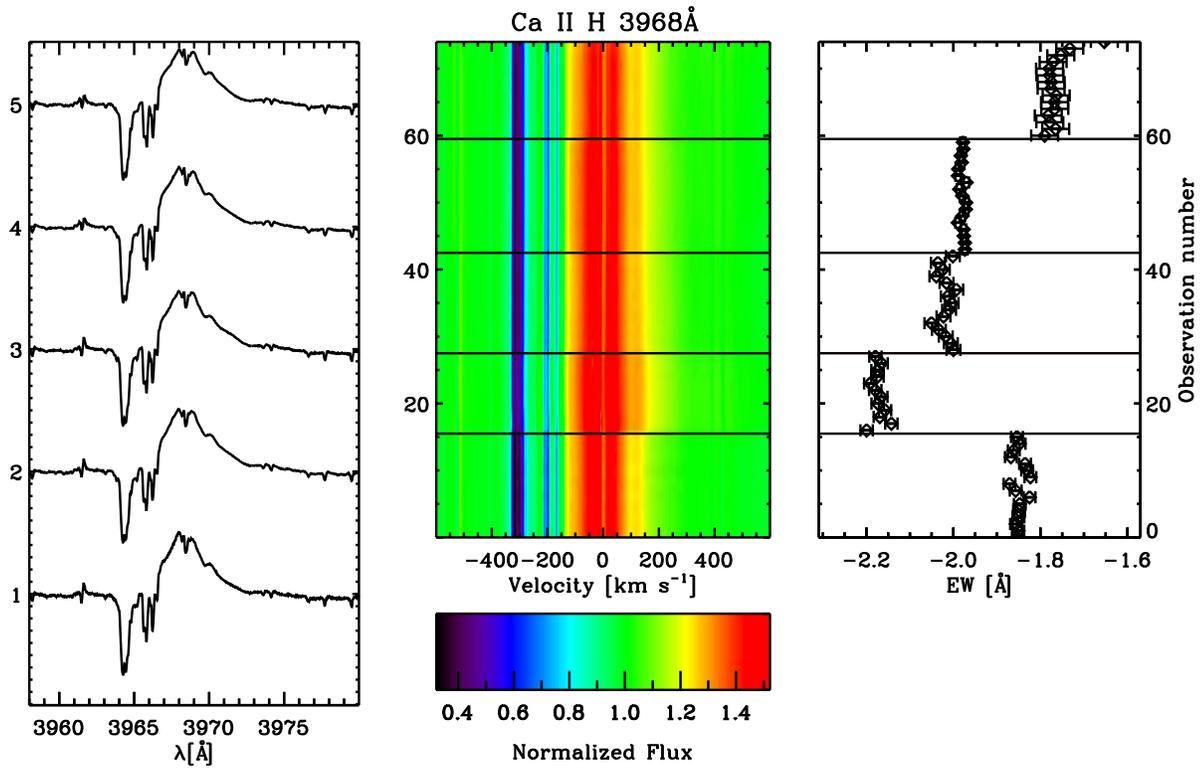}
  \caption{V1295 Aql Ca II 3968\AA. As in Fig. \ref{f_v1295atlas1}.}
\end{figure}

\begin{figure}
  \includegraphics[angle=90,width=0.9\textwidth]{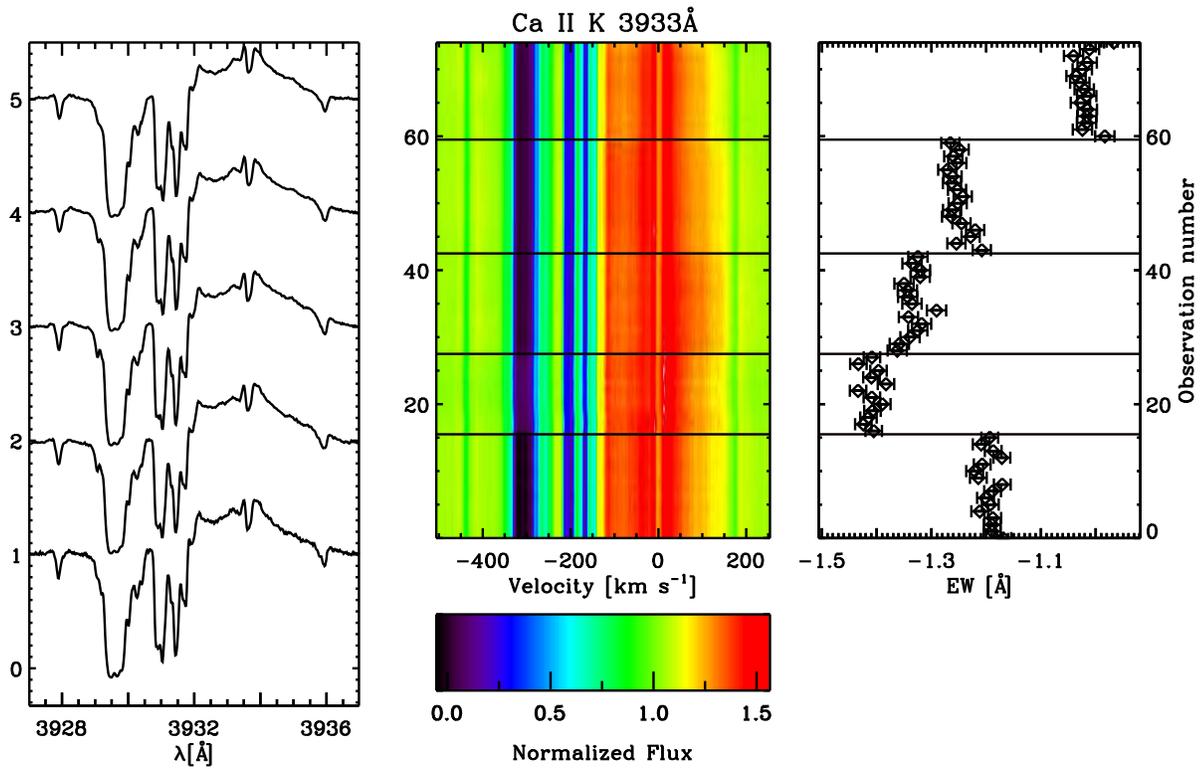}
  \caption{V1295 Aql Ca II 3933\AA. As in Fig. \ref{f_v1295atlas1}.}
\end{figure}
\clearpage

\begin{figure}
  \includegraphics[angle=90,width=0.9\textwidth]{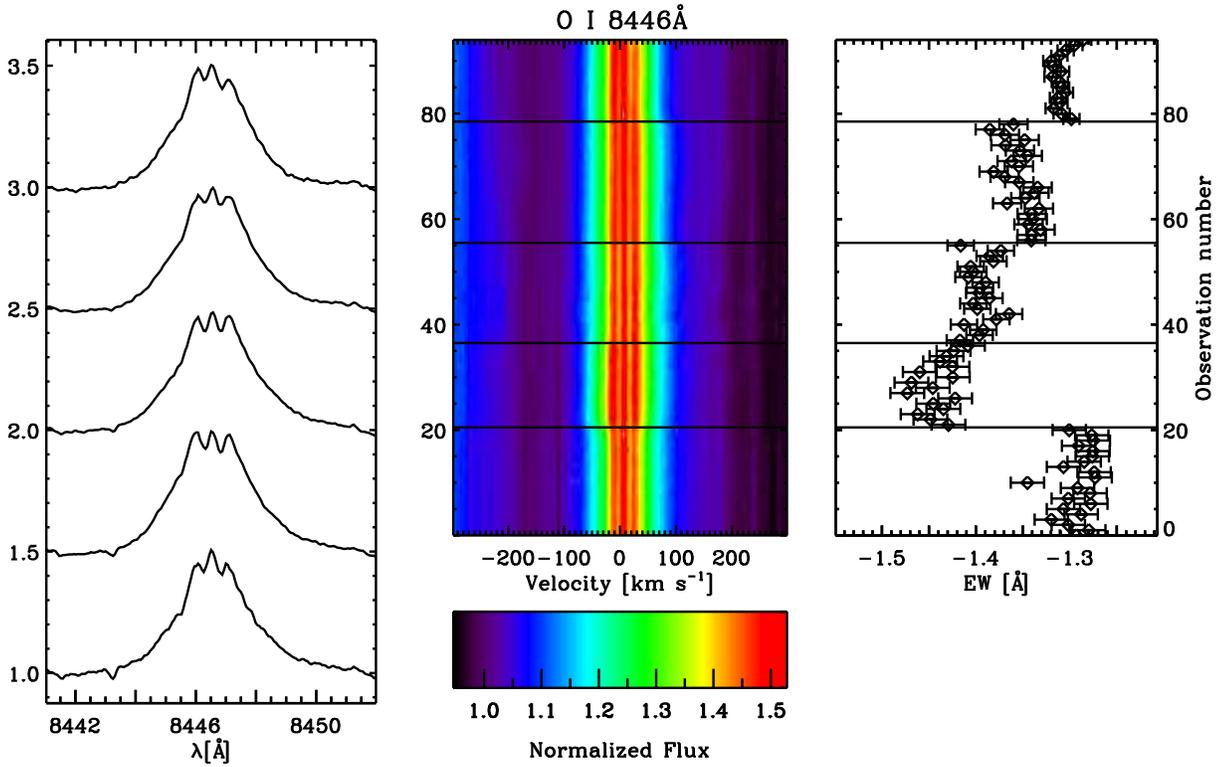}
  \caption{V1295 Aql O I 8446\AA. As in Fig. \ref{f_v1295atlas1}.}
\end{figure}

\begin{figure}
  \includegraphics[angle=90,width=0.9\textwidth]{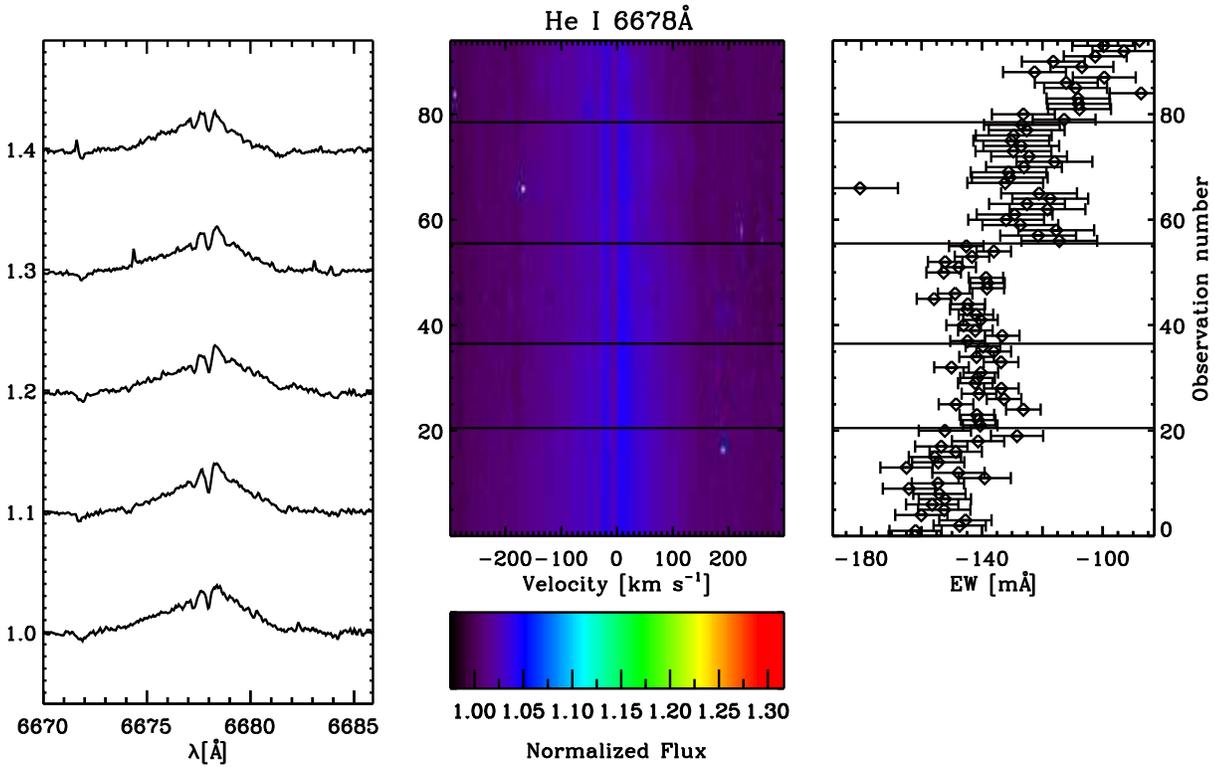}
  \caption{V1295 Aql He I 6678\AA. As in Fig. \ref{f_v1295atlas1}.}
\end{figure}
\clearpage

\begin{figure}
  \includegraphics[angle=90,width=0.9\textwidth]{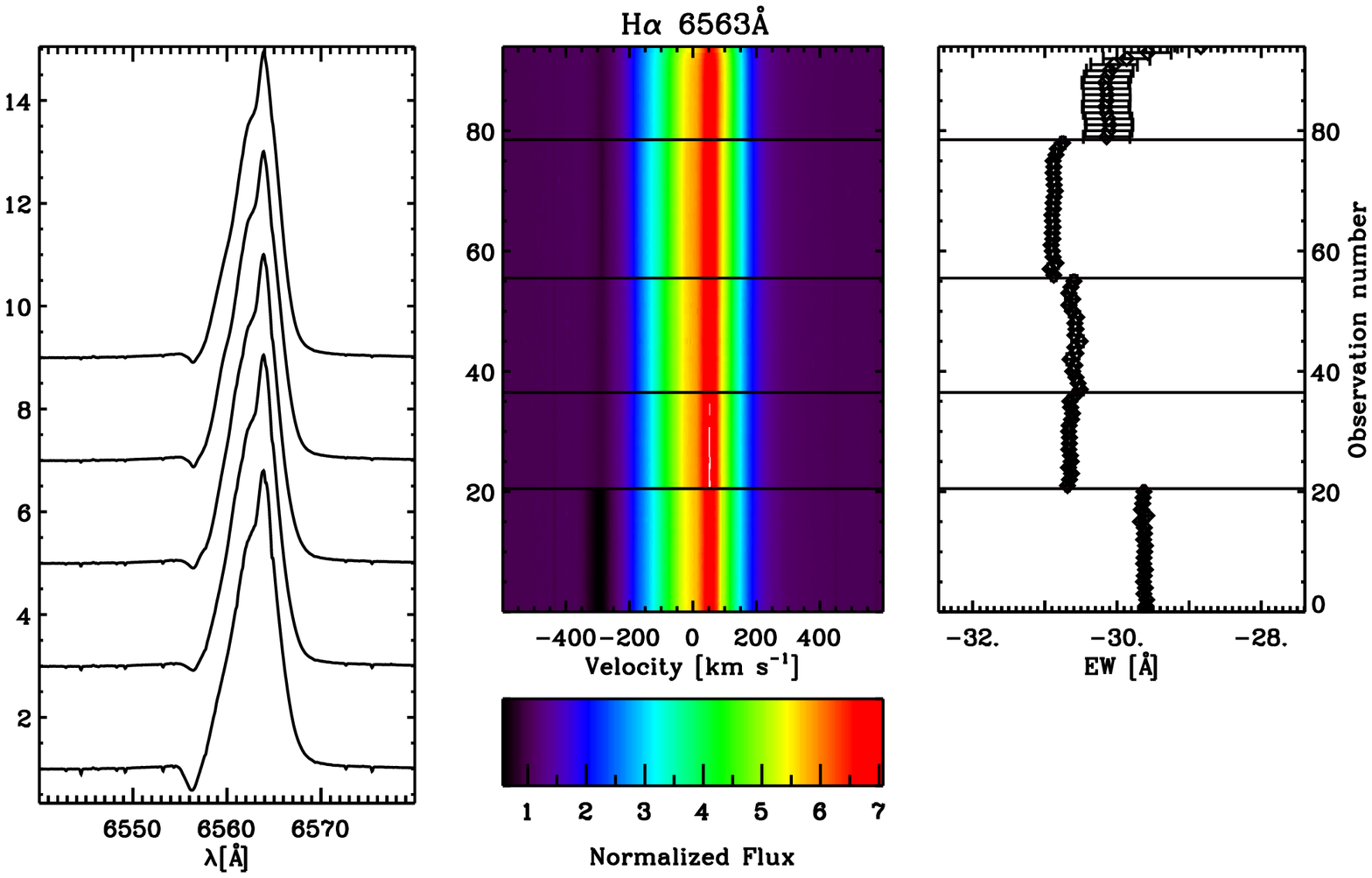}
  \caption{V1295 Aql H$\alpha$. As in Fig. \ref{f_v1295atlas1}. Here, we
  show just the fit emission components' EW for plot scaling purposes.}
\end{figure}

\begin{figure}
  \includegraphics[angle=90,width=0.9\textwidth]{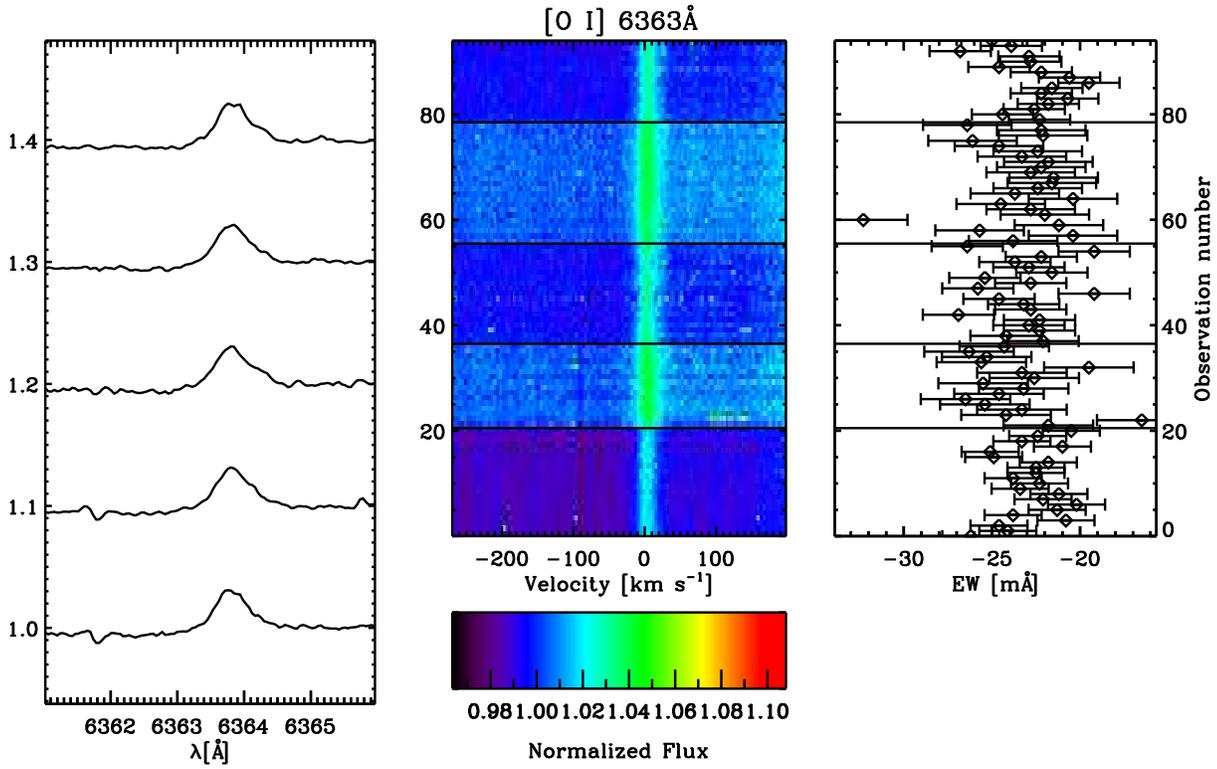}
  \caption{V1295 Aql [O I] 6363\AA. As in Fig. \ref{f_v1295atlas1}.}
\end{figure}
\clearpage

\begin{figure}
  \includegraphics[angle=90,width=0.9\textwidth]{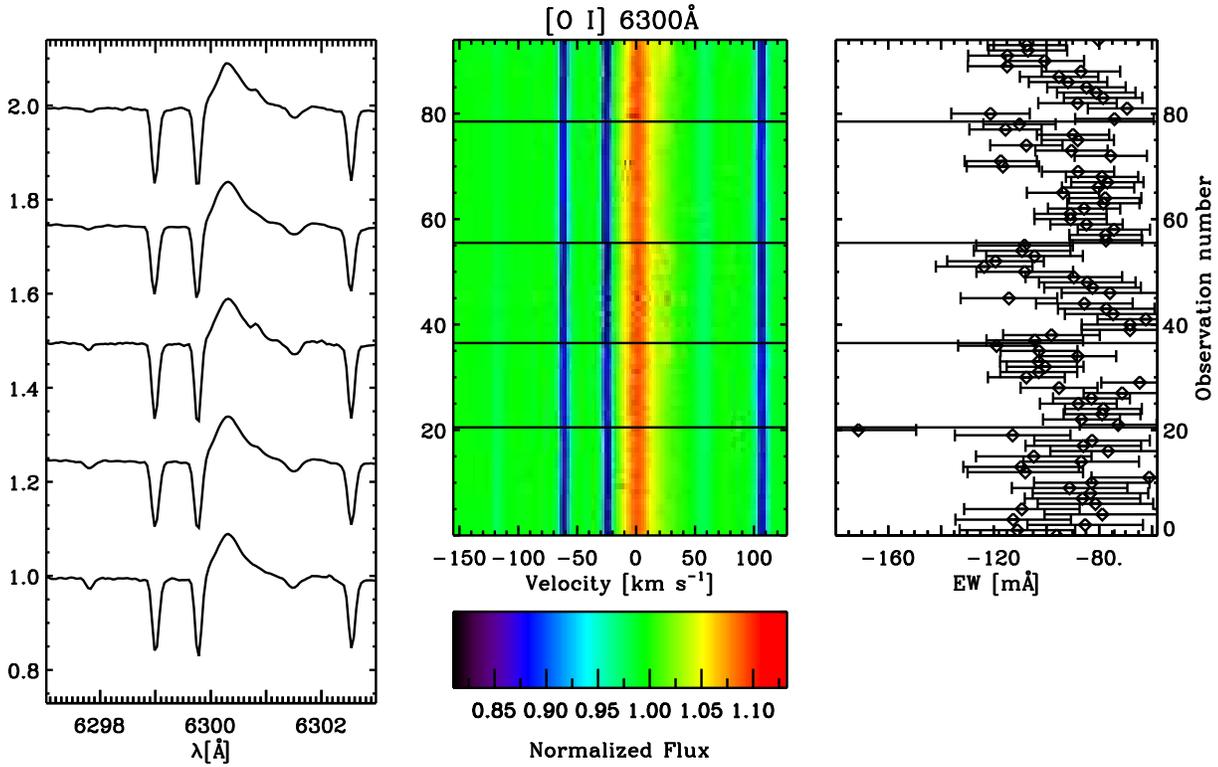}
  \caption{V1295 Aql [O I] 6300\AA. As in Fig. \ref{f_v1295atlas1}.}
\end{figure}

\begin{figure}
  \includegraphics[angle=90,width=0.9\textwidth]{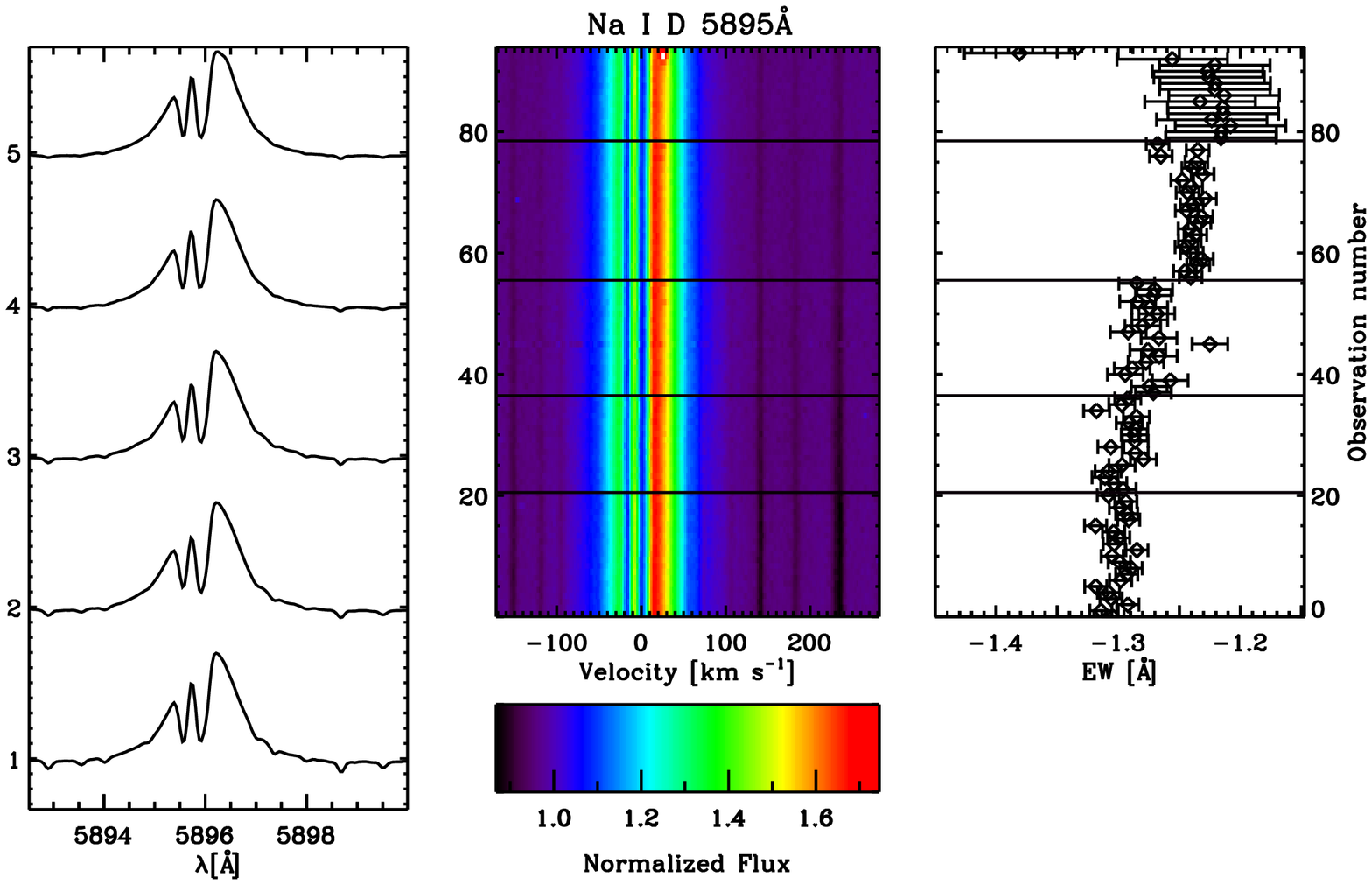}
  \caption{V1295 Aql Na I 5895\AA. As in Fig. \ref{f_v1295atlas1}. Noise
    in the last few observations, visible as hot pixels in the surface plot,
    created the scatter in the last night's EW measurements. We saw this
    occur in both Na I D lines; its origin is unclear.}
\end{figure}
\clearpage

\begin{figure}
  \includegraphics[angle=90,width=0.9\textwidth]{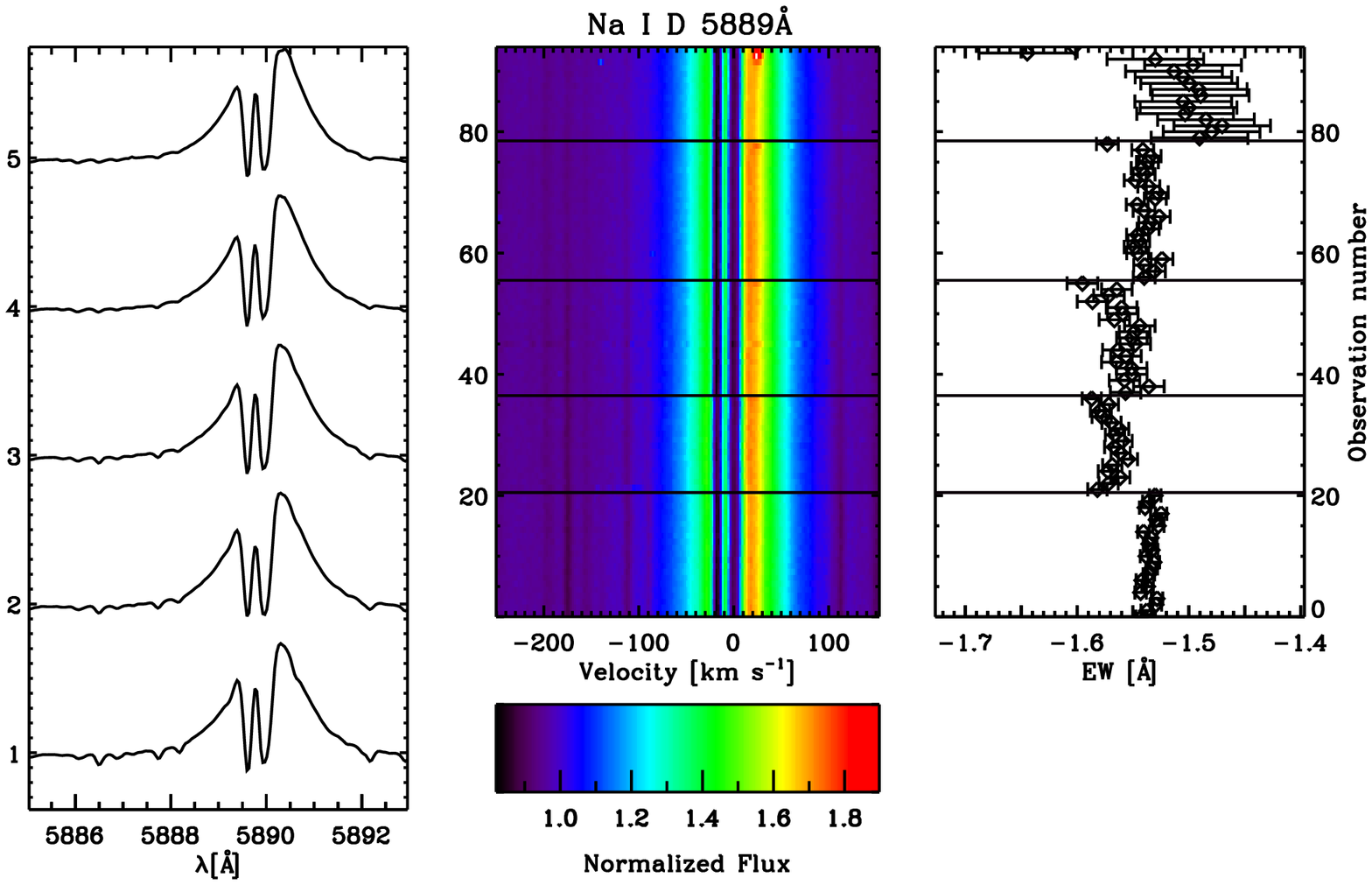}
  \caption{V1295 Aql Na I 5889\AA. As in Fig. \ref{f_v1295atlas1}. Noise
    in the last few observations, visible as hot pixels in the surface plot,
    created the scatter in the last night's EW measurements. We saw this
    occur in both Na I D lines; its origin is unclear.}
\end{figure}

\begin{figure}
  \includegraphics[angle=90,width=0.9\textwidth]{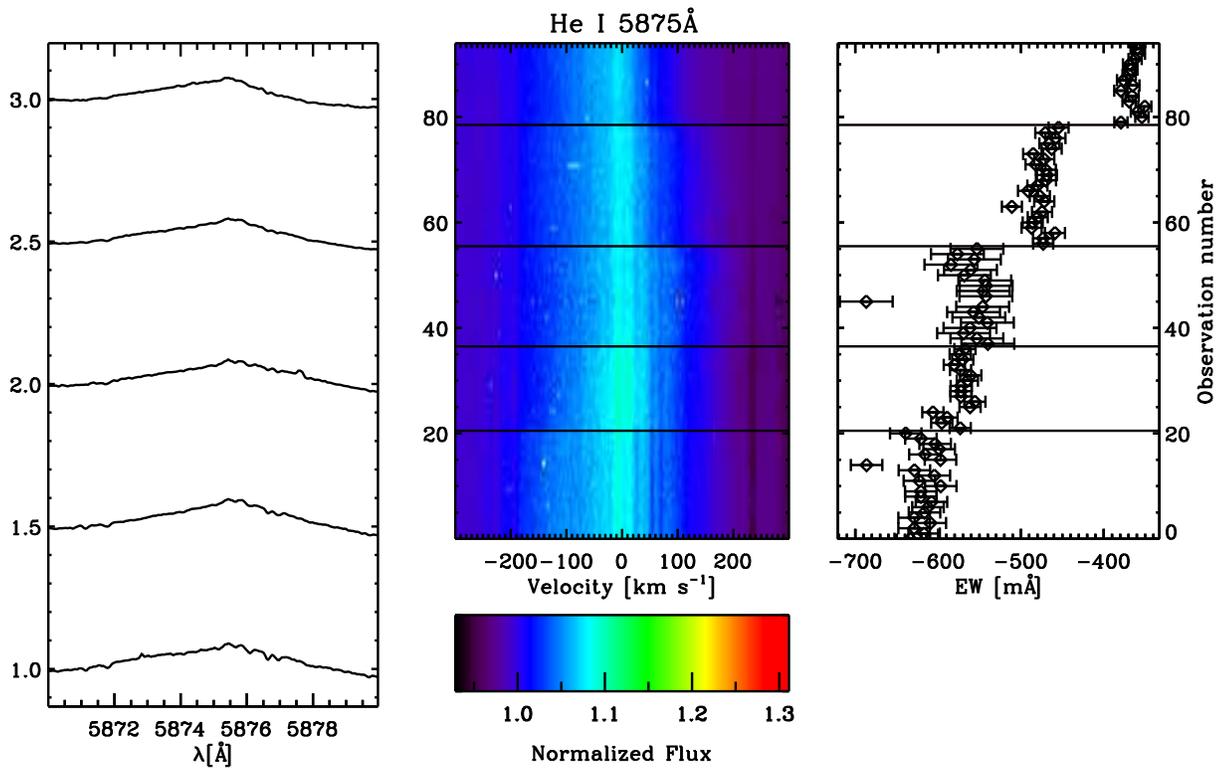}
  \caption{V1295 Aql He I 5875\AA. As in Fig. \ref{f_v1295atlas1}.}
\end{figure}
\clearpage

\begin{figure}
  \includegraphics[angle=90,width=0.9\textwidth]{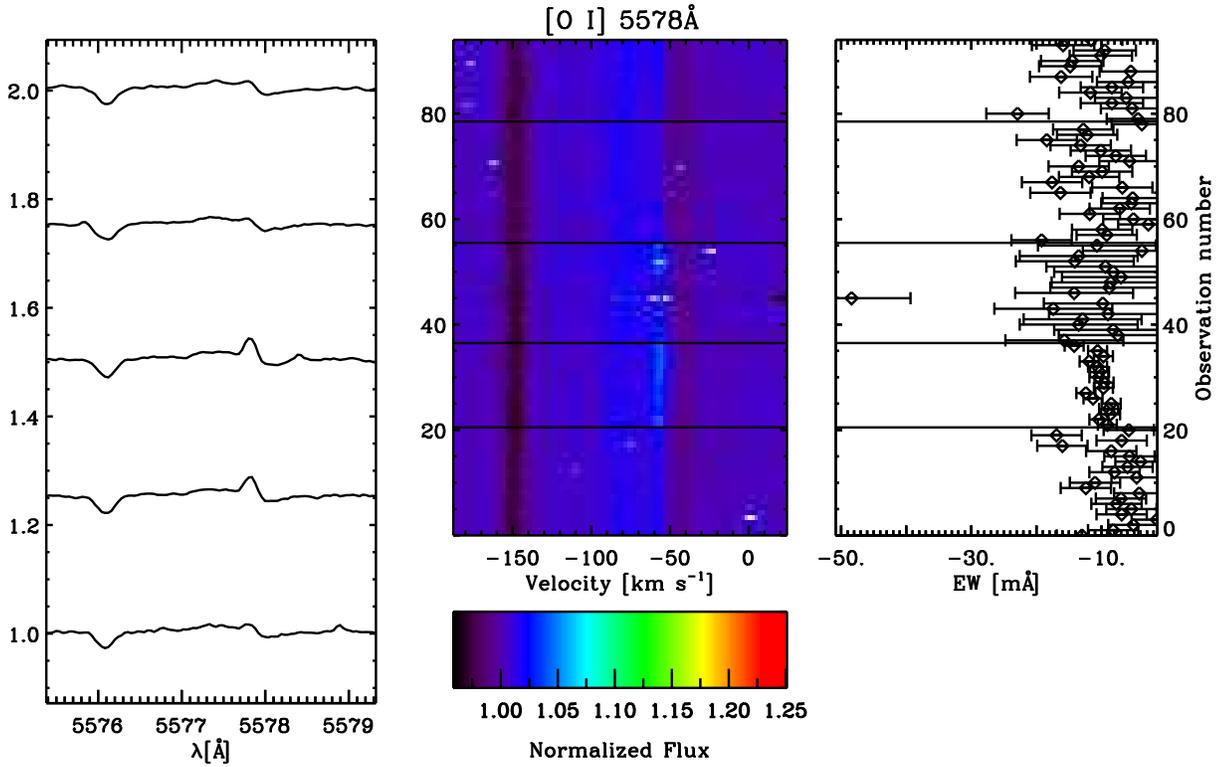}
  \caption{V1295 Aql O I 5578\AA. As in Fig. \ref{f_v1295atlas1}. We suggest this
    line, due to its weakness and night to night variability (unlike other forbidden
    lines we observe which don't vary much), could simply be airglow \citep{Louistisserand:1987}
    and neither stellar nor circumstellar in origin.}
\end{figure}

\begin{figure}
  \includegraphics[angle=90,width=0.9\textwidth]{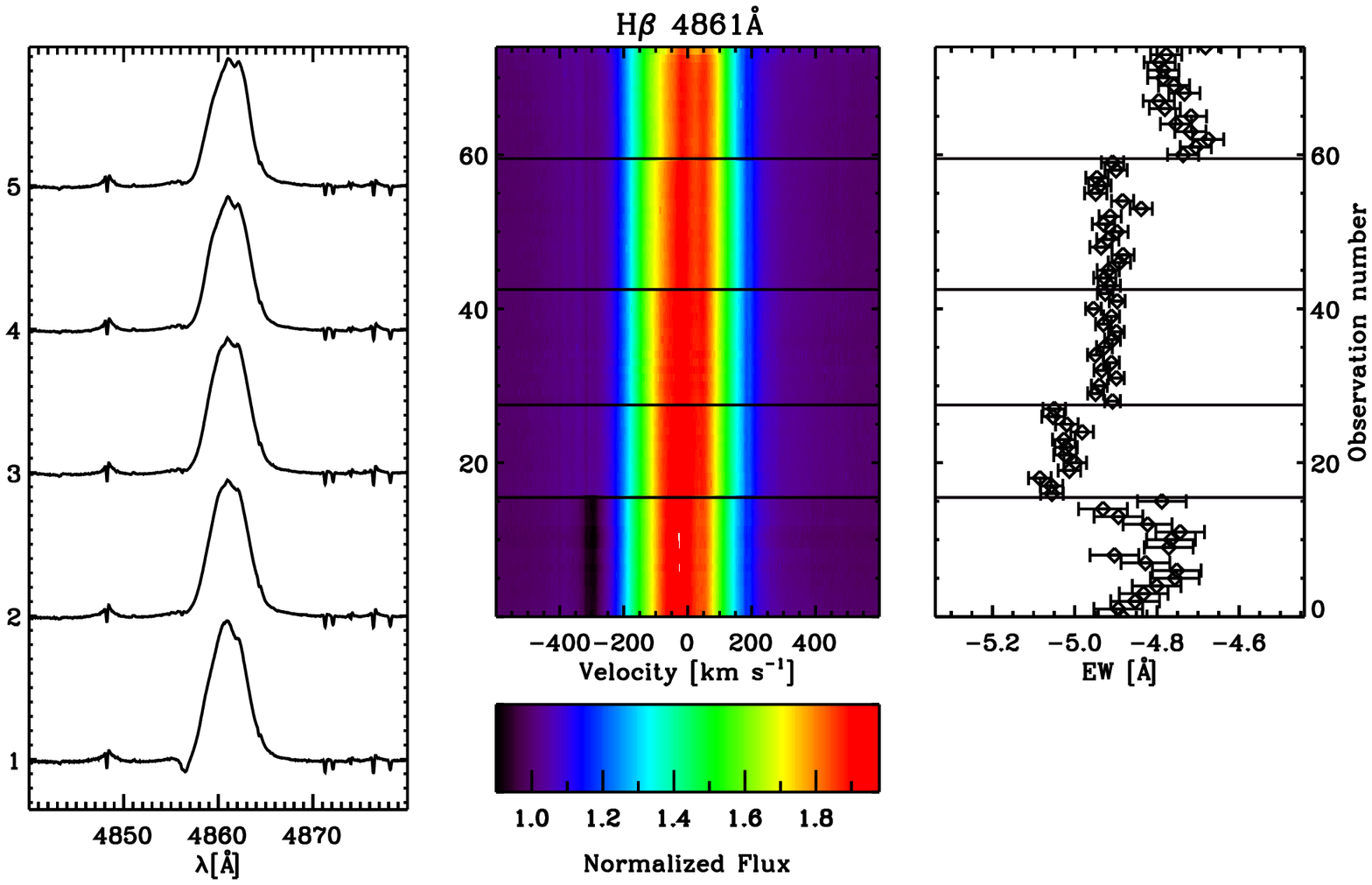}
  \caption{V1295 Aql H$\beta$. As in Fig. \ref{f_v1295atlas1}. Here, we
  show just the fit emission components' EW for plot scaling purposes.}
\end{figure}
\clearpage

\begin{figure}
  \includegraphics[angle=90,width=0.9\textwidth]{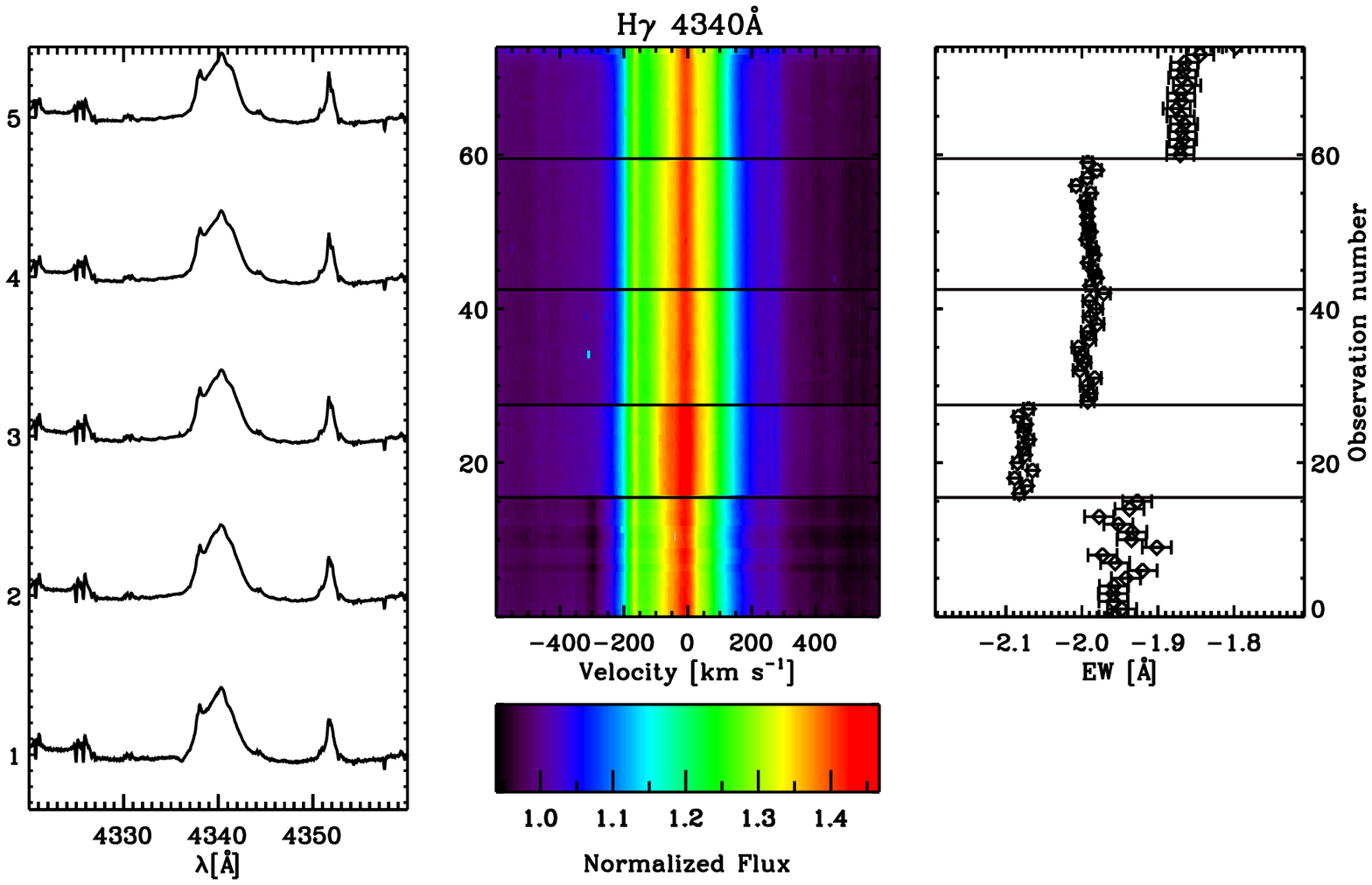}
  \caption{V1295 Aql H$\gamma$. As in Fig. \ref{f_v1295atlas1}. Here, we
  show just the fit emission components' EW for plot scaling purposes.}
\end{figure}

\clearpage

\subsection{HD 98922}

\begin{figure}
  \includegraphics[angle=90,width=0.9\textwidth]{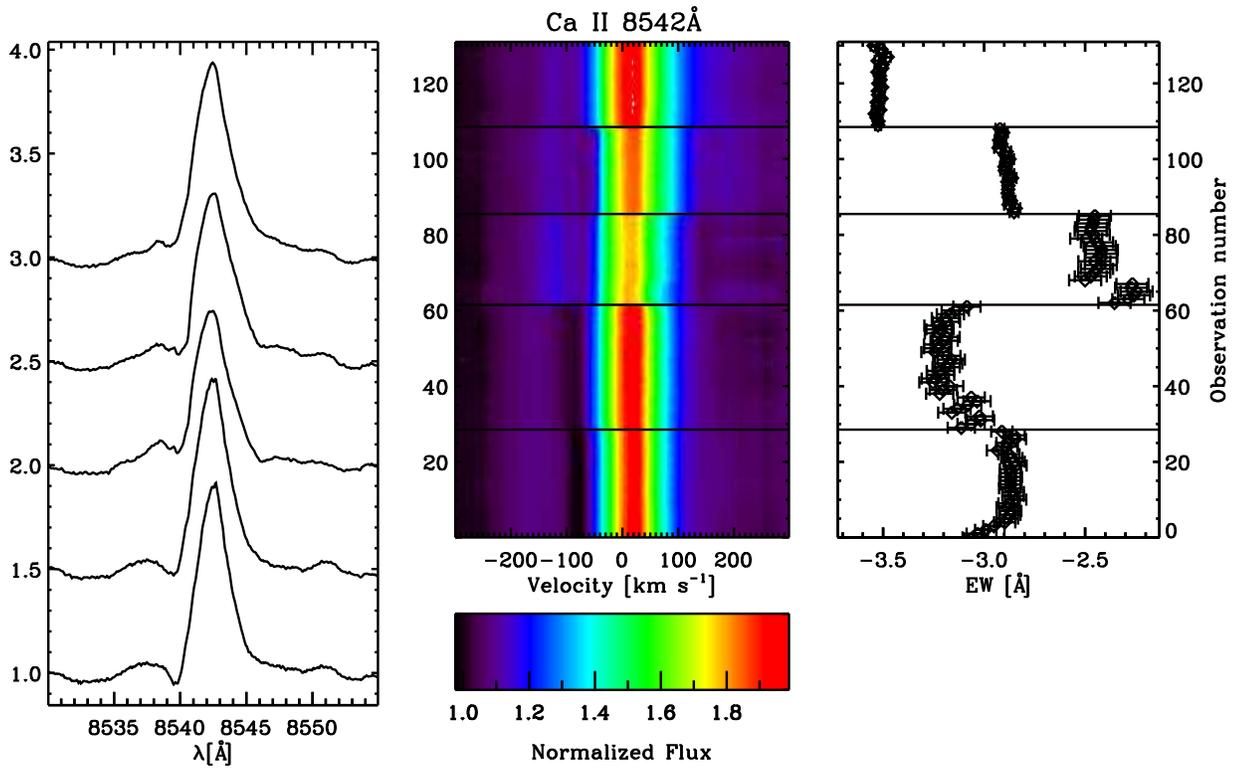}
  \caption{HD 98922 Ca II 8542\AA. As in Fig. \ref{f_hd98922atlas1}.}
\end{figure}
\clearpage

\begin{figure}
  \includegraphics[angle=90,width=0.9\textwidth]{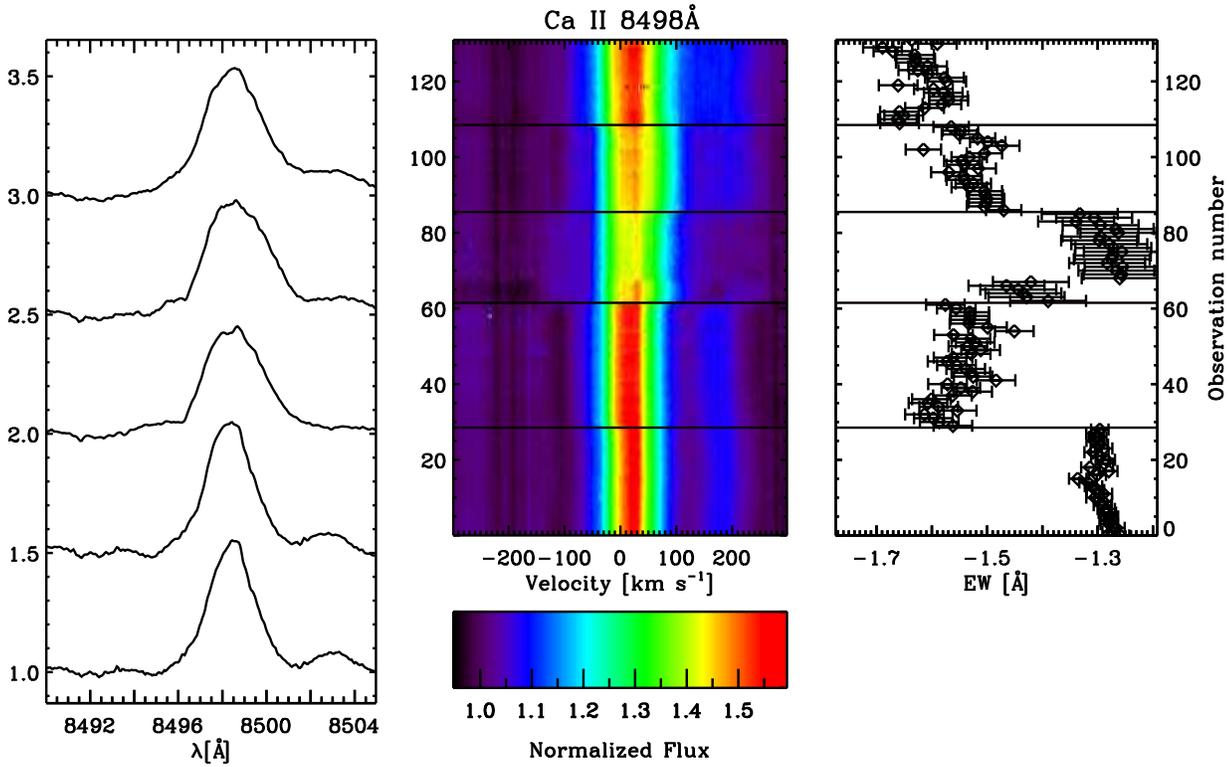}
  \caption{HD 98922 Ca II 8498\AA. As in Fig. \ref{f_hd98922atlas1}.}
\end{figure}

\begin{figure}
  \includegraphics[angle=90,width=0.9\textwidth]{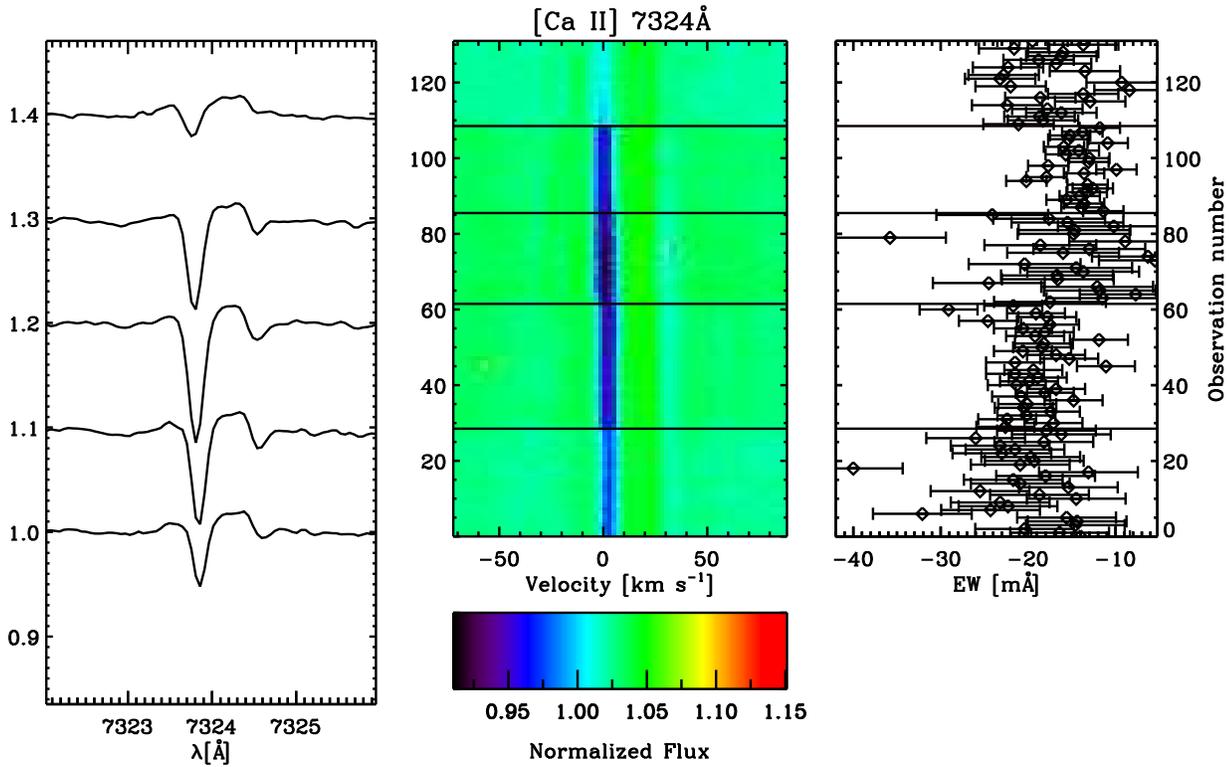}
  \caption{HD 98922 [Ca II] 7324\AA. As in Fig. \ref{f_hd98922atlas1}. Two
    water vapor lines are seen superimposed on the line and redward of
    it \citep{Kurucz:2005}. Here we show only the emission component EW.}
\end{figure}
\clearpage

\begin{figure}
  \includegraphics[angle=90,width=0.9\textwidth]{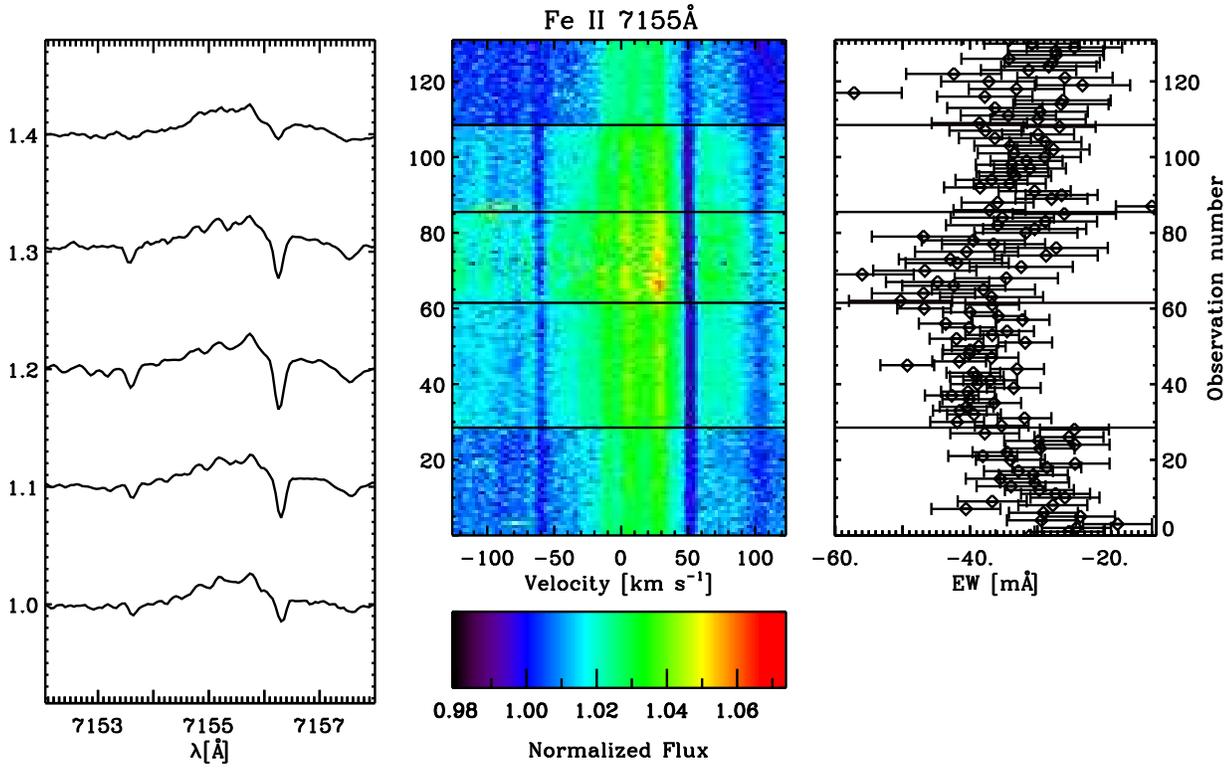}
  \caption{HD 98922 Fe II 7155\AA. As in Fig. \ref{f_hd98922atlas1}. There is a
    telluric absorption line at 7156.27\AA\ \citep{Catanzaro:1997}. Here we
    show only the emission component EW.}
\end{figure}

\begin{figure}
  \includegraphics[angle=90,width=0.9\textwidth]{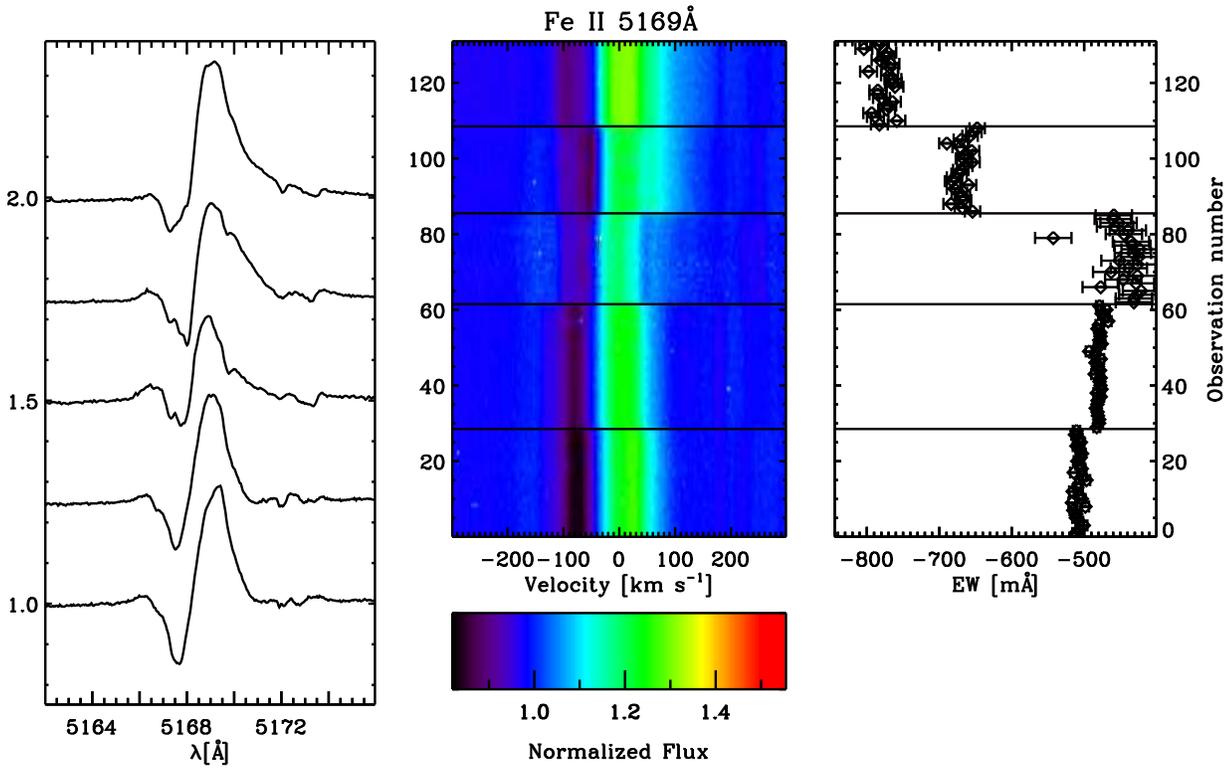}
  \caption{HD 98922 Fe II 5169\AA. As in Fig. \ref{f_hd98922atlas1}.}
\end{figure}
\clearpage

\begin{figure}
  \includegraphics[angle=90,width=0.9\textwidth]{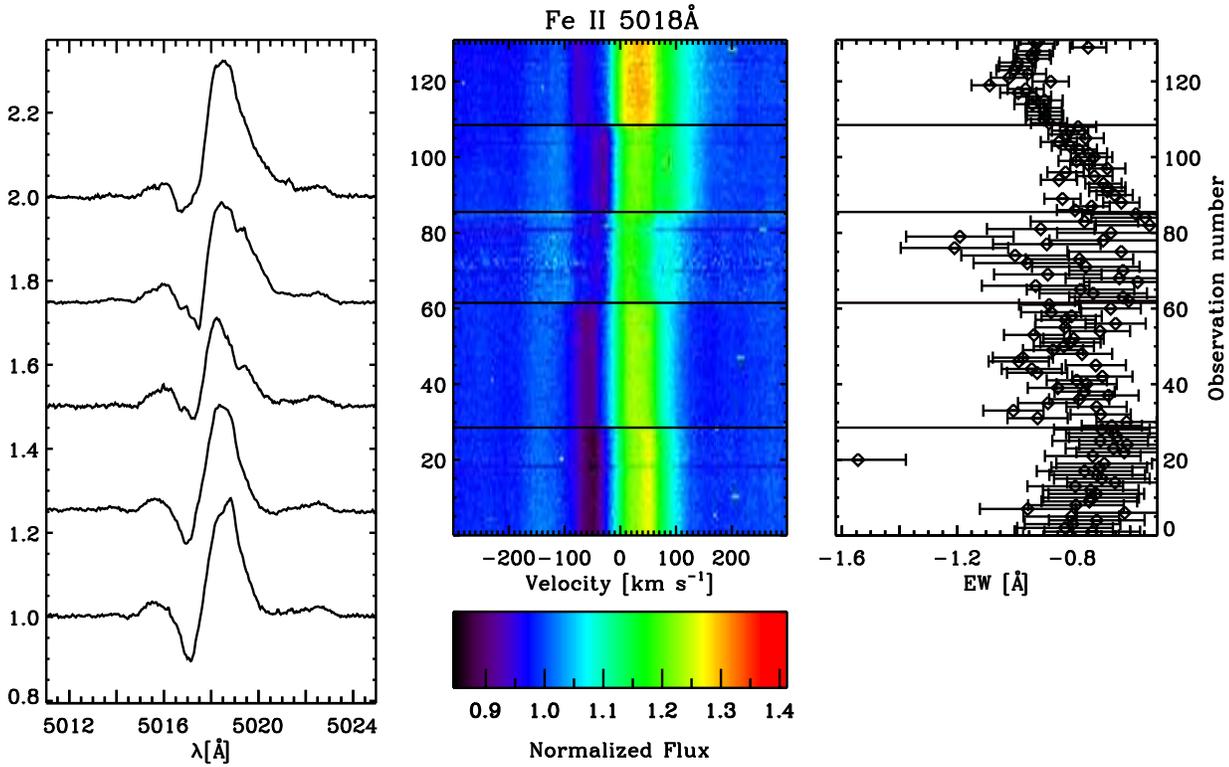}
  \caption{HD 98922 Fe II 5018\AA. As in Fig. \ref{f_hd98922atlas1}.}
\end{figure}

\begin{figure}
  \includegraphics[angle=90,width=0.9\textwidth]{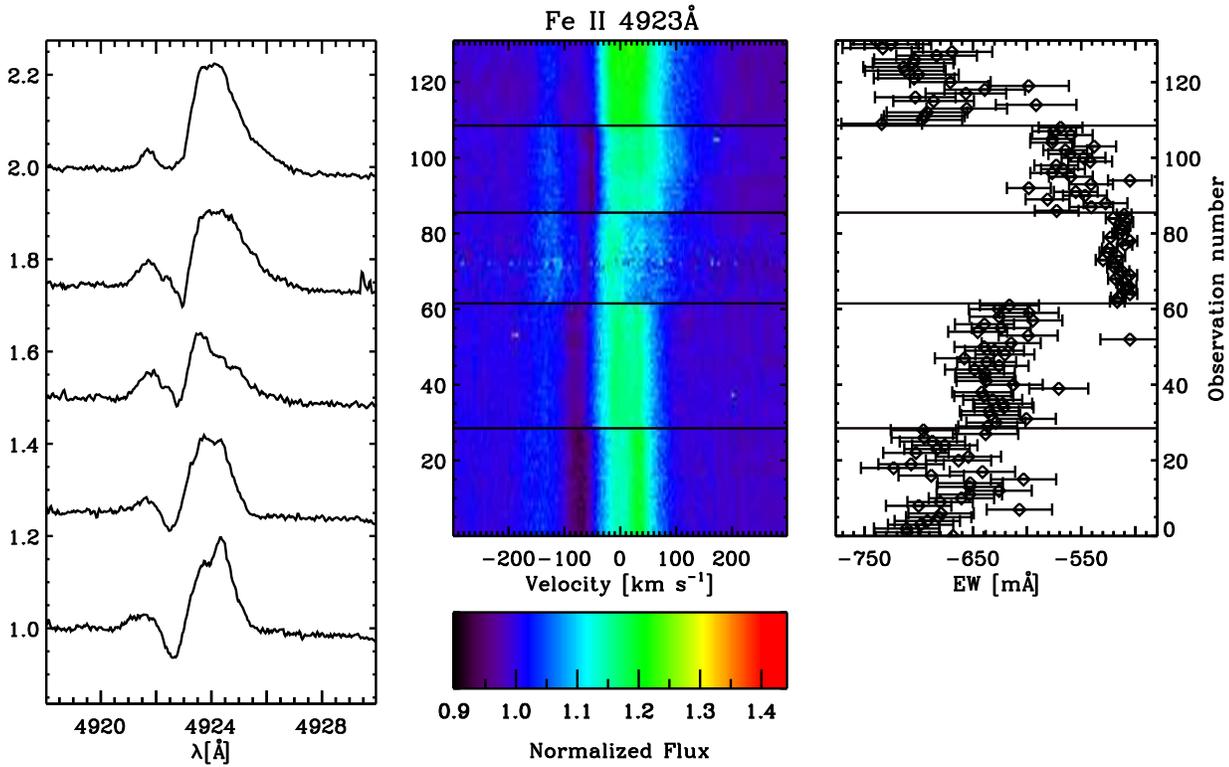}
  \caption{HD 98922 Fe II 4923\AA. As in Fig. \ref{f_hd98922atlas1}.}
\end{figure}
\clearpage

\begin{figure}
  \includegraphics[angle=90,width=0.9\textwidth]{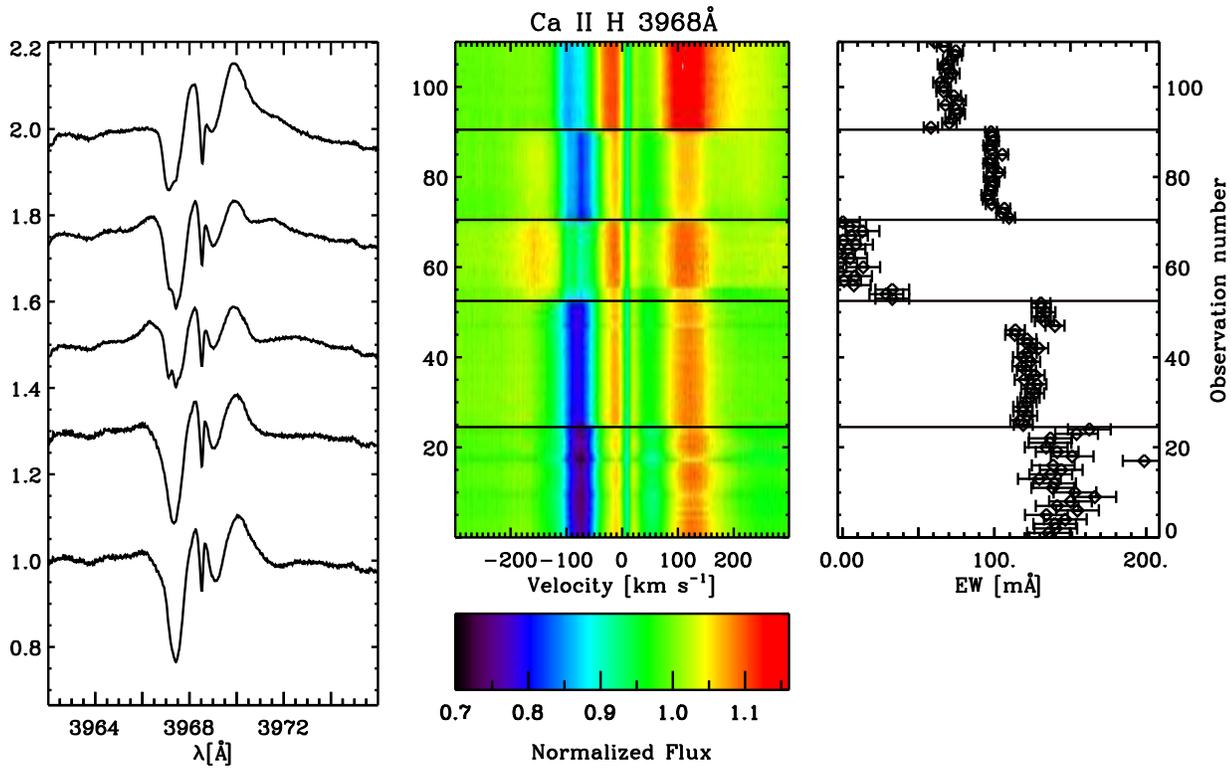}
  \caption{HD 98922 Ca II H 3968\AA. As in Fig. \ref{f_hd98922atlas1}.}
\end{figure}

\begin{figure}
  \includegraphics[angle=90,width=0.9\textwidth]{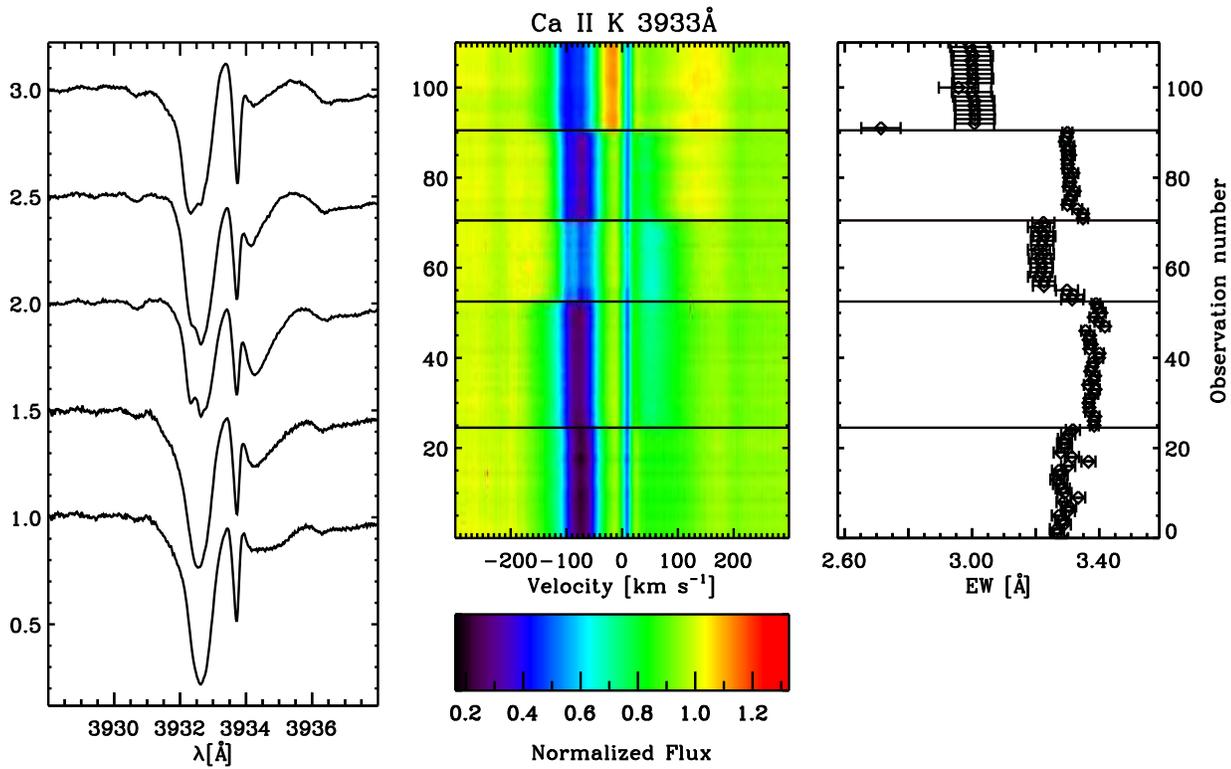}
  \caption{HD 98922 Ca II K 3933\AA. As in Fig. \ref{f_hd98922atlas1}.}
\end{figure}
\clearpage

\begin{figure}
  \includegraphics[angle=90,width=0.9\textwidth]{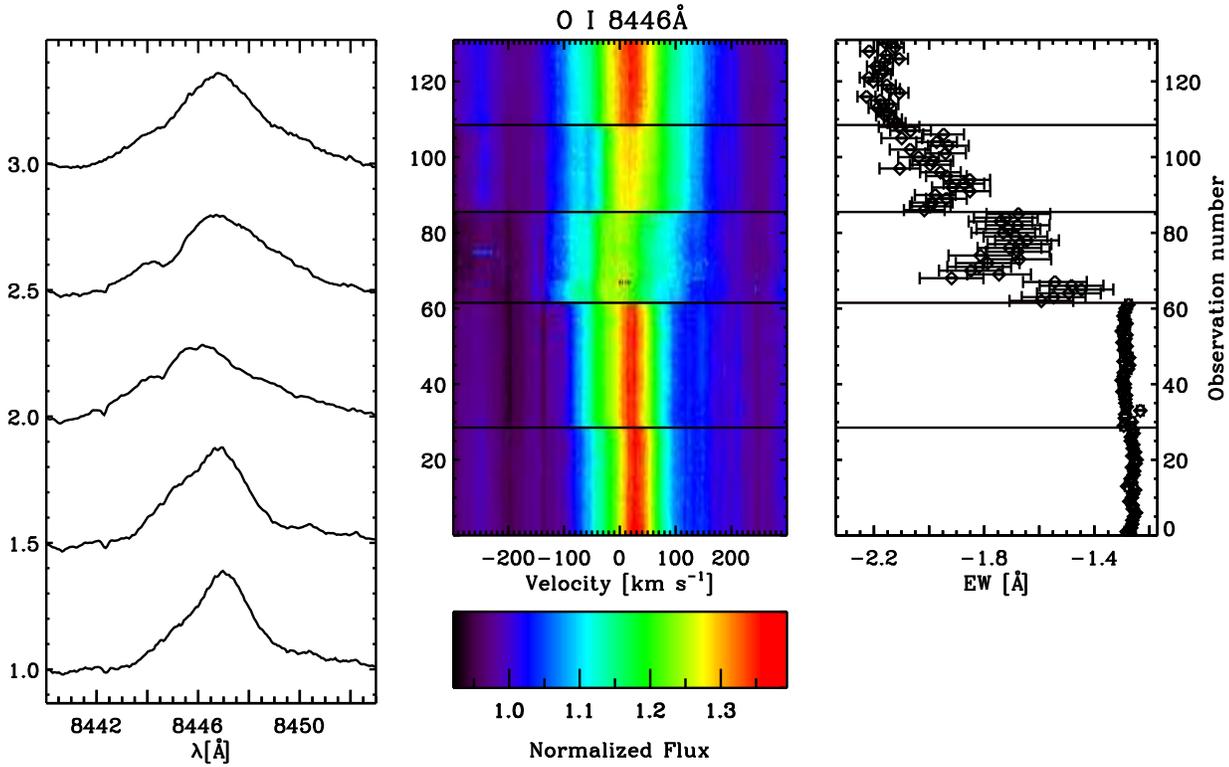}
  \caption{HD 98922 O I 8446\AA. As in Fig. \ref{f_hd98922atlas1}.}
\end{figure}

\begin{figure}
  \includegraphics[angle=90,width=0.9\textwidth]{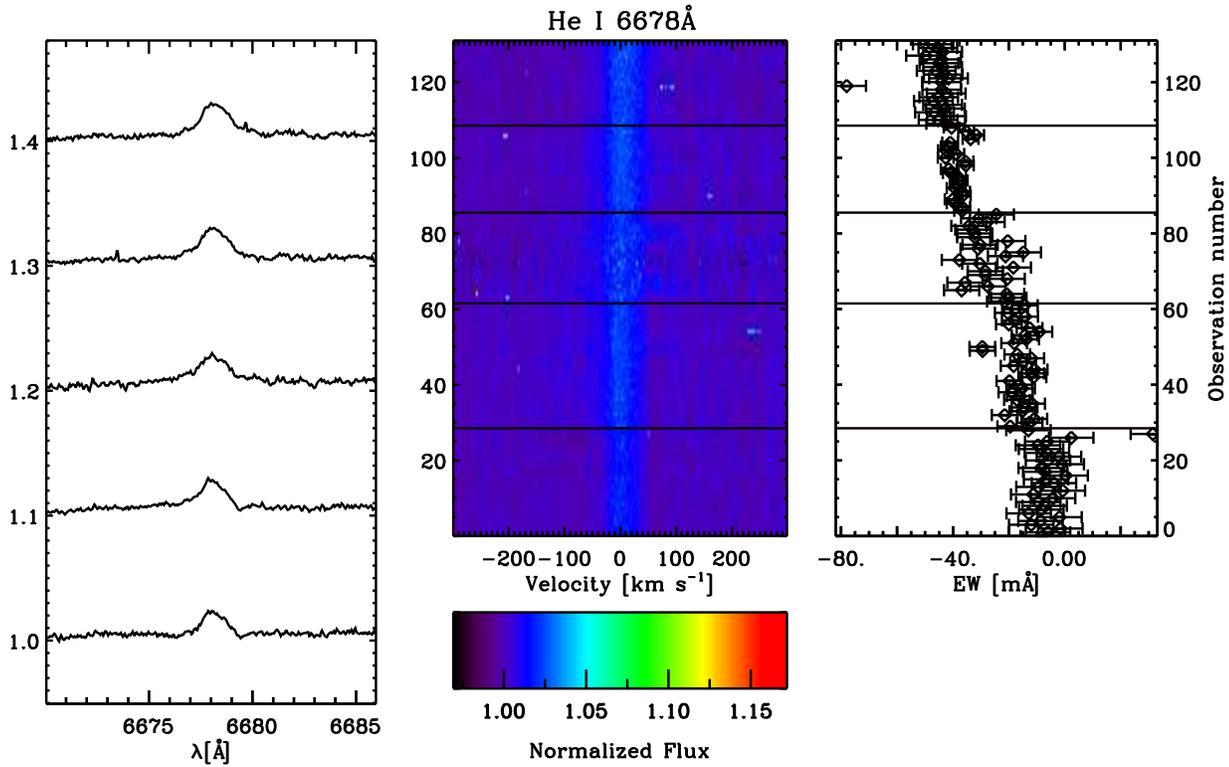}
  \caption{HD 98922 He I 6678\AA. As in Fig. \ref{f_hd98922atlas1}.}
\end{figure}
\clearpage

\begin{figure}
  \includegraphics[angle=90,width=0.9\textwidth]{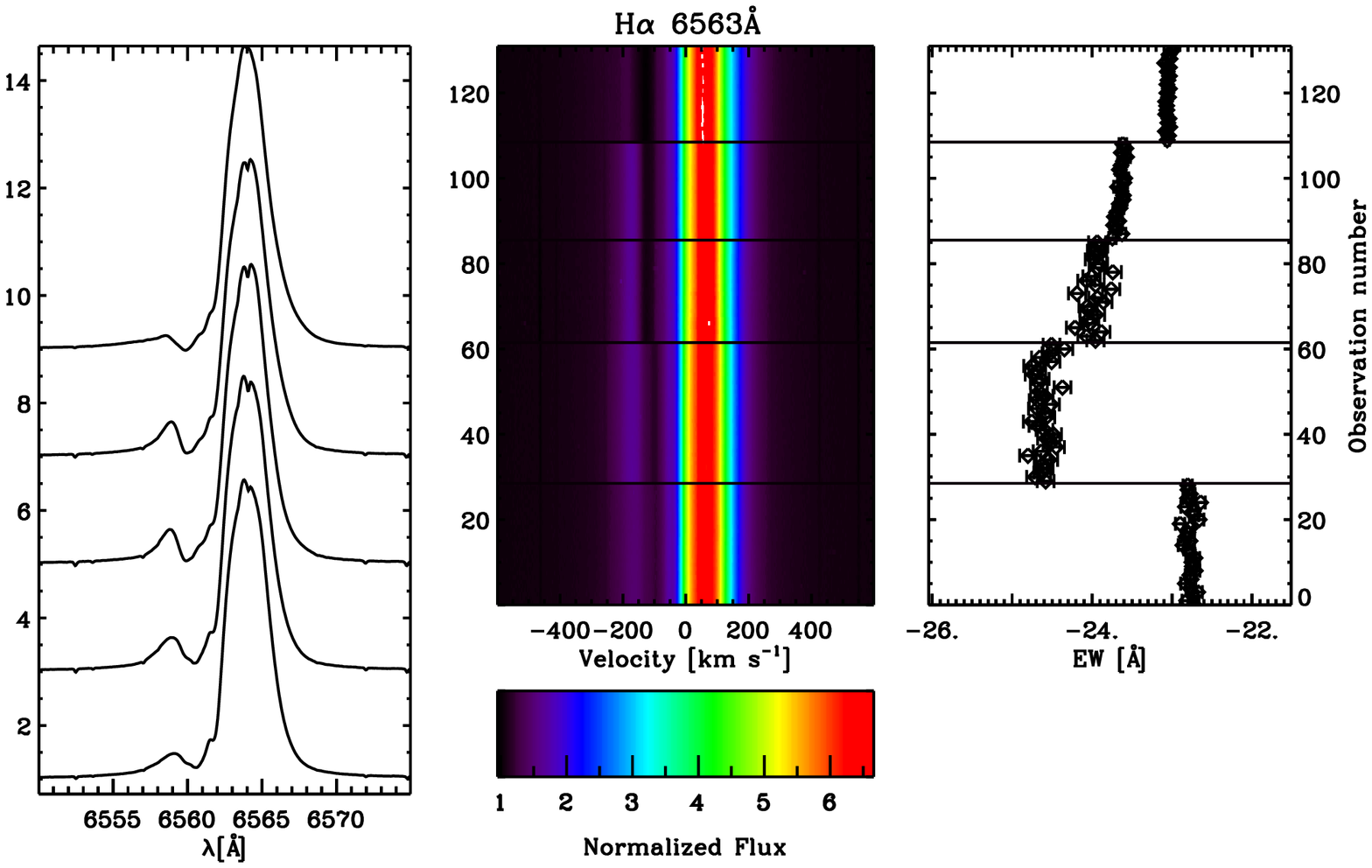}
  \caption{HD 98922 H$\alpha$. As in Fig. \ref{f_hd98922atlas1}. Here, we
  show just the fit emission components' EW for plot scaling purposes.}
\end{figure}

\begin{figure}
  \includegraphics[angle=90,width=0.9\textwidth]{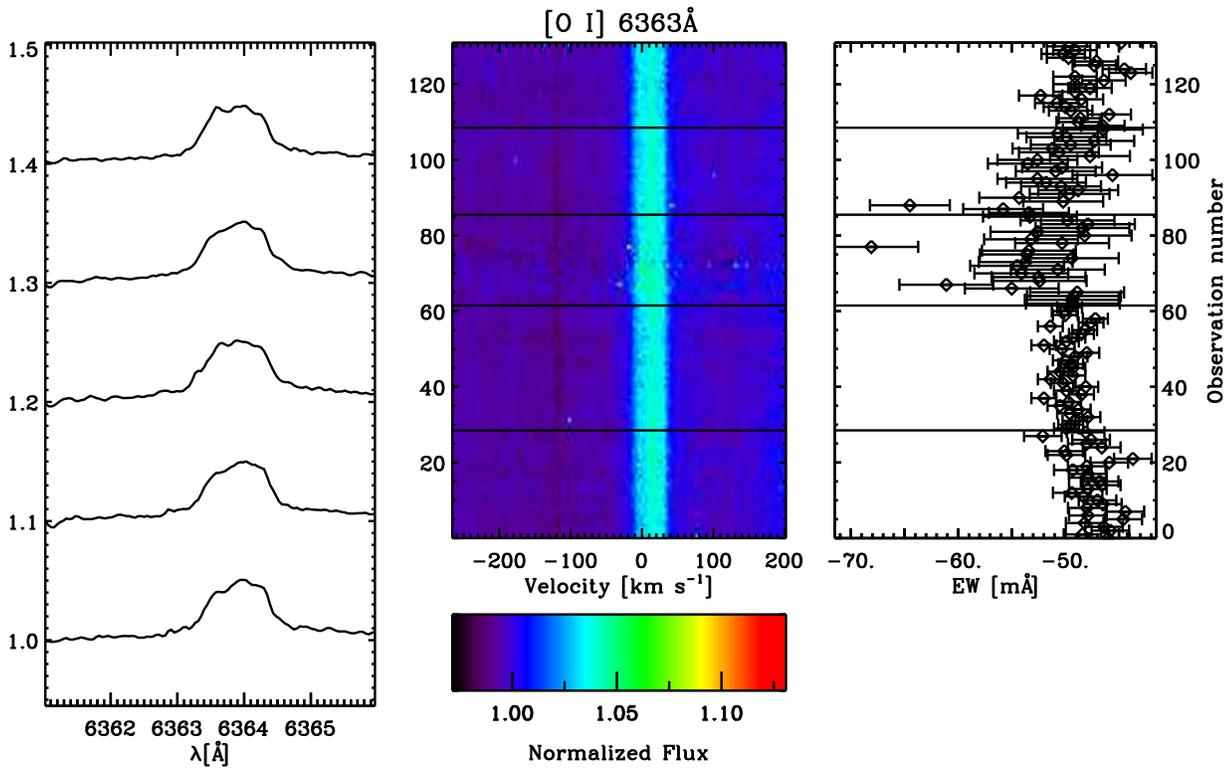}
  \caption{HD 98922 [O I] 6363\AA. As in Fig. \ref{f_hd98922atlas1}.}
\end{figure}
\clearpage

\begin{figure}
  \includegraphics[angle=90,width=0.9\textwidth]{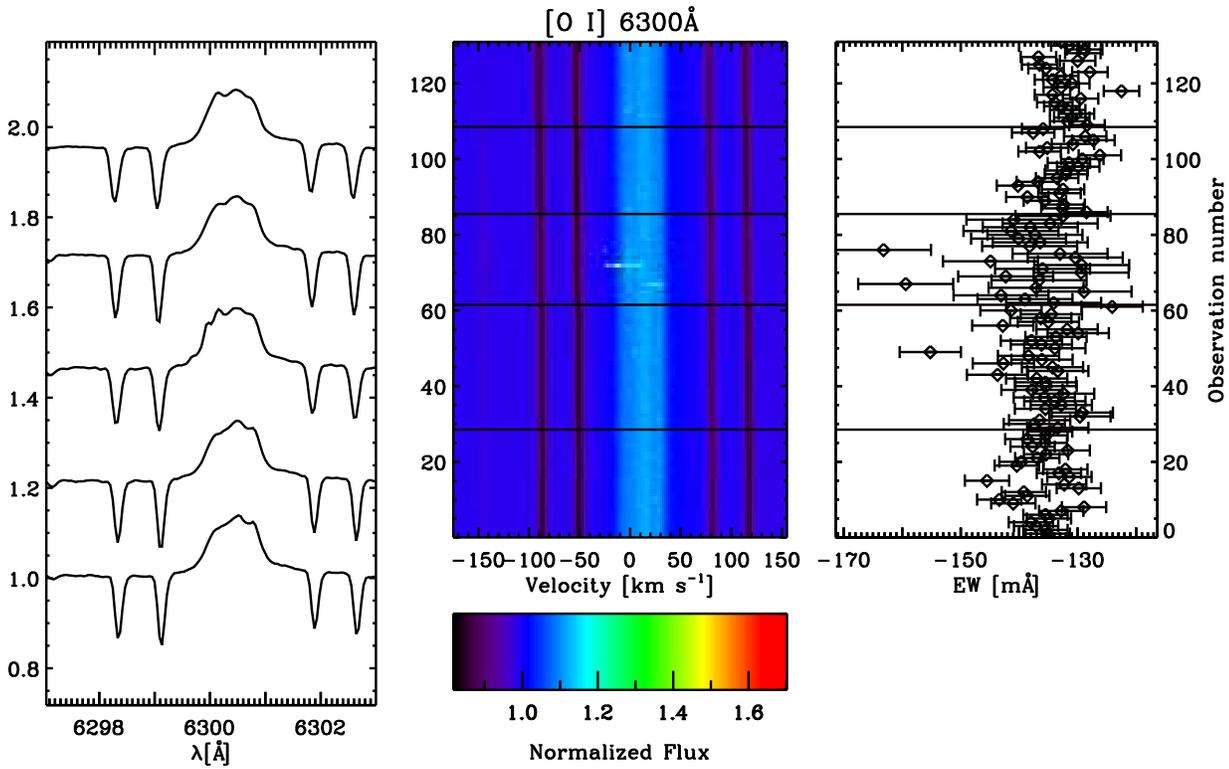}
  \caption{HD 98922 [O I] 6300\AA. As in Fig. \ref{f_hd98922atlas1}.}
\end{figure}

\begin{figure}
  \includegraphics[angle=90,width=0.9\textwidth]{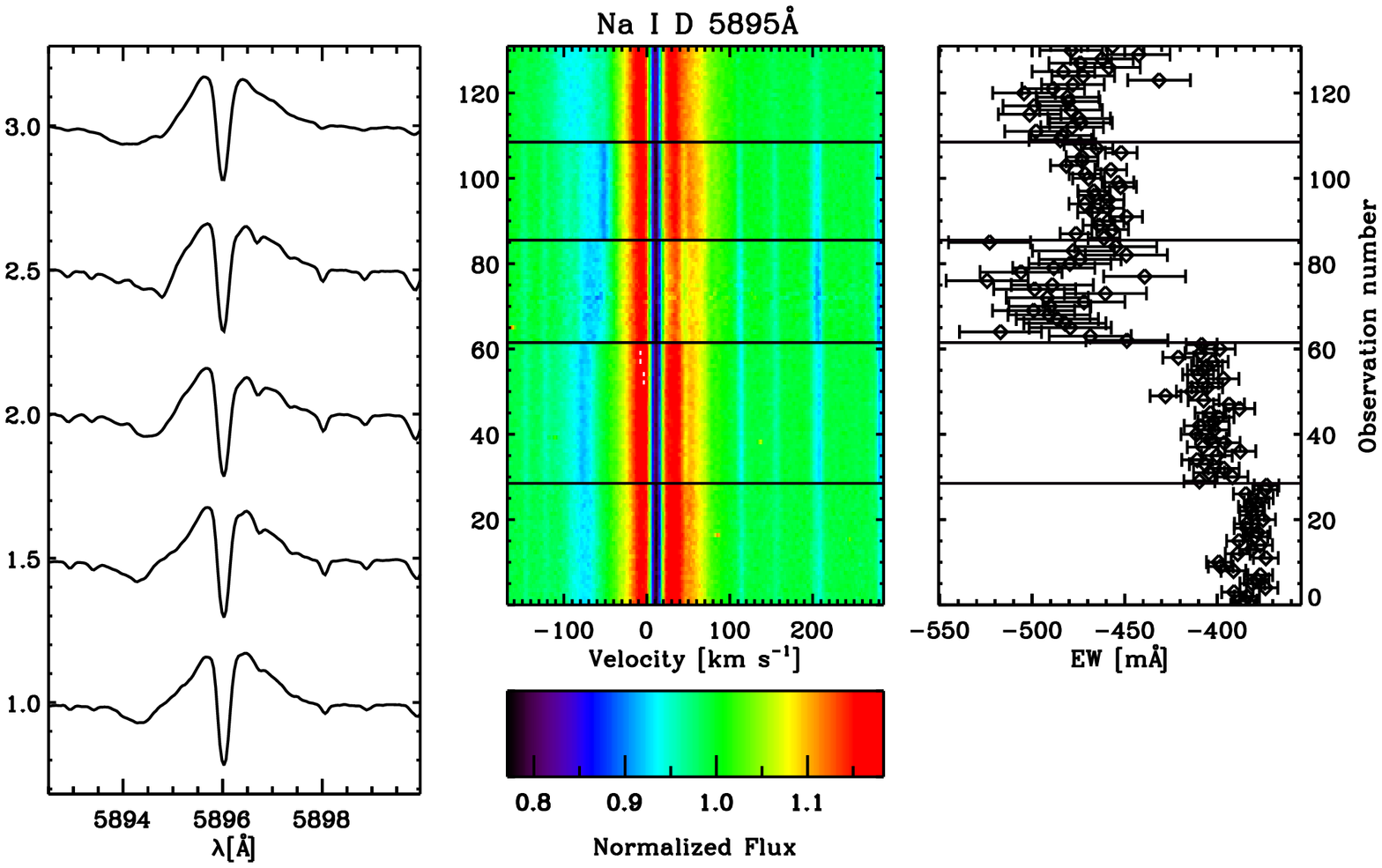}
  \caption{HD 98922 Na I 5895\AA. As in Fig. \ref{f_hd98922atlas1}.}
\end{figure}
\clearpage

\begin{figure}
  \includegraphics[angle=90,width=0.9\textwidth]{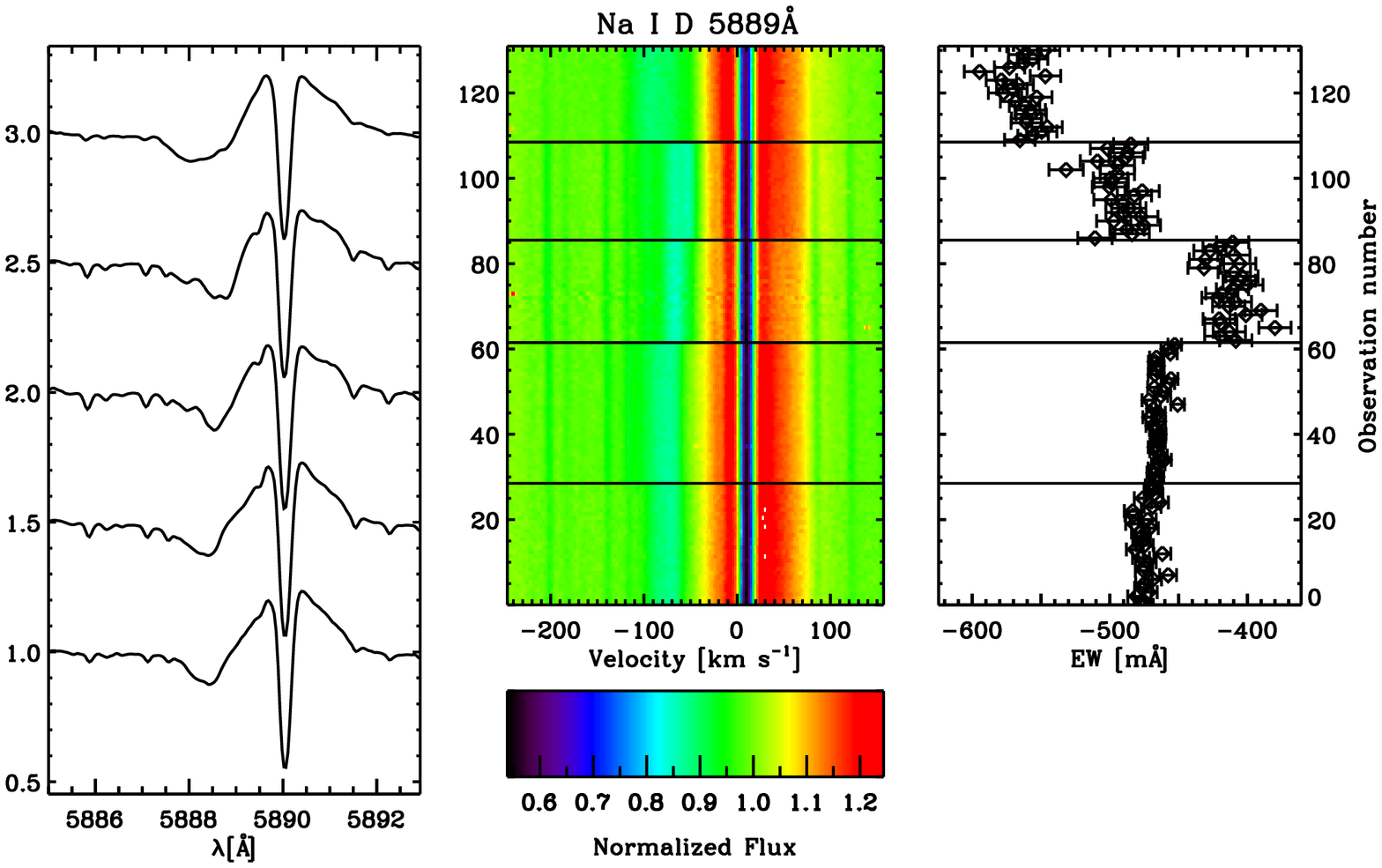}
  \caption{HD 98922 Na I 5889\AA. As in Fig. \ref{f_hd98922atlas1}.}
\end{figure}

\begin{figure}
  \includegraphics[angle=90,width=0.9\textwidth]{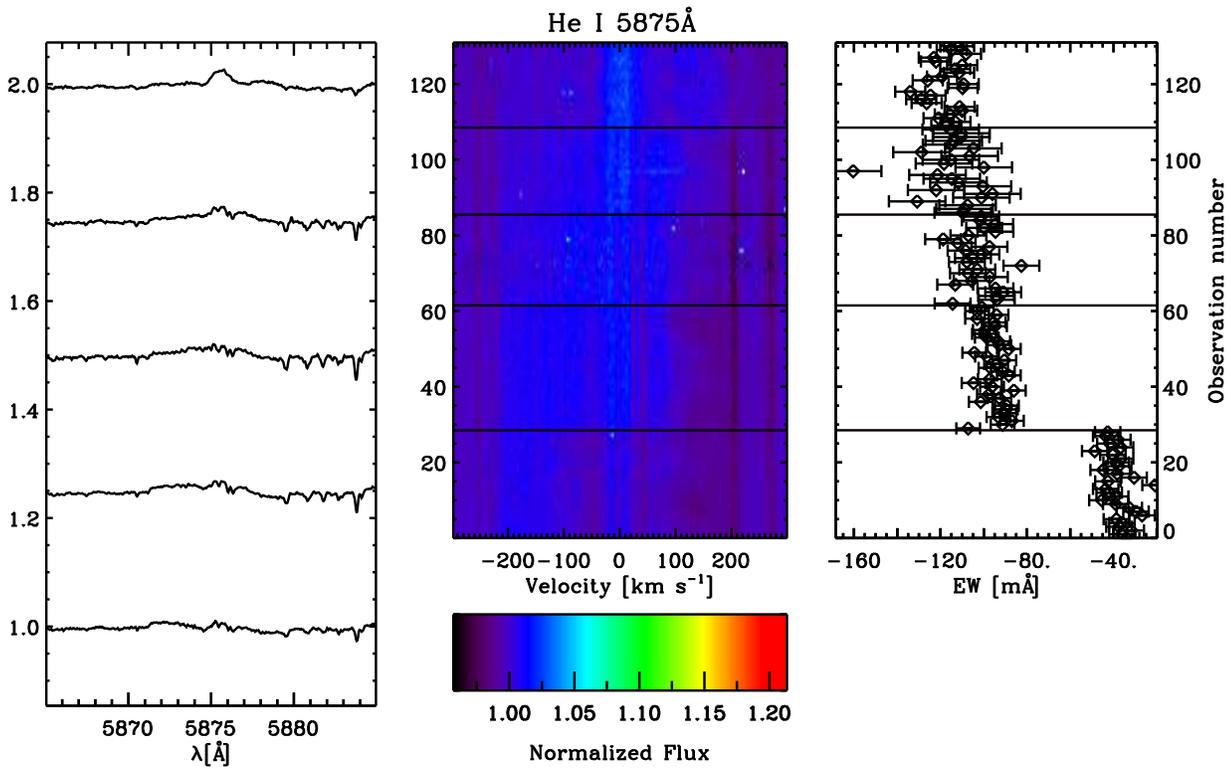}
  \caption{HD 98922 He I 5875\AA. As in Fig. \ref{f_hd98922atlas1}.}
\end{figure}
\clearpage

\begin{figure}
  \includegraphics[angle=90,width=0.9\textwidth]{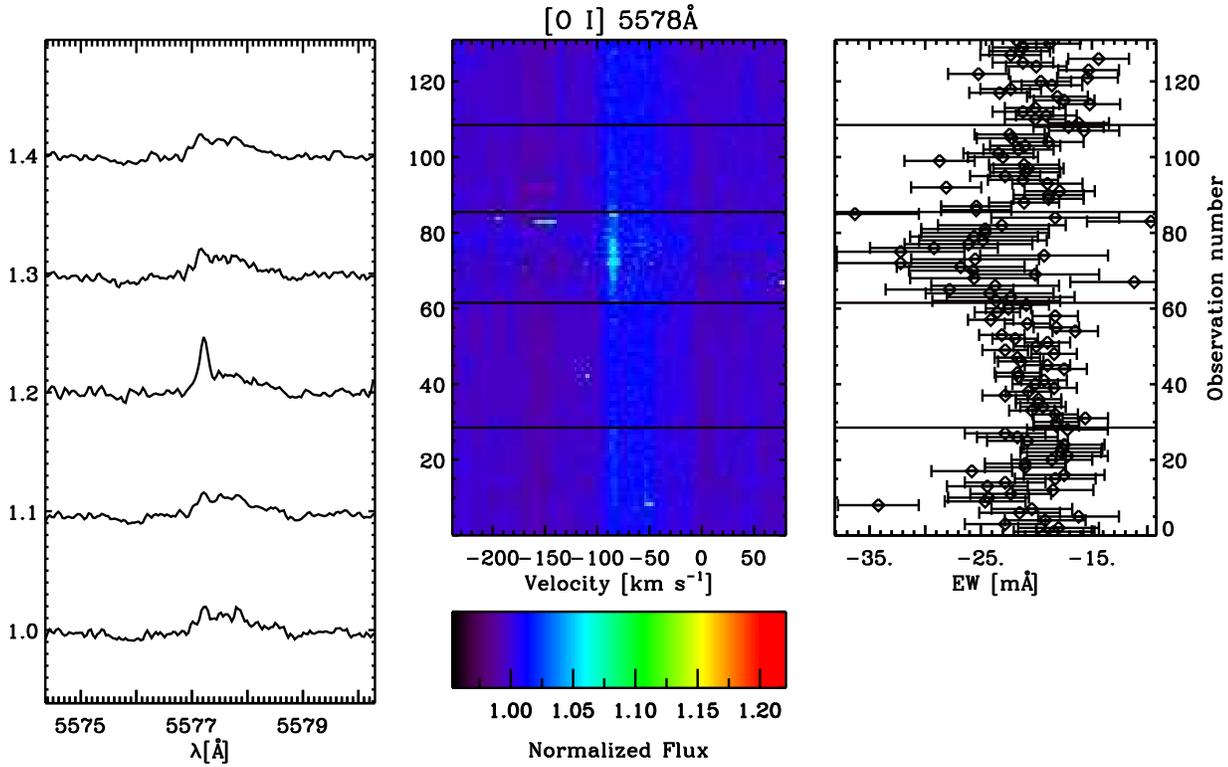}
  \caption{HD 98922 O I 5578\AA. As in Fig. \ref{f_hd98922atlas1}. Due to its
    persistence, we suggest the weak, broad emission line is stellar or circumstellar
    in origin. The narrow component that arises and fluctuates within the third night
    we believe to be airglow \citep{Louistisserand:1987}.}
\end{figure}

\begin{figure}
  \includegraphics[angle=90,width=0.9\textwidth]{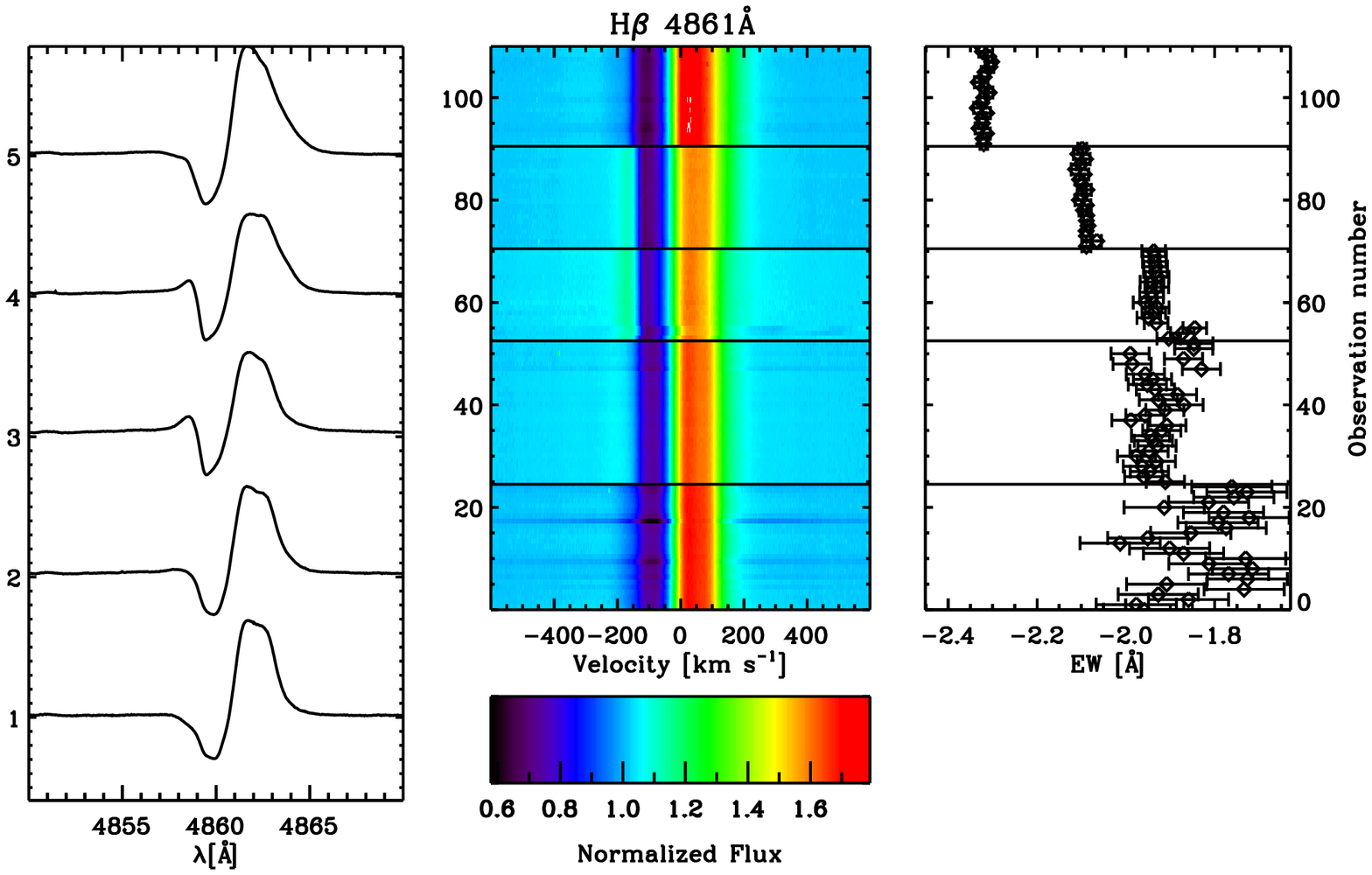}
  \caption{HD 98922 H$\beta$. As in Fig. \ref{f_hd98922atlas1}. Here, we
  show just the fit emission components' EW for plot scaling purposes.}
\end{figure}
\clearpage

\begin{figure}
  \includegraphics[angle=90,width=0.9\textwidth]{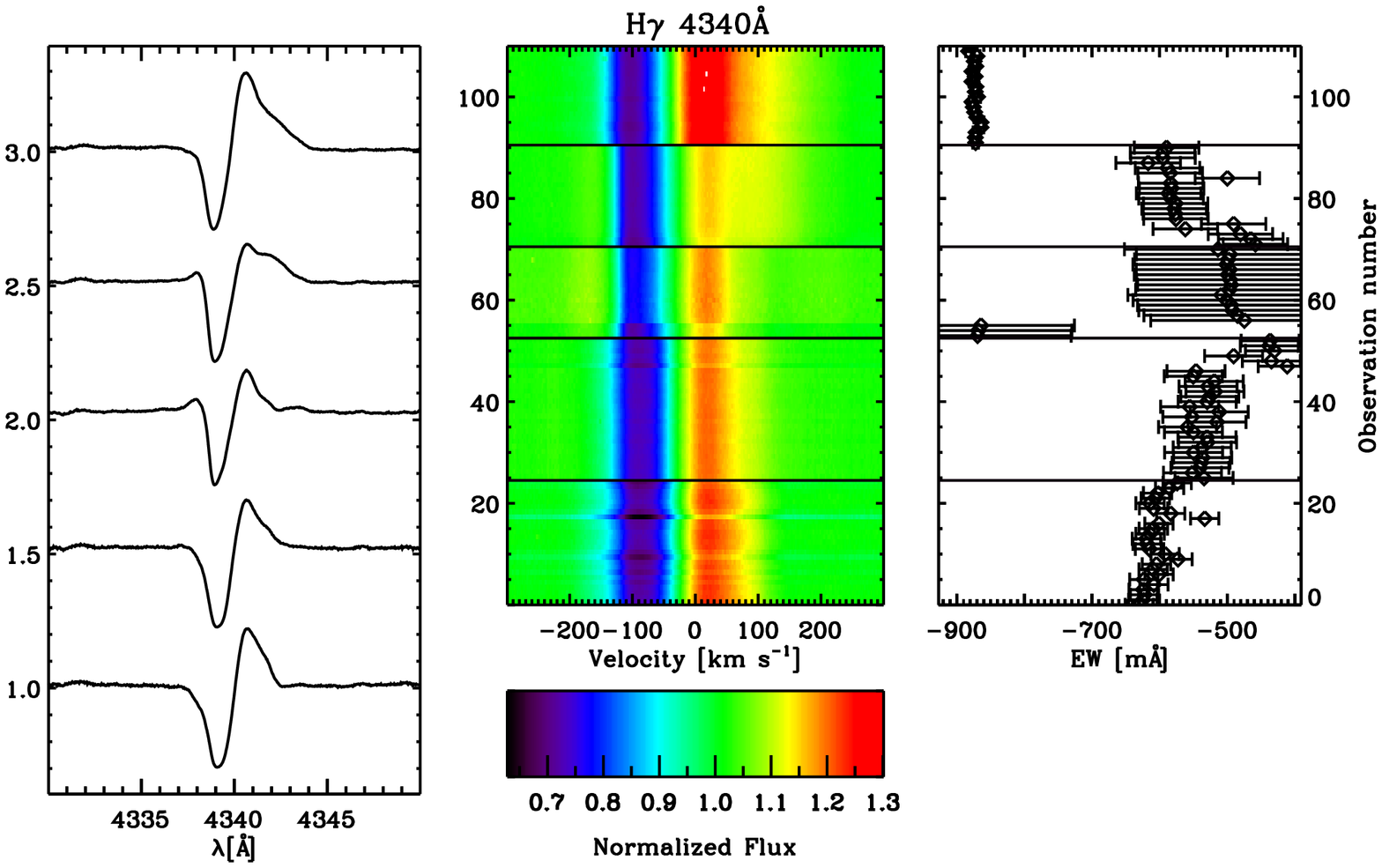}
  \caption{HD 98922 H$\gamma$. As in Fig. \ref{f_hd98922atlas1}. Here, we
  show just the fit emission components' EW for plot scaling purposes.}
\end{figure}

\clearpage

\section{B. TORUS models of V1295 Aquila and HD 98922}\label{s_appendixT}

\subsection{Model setup}

As a means for exploring mechanisms for line formation in the systems we observed, we have calculated line
transfer with TORUS for systems comprised of a star, magnetosphere, disk, and disk wind.
The modular framework of TORUS as applied to star$+$disk systems is 
described in more detail in \citet{Kurosawa:2006}; we highlight salient details for our
Herbig AeBe-specific application below.
Our TORUS models calculate level populations by solving the rate equations for Hydrogen assuming non-LTE conditions.
Doppler broadening is applied to the line profiles assuming turbulent velocity in the flow of 20 km s$^{-1}$; pressure
broadening as formulated in \citet{Muzerolle:2001} is also applied and is the dominant broadening mechanism for the
lines we study here. 

The radial extent at which a dipolar field of a given strength can hold off infall \citep{Konigl:1991}, $r_{T}$, 
is expressed \citep[using the formulation of ][]{Hartmann:1998} as:
\begin{equation}
\frac{r_T}{R_*} = 3.7 B_{3}^{4/7} \dot{M}_{-7}^{-2/7} M_{1/2}^{-1/7} R_{2}^{5/7};
\end{equation}
where $B_{\mathrm 3}$ is the magnetic field strength in units of kG, $\dot{M}_{\mathrm -7}$ is the mass accretion rate
in units of 10$^{-7}$ M$_{\odot}$ yr$^{-1}$, M$_{\mathrm 1/2}$ is the stellar mass in units of 0.5 M$_{\odot}$, and
R$_{\mathrm 2}$ is the stellar radius in units of 2 R$_{\odot}$.

For a fiducial case, we use parameters for the one object we study here with a detected magnetic field, V1295 Aql.
For its measured $\sim$100 G field, stellar parameters as reported in Table \ref{t_params} and an
observationally determined accretion rate of $\sim$10$^{-5}$ M$_{\odot}$ yr$^{-1}$ \citep{Mendigutia:2011Nov} this 
magnetospheric radius is 0.32 R$_{*}$.
To compare this to a T Tauri-like case, we also compute models with magnetospheres truncated where a 1kG
  field would imply. For HD 98922 it is 4.1 R$_{*}$, and 
  for V1295 Aql, we have tested various accretion rates and their corresponding magnetospheric truncation
  radii; we discuss our preferred lower accretion rate (10$^{-7}$ and 10$^{-8}$ M$_{\odot}$ yr$^{-1}$) models 
  with magnetospheres truncated at 4.4 R$_{*}$ below.
For both stars, we set the inner edge of the dust disk to their respective sublimation radii and 
the disk wind half-opening angle to 30$^{\circ}$ with respect to the stellar rotation axis; interior to this is
an optically thick gaseous disk component
that extends to the magnetosphere edge where the disk wind launches.

In Table \ref{t_tmodelparams}, we summarize key parameters adopted in our models and discuss for
each star below how they were
  chosen and the impact on the line profiles of their adjustment.
  In Fig. \ref{f_torusmodels}, we show calculated H$\alpha$ and H$\beta$ line profiles for cases
    we refer to as ``Herbig-like'' (with small/no magnetospheres) and ``T Tauri-like'' (with magnetospheres
    derived assuming 1kG fields). 
  Overall, we found a magnetospheric component
  necessary to produce the observed emission lines but note other choices that could mimic this effect.

\begin{deluxetable*}{l|cc|cc}
\tabletypesize{\scriptsize}
\tablecaption{Adopted TORUS model parameters\label{t_tmodelparams}}
\tablewidth{0pt}
\startdata
                                     & \multicolumn{2}{c}{V1295 Aql}          & \multicolumn{2}{c}{HD 98922} \\ \hline
System inclination [$^{\circ}$]        & \multicolumn{2}{c}{10, 34}            & \multicolumn{2}{c}{45}        \\
                                     & Herbig-like       & T Tauri-like       & Herbig-like & T Tauri-like    \\
\multicolumn{5}{c}{} \\
\multicolumn{5}{l}{Magnetosphere parameters} \\
R$_{inner}$ [R$_{*}$]                   & 1.0               & 3.0                & \nodata     & 3.0       \\
R$_{outer}$ [R$_{*}$]                   & 1.3               & 4.4                & \nodata     & 4.1       \\ 
T$_{max}$ [K]                          & 8000              & 8000               & \nodata     & 8000      \\
$\dot{M}_{acc}$ [M$_{\odot}$ yr$^{-1}$]   & 10$^{-5}$         & 10$^{-7}$, 10$^{-8}$ & \nodata     & 10$^{-6}$  \\ 
\multicolumn{5}{c}{} \\
\multicolumn{5}{l}{Disk wind parameters}                            \\
$\dot{M}_{wind}$ [M$_{\odot}$ yr$^{-1}$]  & \multicolumn{2}{c|}{10$^{-8}$}         & \multicolumn{2}{c}{10$^{-7}$}  \\ 
T$_{wind}$ [K]                          & \multicolumn{2}{c|}{8000}             & \multicolumn{2}{c}{8000}     \\
R$_{inner}$ [R$_{*}$]                   & \multicolumn{2}{c|}{4.4}               & \multicolumn{2}{c}{4.1}         \\
R$_{outer}$ [R$_{*}$]                   & \multicolumn{2}{c|}{10}                & \multicolumn{2}{c}{10, 20}     \\ 
$\theta$ [$^{\circ}$]                  & \multicolumn{2}{c|}{30}                & \multicolumn{2}{c}{30}        \\
\enddata
\tablecomments{Key parameters of the magnetosphere and disk wind for our models of V1295 Aql and
  HD 98922. Adopted stellar parameters appear in Table \ref{t_params}. We have assumed a magnetospheric 
  temperature structure as described in \citet{Hartmann:1994}, defining T$_{max}$ for both stars as
  8000 K. We assume an isothermal disk wind, and $\theta$ is the wind half-opening angle with respect to
  the stellar rotation axis. For all wind models, we set wind acceleration parameter $\beta$ to 0.2 and
    the acceleration length scale to 10 times the inner wind launching radius \citep[cf.][and
      references therein]{Kurosawa:2006}.
  References and derivation of these values discussed in sections devoted to each star.}
\end{deluxetable*}

\subsection{V1295 Aql}

Very narrow photospheric lines and little modulation of photometric and spectropolarimetric signals 
indicate either low inclination or a very long rotation period for
V1295 Aql \citep{Eisner:2004,Isella:2006,Catala:2007}.
Interferometric values reported in the literature poorly constrain system inclination: 
12-65$^\circ$ values are gotten (though the sparse $u-v$ coverage of both 
\citeauthor{Eisner:2004} \citeyear{Eisner:2004} and \citeauthor{Isella:2006} \citeyear{Isella:2006} 
does not permit ruling out a pole-on view or binarity) while \citet{Ragland:2012} derive a
highly inclined 72$^\circ$. For our TORUS models, we adopt 34$^\circ$ 
from analysis of new PIONIER data \citet{Lazareff:2017} and assess the line profile structure's
dependence on inclination.

\begin{figure}
\centering
\includegraphics[width=0.75\textwidth,angle=90]{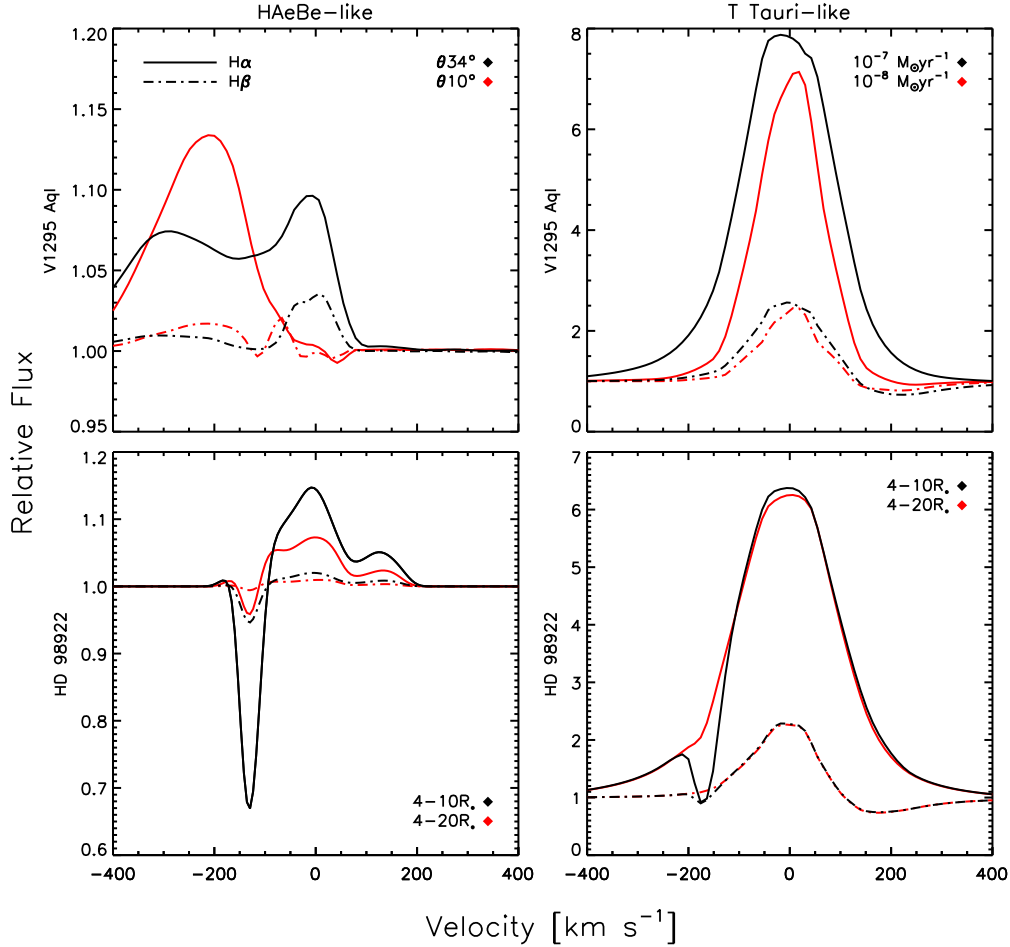}
\caption{Our TORUS model spectra comparing small/no magnetosphere models (``HAeBe-like,'' left
    column) and extended
    magnetospheres assuming 1kG fields (``T Tauri-like,'' right column) for V1295 Aql (top row) and HD 98922
    (bottom row). We present H$\alpha$ (solid) and H$\beta$ (dot-dashed) profiles; for V1295 Aql, we show
    the effect of system inclination and mass accretion rate choice on the line profiles, and for HD 98922
    we demonstrate how the
    size of the wind launching region effects absorption depth. For the T Tauri-like cases, 
    an inverse P Cygni profile is present in H$\alpha$ and becomes readily apparent in the H$\beta$ profiles,
    in disagreement with the observed spectra at all epochs.
Model parameters are summarized in Table \ref{t_tmodelparams}.}
\label{f_torusmodels}
\end{figure}

Assuming an accretion rate of 10$^{-5}$ M$_{\odot}$ yr$^{-1}$ and magnetic field strength of 100 G, we
  derive a magnetospheric extent of 0.32 R$_{*}$. In this ``Herbig-like'' model, we find weak contribution
to the line profile (Figure \ref{f_torusmodels}, upper left panel); emission
and absorption components at $\sim$10\% of the continuum level are
much weaker than observed in our spectra. At 10$^{\circ}$ inclination, higher velocity blueshifted material is all
  that is observed of the magnetospheric emission, its peak
  overlapping with the broad, faster, wind emission in H$\alpha$ but readily visible at $\sim -$70 km s$^{-1}$ in H$\beta$. At an
  inclination of 34$^{\circ}$, more of the accretion stream is visible--notably, the low-velocity material--and a
  distinct emission peak is seen near 0 km s$^{-1}$.
  In Figure \ref{f_torusmodels} (upper left panel), we show a potentially
  unrealistically low disk wind mass loss rate given the high stellar accretion rate adopted; our purpose here is to
  visually isolate the profile of the same wind component used in the T Tauri-like case for demonstration. We note that at
  higher disk wind
  mass loss rates, e.g. at 10\% of the mass accretion rate, the disk wind itself is redshifted from line center and
dominates the overall emission profile, its peak stronger than is observed in our data.

To model a T Tauri-like case, we tested an extended dipolar 
magnetosphere reaching from 3-4.4 R$_{*}$. We performed tests for various inner magnetospheric radii and accretion
rates to then reproduce the overall
  emission strength and velocity width observed in our spectra and assessed line profiles at various potential
  system inclination angles. We found a more modest mass
accretion rate of 10$^{-7}$ M$_{\odot}$ yr$^{-1}$, a couple orders of magnitude lower than that reported in the
literature, and an outer magnetospheric radius of 4.4 R$_{*}$ (calculated assuming 1kG field) was able to
reproduce comparable emission in the modeled lines to our observations (Figure \ref{f_torusmodels}, top
  right panel; black solid/dashed line profiles). A
 higher mass accretion rate as reported in the literature from a study not done in the same epoch
 as our observations combined with a more extended magnetosphere as is modeled here produced stronger emission
 than we observed; interestingly, for the parameter spaces we explored, lower accretion rates 
   of $\sim$10$^{-7}-$10$^{-8}$M$_{\odot}$ yr$^{-1}$ and smaller magnetospheric sizes produce strikingly
   similar profiles to our observations.

   In Figure \ref{f_torusmodels},
   upper right panel, red solid and dashed lines, we show the exact same model with an accretion rate one order of
   magnitude lower (10$^{-8}$ M$_{\odot}$ yr$^{-1}$) and note the line profile peak asymmetry shift from higher on the blue
   side of the line to higher on the red side of the line and an overall decrease in line strength more consistent with
   our observations. 
   Generally, we find in our models that higher accretion rates and more extended magnetospheres (i.e., with inner radii
   set closer to the star) 
   produce more symmetric profiles about the line's peak flux while more compact magnetospheres generate asymmetric peaks. 
   These asymmetries manifest at low accretion rates as 
   the distinctive, narrow, enhanced redward emission peak shape like observed in V1295 Aql.
Uncertainty in the Balmer excess measurement (the difference 
in flux in the U and B bands) of 0.1-0.2 dex can lead to one to two orders of magnitude uncertainty in the 
accretion rate, making our models' low accretion rates consistent with the measurements. The line profiles
observed by us during this epoch indicate it 
is highly unlikely either object observed is in a high mass accretion rate regime, as the line wings are 
relatively narrow ($\pm$200 km s${-1}$) and the continuum opacity is not high enough to have driven the 
lines into absorption \citep[e.g., the high temperature, high $\dot{M}$ models in][]{Muzerolle:2001}.
Finally, we also note the stellar mass cited by 
\citet{Mendigutia:2011Nov} is a factor of 2 higher than our adopted value; a revised stellar mass would likely 
lead to a lower mass accretion rate derived via the accretion shock models.

While the structure in the H$\alpha$ emission profile is highly suggestive, we are only 
able to produce the line profile at relatively low system inclinations ranging from 10$^{\circ}$-34$^{\circ}$.
Even so, the models' redshifted absorption is not present in the observed spectra and we are
  unable to reproduce the blueshifted absorption observed.
Since we have used an unrealistically large magnetosphere as a fiducial model (when the 100 G field 
strength and high measured mass accretion rate implies a $<$1 R$_{*}$ extent), a substantial fraction
of the emission 
could be originating from a hot, inner gaseous disk or a spherically symmetric wind launched from within
a few stellar radii of the star. Indeed, \citet{Catala:2007} were able to model the H$\alpha$ line using
a stellar wind and note the similarity between the observed profile and a familiar P Cygni profile. This
is to say there is a substantial degeneracy in modeling these line profiles: a stellar wind with a
compact launch/emission region can mimic the appearance of a magnetosphere.
Modeling of these systems requires information about the stellar magnetic field to break these
  degeneracies.
Finally, at the low system
  inclination at which a magnetosphere closest matches the emission line profile, in order to produce
  the observed blueshifted absorption, a disk wind launching angle steeper than the critical angle beyond
  which a flow cannot be initiated \citep[e.g., 30$^{\circ}$ from the disk normal vector][]{Blandford:1982}
  would need be invoked. We tested various disk wind geometries, varying the half-opening angle from
    30$^{\circ}$-45$^{\circ}$ and the width of the wind launching region (which then changes the density along
    the line of sight) and even for the most shallow half-opening angle and narrowest case tested (shown in
    Figure \ref{f_torusmodels}), no blueshifted absorption was produced. Thus, we conclude a stellar wind
  scenario is more consistent with all of the data available to us.

\subsection{HD 98922}

In Figure \ref{f_torusmodels} we present ``Herbig-like'' and ``T Tauri-like'' models of HD 98922 in the left
    and right bottom panels respectively; the left panel is simply the same model as in the right, but with the magnetosphere
  component turned off.

For HD 98922, we performed a similar calculation of the magnetospheric truncation radius assuming
a TTS-like, 1kG field and
adopting a mass accretion rate of rate of 10$^{-6}$ M$_{\odot}$ yr$^{-1}$ \citep{GL:2006}; these
parameters set the outer magnetospheric radius to 4.1 R$_{*}$. We initially adopted parameters
from the literature for the disk wind, including an isothermal temperature of 10,000 K
\citep{Caratti:2015}  
and a disk wind mass loss rate of 10$^{-7}$ M$_{\odot}$ yr$^{-1}$, an 
  order of magnitude lower than the accretion rate \citep[consistent with the model of][]{Caratti:2015}.
  In these models, we also initially chose a system inclination of 20$^{\circ}$ and disk wind half opening angle
(with respect to the stellar
  rotation axis, assumed to be perpendicular to the disk)
of 30$^{\circ}$ as done by \citet{Caratti:2015}, but find the disk wind does not provide enough opacity along the line
  of sight and thus doesn't produce blueshifted absorption observed in our spectra. In the models shown here,
  therefore, we adopt the 45$^{\circ}$ system inclination reported by \citet{Blondel:2006}.
  We found at this steeper inclination a 10,000 K disk wind still did not absorb; we adjusted the wind
    temperature until it did begin to absorb at 8000 K and have set the temperature to that value in all models shown
    here. We note an even cooler still 7000 K disk wind produces very deep absorption, to $\sim$0.01\% of the continuum level,
  which we do not observe an any epoch of our spectra.

  At its slightly 
higher system inclination, we are easily able to produce blueshifted absorption in the 
HD 98922 models with a disk wind. In Figure \ref{f_torusmodels} (bottom two panels), we show the effect of
  changing the disk wind launching region width on the depth of the absorption profile. A narrower launch region
  increases density along the line of sight and produces deeper absorption. This effect can be mimicked by altering
the system inclination or the mass loss rate.
We have calculated observationally and physically motivated models without a magnetosphere, as
would be suggested by the non-detection of this
  object in spectropolarimetric surveys, but found no combination of disk wind or hot inner disk components
  able to produce substantial emission above the continuum level; indeed, the disk wind itself produces a
  multiply-peaked profile not observed in any line in our spectra
  (Figure \ref{f_torusmodels}, lower left panel).

Our T Tauri-like line profile models for HD 98922 (Fig. \ref{f_torusmodels}) are better able to reproduce
  observations than in the case of V1295 Aql,
though we find the depth of the disk wind absorption highly dependent on the disk wind launching region extent.
  Additionally, with a symmetric disk wind, we are unable to reproduce the multiple absorption components and substructure
  observed in our spectra. This model, while it coarsely resembles our data, includes a
  magnetosphere and its attendant inverse P Cygni absorption which is not physically realistic given the magnetic
  field non-detection status of HD 98922. It is unclear what could be producing the observed emission, though a source
  spatially compact and close to the star that is unlikely to be a magnetosphere could be invoked. A more complex disk
  wind, incorporating asymmetric regions and temperature stratification, could be imagined.

A combination of factors suggest a disk wind is producing the observed features in HD 98922: 
the interferometrically determined system inclination provides a more favorable geometry for a disk wind to 
cross our line of sight, and the escape velocity in the blueward absorption component of the spectral lines 
is lower than the stellar escape velocity ($v_{esc}$sin$i \sim$295 km s$^{-1}$) and thus more consistent with a radius 
farther from the star.

\section{C. Time evolution of line component equivalent widths and absorption feature substructure}\label{s_appendixS}

\begin{figure}
    \includegraphics[width=0.3\textwidth,clip,trim=0cm 21cm 0cm 0.7cm]{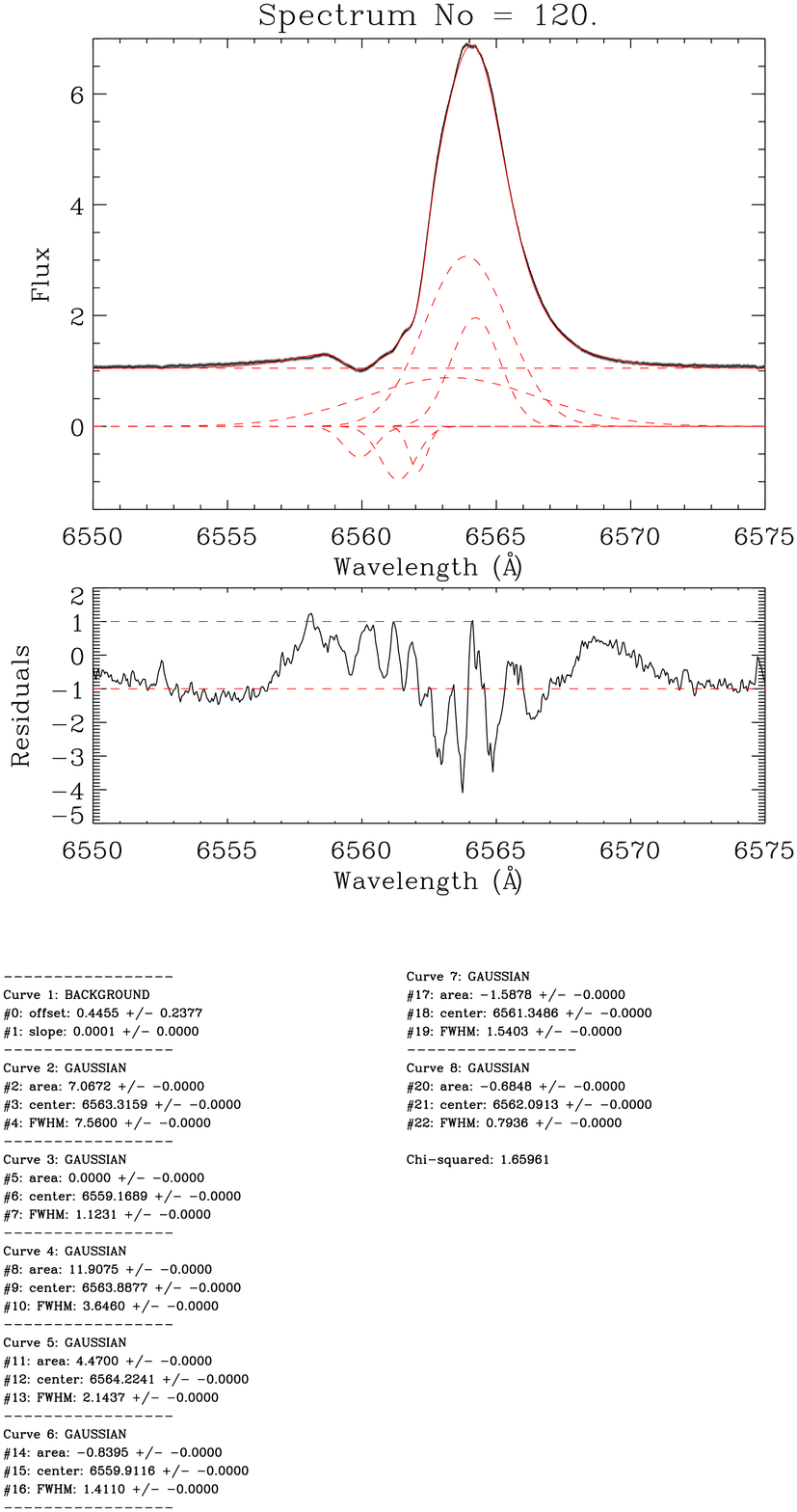}
    \includegraphics[width=0.3\textwidth,clip,trim=0cm 19.5cm 0cm 0.7cm]{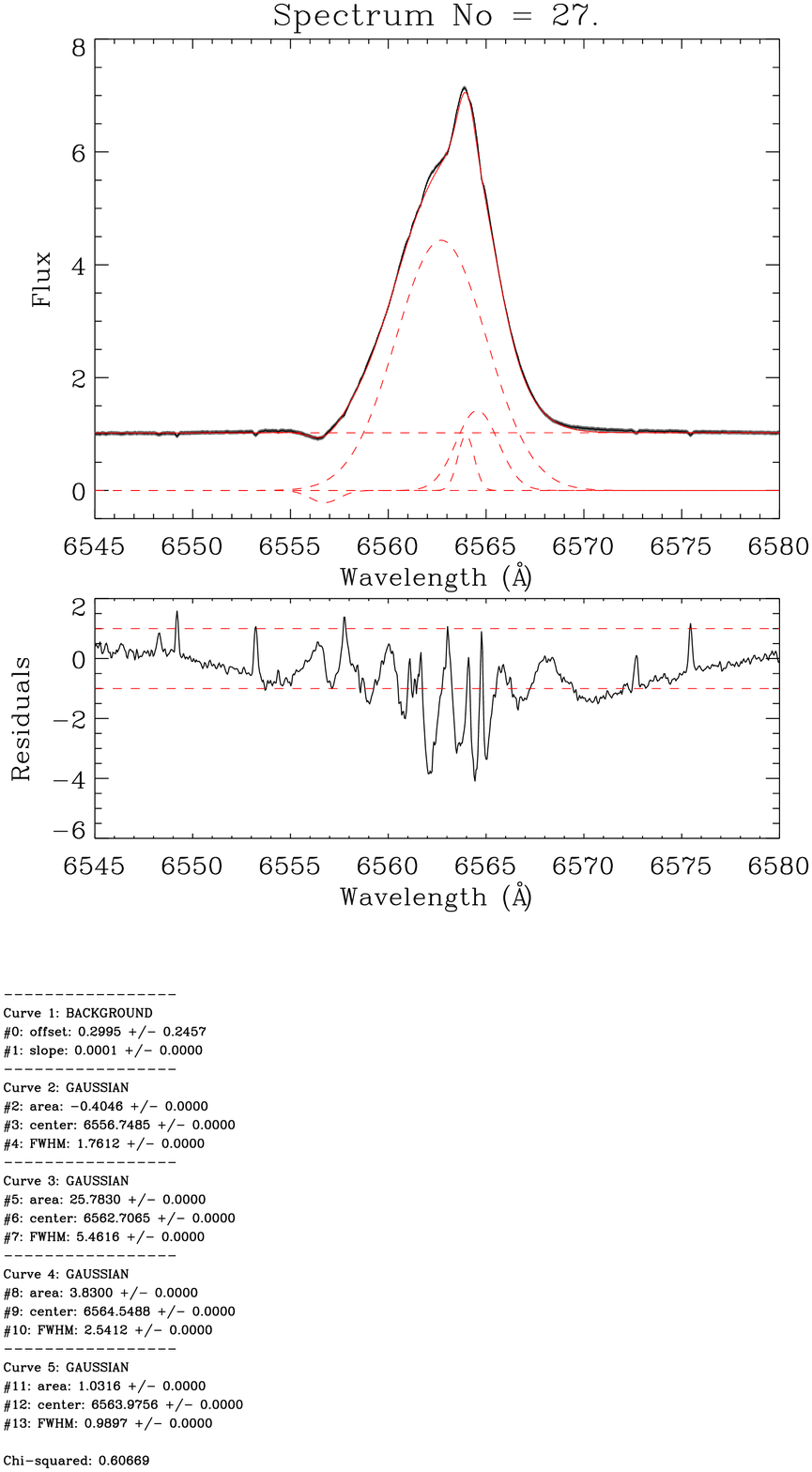}
    \caption{Gaussian profile fits to observed H$\alpha$ lines of HD 98922 (left) and V1295 Aql (right) using
      Peak ANalysis (PAN). Individual fit components (red, dashed lines) and total fit (red, solid line) are
    shown along with a single observed line profile (black, solid line).}
    \label{f_hagaussfits}
  \end{figure}

As described in Section \ref{s_analysis}, we performed line profile analysis by fitting Gaussian and
  background functions to our spectral line profiles in order to determine the overall line equivalent
  width and to assess potential substructures contributing to the line profile. In Figure \ref{f_hagaussfits}
  we show an example of this technique for H$\alpha$ line profiles of HD 98922 and V1295 Aql.

\begin{figure}
\includegraphics[width=0.26\textwidth,angle=90]{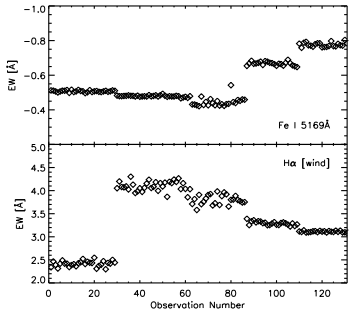}
\caption{Comparison of the time evolution of the Fe \textsc{II} (42) 5169\AA\ emission line
  compared to the wind absorption component of H$\alpha$ in HD 98922. As the Fe II emission strengthens
  (increasingly negative equivalent width, per convention), the H$\alpha$ wind absorption is growing
  weaker. Here (and in Figures \ref{f_v1295corrmap} and \ref{f_hd98922corrmap}), we have accounted for
  sign convention and see this represented as anticorrelation with a coefficient of -0.53.}
\label{f_hd98922ewcomp}
\end{figure}

Here, in Figure \ref{f_hd98922ewcomp}, we supply an example demonstrating the time-evolving equivalent width correlation calculations as seen in Figs. \ref{f_v1295corrmap}
 and \ref{f_hd98922corrmap}. The H$\alpha$ data are non-monotonic between the first two nights, but our test statistic detects and returns that the bulk of the EW evolution is monotonic and anticorrelated with Fe \textsc{II} (42) 5018\AA.
 
\begin{figure}
\centering
\includegraphics[width=0.3\textwidth]{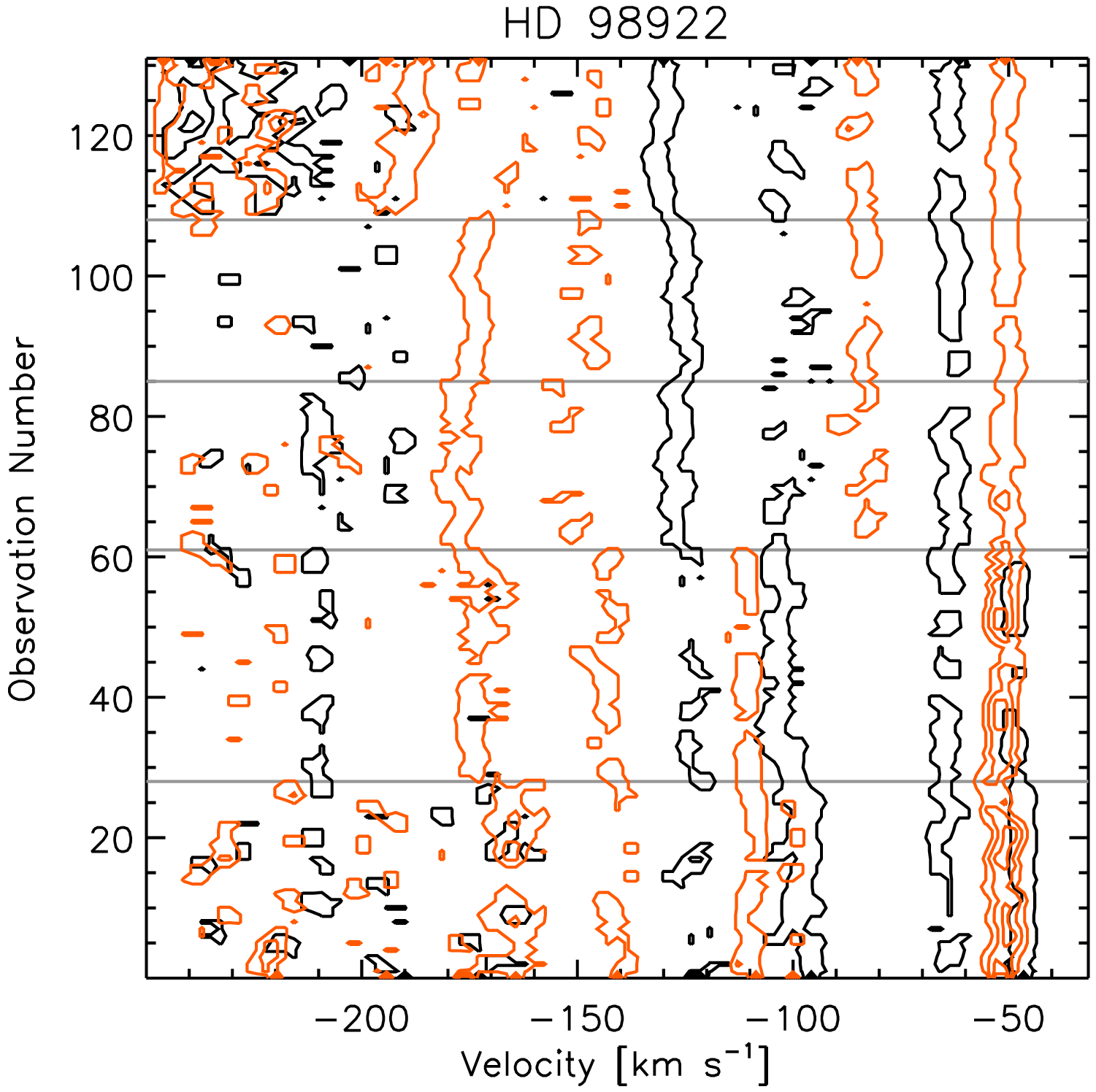}
\includegraphics[width=0.3\textwidth]{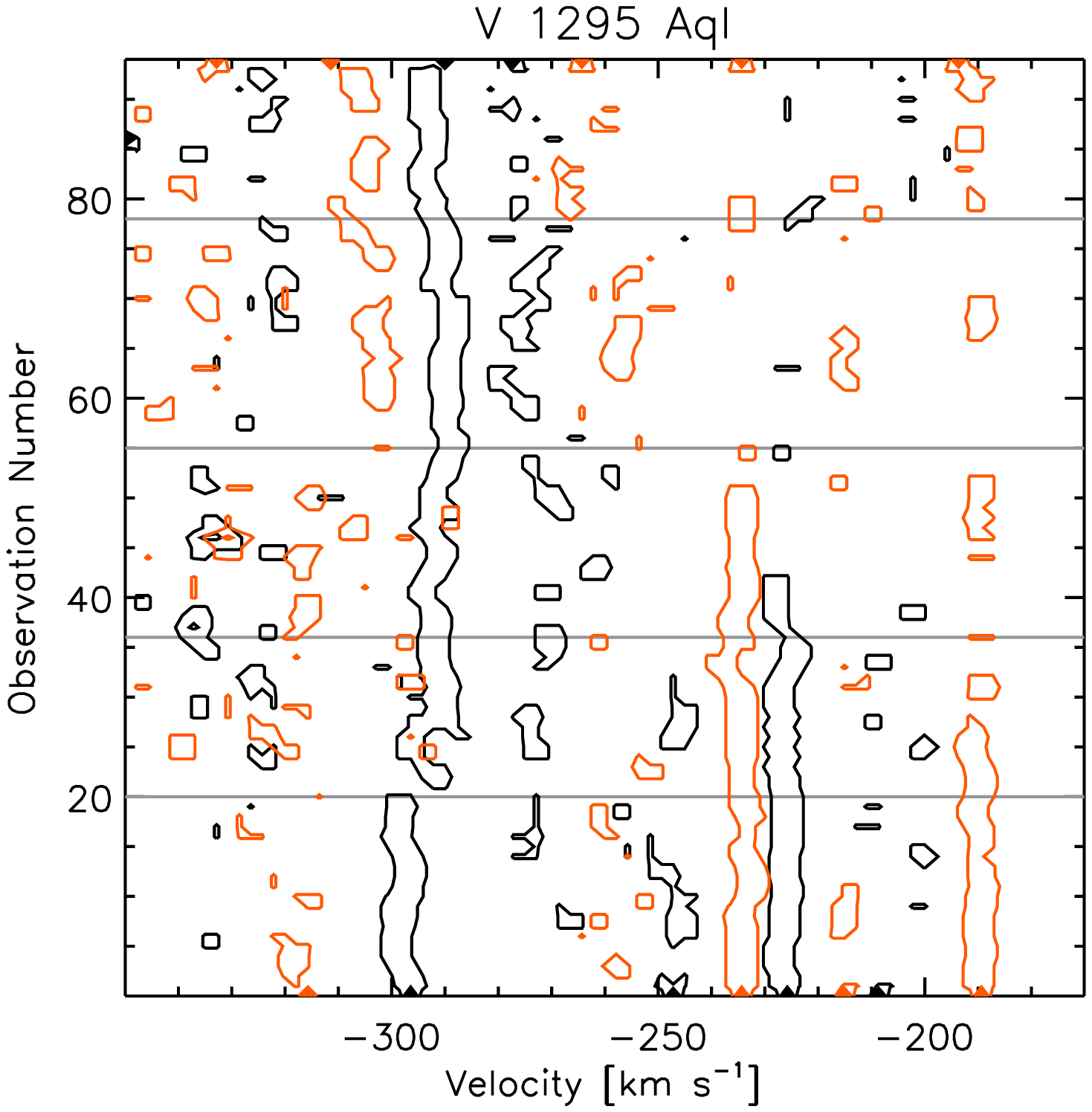}
\caption{Tracing weak absorption (black contours) and emission components (orange contours) within
  the blueshifted absorption of HD 98922 (left) and V1295 Aql (right). While noise does interfere
  with the tracing of these weak features, several distinct velocity components persist over multiple
  nights of observation (demarcated by horizontal lines), either remaining static at one velocity or
  evolving over the course of our observations.}\label{f_haabslmm}
\end{figure}

Both static and evolving features are observed within the absorption components
of H$\alpha$ in V1295 Aql and HD 98922 (Fig. \ref{f_haabslmm}). These features, seen
both in emission and absorption, are not telluric in origin. We have employed a local
minimum and maximum finding algorithm in order to quantify the velocity evolution of
these features.
Potentially, static components could be associated with slower-evolving disk wind
  structure, while the higher velocity accelerating and decelerating features could 
  be structures in faster-moving flow like a stellar wind.

\bibliographystyle{apj}
\bibliography{refs}

\end{document}

%% file: table_obslog_cropped.txt
\begin{deluxetable*}{lcccc}
\tabletypesize{\scriptsize}
\tablecaption{Time series observing log\label{t_timeseries}}
\tablewidth{0pt}
\tablehead{\colhead{Object} & UT Start   & Blue/Red & Integration & Airmass \\
           \colhead{ID}     & [hh:mm:ss] & Side     & Time [s]    &          }
\startdata
\multicolumn{5}{c}{2005 September 18, UVES archival spectra} \\ \hline
V1295 Aql   & 01:25:49.04  &  6650$-$10250\AA\ & 200  & \nodata	  \\
V1295 Aql   & 01:25:53.13  &  3044$-$3916\AA\ & 360   & \nodata          \\  
V1295 Aql   & 01:25:53.13  &  3732$-$4999\AA\ & 175   & \nodata	  \\
V1295 Aql   & 01:34:59.91  &  4726$-$6835\AA\ & 160   & \nodata	  \\
\multicolumn{5}{c}{2006 August 01, 1'' slit, 1x3 binning, slow readout} \\ \hline
V1295 Aql   & 07:53:32	    &  B    &  50    & 2.325   \\   
V1295 Aql   & 07:57:19	    &  B    &  25    & 2.392   \\
V1295 Aql   & 07:58:38	    &  B    &  25    & 2.416   \\
V1295 Aql   & 07:53:33	    &  R    &  50    & 2.325   \\   
V1295 Aql   & 07:57:19	    &  R    &  25    & 2.392   \\
V1295 Aql   & 07:58:45	    &  R    &  25    & 2.418   \\
V1295 Aql   & 08:03:22	    &  R    &  50    & 2.509   \\
\multicolumn{5}{c}{2008 November 10, HARPS archival spectra} \\ \hline
V1295 Aql   & 23:48:09	    & 3781$-$6912\AA\ & 600  & 1.527  \\ 
V1295 Aql   & 23:58:42	    & 3781$-$6912\AA\ & 600  & 1.586  \\ 
\multicolumn{5}{c}{2008 November 11, HARPS archival spectra} \\ \hline
V1295 Aql   & 00:09:15	    & 3781$-$6912\AA\ & 600  & 1.653  \\ 
V1295 Aql   & 00:19:47	    & 3781$-$6912\AA\ & 600  & 1.730  \\
V1295 Aql   & 00:30:19	    & 3781$-$6912\AA\ & 600  & 1.819  \\
V1295 Aql   & 00:40:52	    & 3781$-$6912\AA\ & 600  & 1.922  \\
V1295 Aql   & 00:51:25	    & 3781$-$6912\AA\ & 600  & 2.042  \\
V1295 Aql   & 01:01:58	    & 3781$-$6912\AA\ & 600  & 2.185  \\
\multicolumn{5}{c}{2010 July 30, 1x1 binning, fast readout} \\ \hline
HD 98922     & 23:40:52     & B      & 15   & 1.564  \\   
HD 98922     & 23:45:49     & B      & 80   & 1.588  \\
HD 98922     & 23:48:49     & B      & 80   & 1.604  \\
HD 98922     & 23:51:56     & B      & 80   & 1.621  \\
HD 98922     & 23:54:53     & B      & 80   & 1.637  \\
HD 98922     & 23:40:54     & R      & 15   & 1.564  \\	  
HD 98922     & 23:45:50     & R      & 40   & 1.588  \\
HD 98922     & 23:47:52     & R      & 40   & 1.599  \\
HD 98922     & 23:50:12     & R      & 40   & 1.611  \\
HD 98922     & 23:53:19     & R      & 40   & 1.628  \\
HD 98922     & 23:56:20     & R      & 40   & 1.645  \\
\multicolumn{5}{c}{\nodata Table can be viewed in full in online data \nodata} \\
\enddata
\tablecomments{Multi-epoch and time-series observations of HD 98922 and V1295 Aquila. 
All data are MIKE spectra, unless otherwise indicated. Time-series spectra were generally 
taken in sets of four integrations, with ThAr lamps taken between sets.}
\end{deluxetable*}